\documentclass[prd,nofootinbib,onecolumn,superscriptaddress,preprintnumbers]{revtex4-2}
\usepackage[T1]{fontenc}
\usepackage{lmodern}
\usepackage{amsmath,amssymb,amsthm,mathtools,cancel}
\usepackage{bm,mathrsfs,slashed}
\usepackage{graphicx,epsfig,float,adjustbox}
\usepackage[normalem]{ulem}
\usepackage{subcaption}
\usepackage{multirow}
\usepackage{tikz,xcolor}
\usepackage{natbib}
\usepackage{booktabs}
\usepackage[colorlinks=true,citecolor=black,linkcolor=black,urlcolor=black]{hyperref}
\usepackage[nameinlink]{cleveref}
\usepackage[includemp,
paperwidth=21.5cm,
paperheight=29.7cm,
top=2.30cm,
bottom=2.3cm,
inner=3cm,
outer=2.0cm,
marginparwidth=0.3cm,
marginparsep=0.3cm]{geometry}

\definecolor{changebrown}{RGB}{120,72,24}

\Crefname{figure}{Fig.}{Figs.}

\definecolor{Gray}{gray}{0.9}
\definecolor{lime}{HTML}{A6CE39}
\DeclareRobustCommand{\orcidicon}{%
  \begin{tikzpicture}
    \draw[lime, fill=lime] (0,0) circle [radius=0.2]
      node[white] {{\fontfamily{qag}\selectfont \tiny ID}};
    \draw[white, fill=white] (-0.0625,0.095) circle [radius=0.007];
  \end{tikzpicture}%
  \hspace{-2mm}%
}
\foreach \x in {A, ..., Z}{\expandafter\xdef\csname orcid\x\endcsname{\noexpand\href{https://orcid.org/\csname orcidauthor\x\endcsname}{\noexpand\orcidicon}}}

\newcommand{\be}{\begin{equation}}
\newcommand{\ee}{\end{equation}}
\newcommand{\bea}{\begin{eqnarray}}
\newcommand{\eea}{\end{eqnarray}}
\newcommand{\Beq}{\begin{align}}
\newcommand{\Eeq}{\end{align}}
\newcommand{\beq}{\begin{equation}\begin{aligned}}
\newcommand{\eeq}{\end{aligned}\end{equation}}

\newcommand{\bse}{\begin{subequations}}
\newcommand{\ese}{\end{subequations}}

\newcommand{\Pg}{\mathcal P_g}
\newcommand{\PS}{\mathcal P^{S}}
\newcommand{\Om}{\Omega_{\rm GW}}
\newcommand{\fbar}{\bar f_{\rm NL}}

\begin{document}
\preprint{TTK-26-27}

\title{Impact of Scale-dependent Primordial Non-Gaussianity \\ on Scalar-induced Gravitational Waves}

\author{Alisha Marriott-Best\orcidB{}}
\email{a.marriottbest@gmail.com}
\affiliation{Physics Department, Swansea University, SA2 8PP, United Kingdom}

\author{Devanshu Sharma\orcidC{}}
\email{drsharma@physik.rwth-aachen.de}
\affiliation{Institute for Theoretical Particle Physics and Cosmology (TTK), RWTH Aachen University,
D-52056 Aachen, Germany}

\author{Anish Ghoshal\orcidB{}}
\email{a.ghoshal@sussex.ac.uk}
\affiliation{Department of Physics and Astronomy, University of Sussex, \\
Brighton, BN1 9RH, United Kingdom }

\author{Gianmassimo Tasinato}
\affiliation{Physics Department, Swansea University, SA2 8PP, United Kingdom}
\affiliation{ Dipartimento di Fisica e Astronomia, Universit\`a di Bologna,\\
 INFN, Sezione di Bologna,  viale B. Pichat 6/2, 40127 Bologna,   Italy}

\begin{abstract}
Scalar-induced gravitational waves (SIGWs) provide a frequency-space probe of primordial curvature perturbations on scales far smaller than those accessible to the cosmic microwave background, while primordial black holes (PBHs) probe the rare tail of the same underlying statistics. We investigate whether the spectral morphology of the induced gravitational-wave background can reveal scale dependence in primordial non-Gaussianity. We adopt a specified internal-leg separable quadratic kernel,  $\mathcal F_{\rm NL} (\mathbf{k};\mathbf q,\mathbf{k}-\mathbf q) = \bar f_{\rm NL}\, f(q)f(\lVert\mathbf{k}-\mathbf q\rVert),$ and compute the Gaussian, reducible, and connected primordial-statistics contributions through $\mathcal O(\bar f_{\rm NL}^{4})$ within the second-order tensor-source approximation. We compare two phenomenologically distinct ultraviolet behaviors. A power-law leg weight describes persistent running over the momentum range sampled by the scalar source, whereas a smooth tanh weight models a finite transition followed by ultraviolet saturation. Across narrow, finite-width, asymmetric, and multi-slope scalar spectra, we find that scale dependence is not equivalent to an overall amplitude renormalization. For sources with finite width or extended momentum support, power-law running shifts the effective scalar support sampled by the non-Gaussian convolutions and can produce peak displacement, asymmetric shoulders, and a persistent deformation of the ultraviolet flank. The UV-matched tanh template instead generates a transition-localized modification and approaches a finite momentum-independent plateau at sufficiently large scalar momentum. The contrast between these templates shows that ultraviolet-tail deformations can retain information about the momentum dependence of the primordial interaction and are not a generic consequence of non-Gaussianity alone. We support the numerical results with analytical estimates for the perturbative hierarchy, effective local-slope shifts, log-normal saddle displacement, peak shifts, and deep-infrared scaling. These estimates explain why the $\mathcal O(\bar f_{\rm NL}^{4})$ sector is especially sensitive to running and why, for compact or sufficiently localized scalar sources with finite weighted moments, smooth scale dependence preserves the leading $k^{3}$ infrared behavior up to diagram-dependent logarithmic corrections. Finally, we formulate a conditional connection to PBH phenomenology. Since the characteristic PBH mass scales approximately as $M\propto k^{-2}$, enhanced high-wavenumber weighting in the SIGW convolution corresponds, within a narrow-support interpretation, to stronger positive-PNG weighting toward lower PBH masses. Our results identify spectral curvature, peak displacement, shoulder structure, and ultraviolet slopes as promising template-level observables for future multi-band searches with PTAs, LISA, and third-generation ground-based detectors, and provide controlled templates for testing primordial interactions on scales inaccessible to conventional cosmological probes.
\end{abstract}

\maketitle

\section{Introduction}
Scalar-induced gravitational waves (SIGWs) convert otherwise inaccessible small-scale primordial fluctuations into an observable stochastic background. {Unlike cosmic microwave background and large-scale structure measurements, which constrain comparatively large comoving scales, SIGWs can probe enhanced scalar perturbations associated with primordial black hole production and other strongly scale-dependent early-Universe dynamics.} Their scientific value lies not only in the possible detection of a stochastic signal, but in the possibility of reconstructing the statistics and momentum dependence of the primordial source from the frequency dependence of $\Omega_{\rm GW}$. The Gaussian SIGW calculation is well established~\cite{Kohri:2018awv,Espinosa:2018eve,Domenech:2021ztg}; the next challenge is to determine which spectral information survives when the primordial fluctuations are non-Gaussian.

This question has become observationally very timely. Pulsar-timing arrays probe the nanohertz band~\cite{NANOGrav:2023gor,EPTA:2023fyk}, while LISA, DECIGO, BBO, Einstein Telescope, and related experiments will extend stochastic-background searches across many additional decades in frequency~\cite{LISACosmologyWorkingGroup:2022jok,LISA:2024hlh,Baker:2019nia,Punturo:2010zz,Hild:2010id,ET:2025xjr}. If a primordial background is observed in more than one frequency interval, peak position, spectral curvature, shoulders, and ultraviolet fall-off will carry substantially more information than a single amplitude measurement.

Primordial non-Gaussianity (PNG) is essential in this regime. Constant local-type non-Gaussianity modifies both disconnected and connected contributions to the induced tensor spectrum~\cite{Adshead:2021hnm,Li:2023qua,Yuan:2020iwf}. The same mode coupling reshapes the rare tail controlling PBH production~\cite{Young:2013oia,Franciolini:2018vbk,Atal:2019cdz,Young:2022phe,Ferrante:2022mui}. For sharply peaked scalar spectra, perturbative treatments have organized the bispectrum and trispectrum corrections into effective cumulant combinations relevant to the PBH mass fraction~\cite{Matsubara:2022nbr}. The sign of local non-Gaussianity is also physically important: positive $f_{\rm NL}$ enhances the abundance in the usual local mapping, and non-attractor single-field realizations have been argued to yield positive $f_{\rm NL}$ at the power-spectrum peak~\cite{Firouzjahi:2023ahg}. Complementary nonperturbative constructions can instead generate heavy tails and copious PBHs without a large scalar power spectrum and with specific characteristics (see e.g.~\cite{Ezquiaga:2018gbw,Franciolini:2018vbk,Passaglia:2018ixg, Ezquiaga:2019ftu,Young:2019yug,DeLuca:2019qsy,Yoo:2019pma,Atal:2019erb,Franciolini:2023pbf}), although such scenarios can face strong clustering and isocurvature constraints~\cite{Chen:2025alpPBH}. SIGWs and PBHs are therefore complementary: SIGWs probe weighted convolution integrals of primordial correlators, whereas PBHs probe exponentially rare threshold-crossing configurations. A distinct observable channel arises from GWs sourced by the density fluctuations of an ultralight PBH population: scale-dependent local non-Gaussianity can modify PBH clustering and generate characteristic multi-feature spectra in that setting~\cite{papanikolaou2024new}. This mechanism is complementary to, but physically different from, the scalar-induced background calculated in the present work. A consistent joint description must use the same primordial kernel for both observables.

There is no reason for the effective nonlinear coupling on small scales to be constant. Multi-field evolution, spectator dynamics, transient inflationary phases, and post-horizon conversion can generate scale-dependent PNG, see e.g.~\cite{Chen:2005fe,Byrnes:2008zy,Byrnes:2009pe,Sefusatti:2009xu,Byrnes:2010ft}. Running then does more than rescale the SIGW amplitude: it changes which scalar momenta dominate the non-Gaussian convolutions and can shift peaks, create shoulders, and alter ultraviolet slopes. Recent works already indicate that scale-dependent non-Gaussianity can be relevant in PBH and SIGW physics, see e.g. \cite{Bartolo:2019zvb,Escriva:2025ftp,Namjoo:2025hrr}. This motivates the central question of this paper: \emph{can SIGW morphology discriminate a running primordial kernel from constant non-Gaussianity and from a change in the scalar source itself?}

Three strands of the literature provide the relevant baselines. Phenomenological power-law running of the bispectrum amplitude was introduced and forecast for CMB and large-scale-structure measurements in Refs.~\cite{Sefusatti:2009xu,Becker:2010nm}, while {scale dependence in quasi-local multifield models} from multifield evolution and self-interacting spectator sectors was developed in Refs.~\cite{Byrnes:2010ft,Byrnes:2008zy}. For SIGWs, early work identified the possibility that a subdominant non-Gaussian scalar component can generate a dominant induced-GW contribution and shift spectral features, see e.g.~\cite{Unal:2018yaa,Cai:2018dig}. Constant local non-Gaussianity and its connected diagrams have subsequently been studied systematically, see e.g.~\cite{Adshead:2021hnm,Yuan:2020iwf,Ragavendra:2021qdu,Ferrante:2022mui,Li:2023qua,Li:2023tup,Li:2023xtl,Yuan:2023ofl,LISACosmologyWorkingGroup:2024hsc,Perna:2024ehx,Zhou:2024ncc,Picard:2024ekd,Zhao:2024gan,Iovino:2024sgs,LISACosmologyWorkingGroup:2025vdz,Li:2025met,Zeng:2025cer,Iovino:2025cdy,Veermae:2026yzz}, and recent work has emphasized both the rich scale and shape dependence generated in transient non-attractor dynamics~\cite{Tasinato:2023ukp} and the importance of higher-order non-Gaussianity beyond a finite diagrammatic truncation~\cite{Zeng:2025allorder}. The novelty here is complementary to these developments: we isolate a definite internal-leg separable kernel, propagate its momentum dependence through the complete Gaussian, reducible, and connected sectors up to $\mathcal O(\bar f_{\rm NL}^4)$, compare growing and saturating scale weights over a wide family of scalar-source morphologies, and derive analytic diagnostics that explain when the resulting SIGW deformation is a genuine shape effect rather than an amplitude shift.

We address this question using the internal-leg separable ansatz
\[
\zeta(\mathbf{k})=\zeta_g(\mathbf{k})+\bar f_{\rm NL}\int\frac{d^3q}{(2\pi)^{3/2}}f(q)f(\lVert\mathbf{k}-\mathbf q\rVert)\zeta_g(\mathbf q)\zeta_g(\mathbf{k}-\mathbf q).
\]
The overall nonlinear amplitude is $\bar f_{\rm NL}$, while $f(p)$ is a dimensionless weight assigned to each Gaussian input leg. We denote the complete quadratic kernel by $\mathcal F_{\rm NL}(\mathbf{k};\mathbf q,\mathbf{k}-\mathbf q)=\bar f_{\rm NL}f(q)f(\lVert\mathbf{k}-\mathbf q\rVert)$ and introduce $\mathcal P^S(p)=f^2(p)\mathcal P_g(p)$ only as a contraction-level bookkeeping device. This prescription is less general than an arbitrary triangle-dependent kernel, but it is explicit, computationally tractable, and implemented consistently in every numerical contribution. We calculate the Gaussian sector together with the reducible and connected primordial-statistics terms through $\mathcal O(\bar f_{\rm NL}^4)$ within the second-order tensor source.

We study power-law and UV-matched tanh leg weights across monochromatic, log-normal, ultra-slow-roll-inspired, Gaussian scalar bump, power-law-cutoff, broken-power-law, and double-broken-power-law sources. The two weights are chosen to bracket two ultraviolet behaviors rather than to claim a unique microscopic completion. The power law $f(k)=(k/k_\star)^{n_f}$ is the standard leading-log parametrization of a slowly running nonlinear amplitude: over a finite range of scales, renormalization-group evolution, slowly varying sound speed, or slowly evolving multifield transfer coefficients exponentiate a logarithmic correction into a power law~\cite{Sefusatti:2009xu,Byrnes:2010ft,Becker:2010nm}. It is therefore an appropriate local-in-$\ln k$ description as long as the range of validity and perturbative hierarchy are monitored. The tanh weight, by contrast, is a bounded smooth-step completion of a low-momentum power law. It represents a coupling or transfer efficiency that turns on across a finite interval and saturates once the relevant heavy-field, feature, or non-attractor transition has completed. Inflationary features and transient departures from attractor evolution generically produce localized and scale-dependent bispectra~\cite{Chen:2010xka,Tasinato:2023ukp}. More concretely, the effective field theory of inflation with a step in the potential or sound speed gives a self-consistent setting in which a finite-duration feature enhances and localizes the bispectrum~\cite{Bartolo:2013exa}, while transient reductions of the sound speed provide a theoretically controlled effective single-field realization with correlated power-spectrum and bispectrum features~\cite{Achucarro:2014msa}. These models motivate the physical picture of a coupling that changes over a finite interval of scales and then approaches a new asymptotic regime. The precise function $\tanh[(k/k_\star)^{n_f}]$ is nevertheless used here as a phenomenological smooth interpolation, not as the exact prediction of any  ultraviolet Lagrangian. This distinction is important: the power-law and tanh cases test, respectively, persistent running and ultraviolet saturation. The benchmark set isolates how width, asymmetry, and multiple local slopes control the response to running. To our knowledge, the combined comparison of these two ultraviolet behaviors, with all reducible and connected contributions retained through $\mathcal O(\bar f_{\rm NL}^4)$ and tested across this breadth of scalar-source shapes, has not previously been carried out for the internal-leg separable kernel. The source survey determines which signatures of running remain robust under changes in the underlying scalar morphology. Our principal finding is that a power-law leg weight changes the effective local slopes and moves the weighted support of a broad source, leading to an asymmetric enhancement of the peak and ultraviolet flank. The UV-matched tanh weight instead produces a transition-localized deformation and saturates toward the constant-weight normalization in the ultraviolet. The ultraviolet uplift is therefore not a generic consequence of non-Gaussianity, but a diagnostic of the momentum dependence assigned to the primordial interaction.

For an exact monochromatic source, every weighted Gaussian leg is fixed at $k_\star$, so a pivot-normalized power law cannot change the normalized sector shapes. The nontrivial narrow-source curves shown below arise from the finite-width regularization used numerically. This limit provides an important analytical check on both the kernel assignment and the code. More generally, analytical estimates for the hierarchy parameter, local-slope shift, log-normal saddle displacement, apparent peak shift, and deep-infrared scaling explain the numerical trends and motivate a normalized shape statistic that separates morphology from amplitude.

The phenomenological implication is that amplitude-only templates are insufficient. Searches for a primordial stochastic background should fit turnover, curvature, shoulders, and ultraviolet slope together with the amplitude. Such observables can reduce degeneracies between scalar-source morphology and primordial momentum dependence. In a narrow-support PBH interpretation, positive power-law running also strengthens the effective positive-PNG weighting toward lower horizon masses because $M\propto k^{-2}$. This is a conditional consistency relation, not a PBH mass-function calculation.

{The paper is organized as follows. Section~\ref{sec:formalism} defines the SIGW calculation and the internal-leg separable kernel. Section~\ref{sec:templates} develops the dynamical motivation, running templates, and analytical estimates. Section~\ref{sec:results} presents the numerical benchmark survey, including the additional source classes collected in Sec.~\ref{sec:extra_templates}. Section~\ref{sec:phenom} discusses detector-facing shape observables, the conditional PBH connection of Sec.~\ref{sec:pbh_running_estimates}, and infrared behavior. Section~\ref{sec:concl} summarizes the implications. The appendices provide the detailed multifield realization (Appendix~\ref{app:toy_internal_leg_realization}), the Gaussian SIGW conventions (Appendix~\ref{sec:A}), the non-Gaussian diagrammatic decomposition (Appendix~\ref{app:ng_4_pt_diagram}), the numerical variables (Appendix~\ref{app:variables}), the full scale-dependent integral expressions (Appendix~\ref{app:scale_dependent_extension}), the exact monochromatic check (Appendix~\ref{app:exact_monochromatic_limit}), and the convergence and reproducibility tests (Appendix~\ref{app:numerical_stability}, including Sec.~\ref{app:reproducibility}).}

\section{Non-Gaussian Scalar-Induced Gravitational Waves}
\label{sec:formalism}

{We consider a spatially flat Friedmann-Lema\^itre-Robertson-Walker (FLRW) background.} Scalar-induced gravitational waves (SIGWs) are sourced at second order by the quadratic interactions of first-order scalar curvature perturbations \(\zeta\). {We adopt conformal Newtonian gauge and neglect first-order tensor and vector perturbations as well as anisotropic stress:}
\begin{equation}
ds^2 = a(\tau)^2 \left[ -(1+2\Phi)d\tau^2 + \left( (1-2\Phi)\delta_{ij} + \frac12 h_{ij} \right) dx^i dx^j \right],
\end{equation}
where \(\Phi = \Psi\) and \(h_{ij}\) is the transverse-traceless tensor perturbation at second order. We expand \(h_{ij}\) in Fourier space using the two polarization tensors \(\epsilon_{ij}^\lambda\) (\(\lambda = +, \times\)):
\begin{equation}
h_{ij}(\tau,\mathbf{x}) = \sum_{\lambda=+, \times} \int \frac{d^3k}{(2\pi)^{3/2}} e^{i\mathbf{k}\cdot\mathbf{x}} \epsilon_{ij}^\lambda(\mathbf{k}) h_\lambda(\tau,\mathbf{k}).
\end{equation}

The dimensionless GW energy-density spectrum at conformal time $\tau$ is
\begin{equation}
\Omega_{\rm GW}(\tau,k)=\frac{1}{48}\left(\frac{k}{\mathcal H(\tau)}\right)^2\sum_{\lambda=+,\times}\overline{\Delta_\lambda^2(\tau,k)}.
\end{equation}
where the overline denotes oscillation averaging and \(\Delta_\lambda^2(k)\) is the dimensionless power spectrum of the tensor mode \(h_\lambda\).

The equation of motion for the tensor modes sourced by scalars reads
\begin{equation}
h_\lambda''(\tau,\mathbf{k}) + 2\mathcal H h_\lambda'(\tau,\mathbf{k}) + k^2 h_\lambda(\tau,\mathbf{k}) = 4 S_\lambda(\tau,\mathbf{k}),
\end{equation}
with source
\begin{equation}
S_\lambda(\tau,\mathbf{k}) = \int \frac{d^3q}{(2\pi)^{3/2}} Q_\lambda(\mathbf{k},\mathbf{q}) {\mathcal K_{\rm src}(|\mathbf{k}-\mathbf{q}|,q,\tau)} \zeta(\mathbf{q}) \zeta(\mathbf{k}-\mathbf{q}).
\end{equation}
Here \(Q_\lambda(\mathbf{k},\mathbf{q}) = \epsilon_{ij}^\lambda(\mathbf{k}) q^i q^j\) is the projection factor, and {\(\mathcal K_{\rm src}(p,q,\tau)\) denotes the radiation-era source kernel built from scalar transfer functions \(\Phi(k\tau)\)} (see \cite{Kohri:2018awv,Espinosa:2018eve,Domenech:2021ztg} for explicit expressions in radiation domination).

{The retarded solution is obtained using the Green function:}
\begin{equation}
h_\lambda(\tau,\mathbf{k}) = \frac{4}{a(\tau)} \int_{\tau_0}^\tau d\tau' G_k(\tau,\tau') a(\tau') S_\lambda(\tau',\mathbf{k}),
\end{equation}
where \(G_k\) satisfies the homogeneous wave equation. In radiation domination, analytic expressions for \(G_k\) and \(\Phi\) are available \cite{Kohri:2018awv,Adshead:2021hnm}.

The GW power spectrum follows from the four-point function of \(\zeta\):
\begin{equation}
\langle h_{\lambda_1}(\mathbf{k}_1) h_{\lambda_2}(\mathbf{k}_2) \rangle = 16 \int \frac{d^3q_1 d^3q_2}{(2\pi)^3} \langle \zeta(\mathbf{q}_1) \zeta(\mathbf{k}_1-\mathbf{q}_1) \zeta(\mathbf{q}_2) \zeta(\mathbf{k}_2-\mathbf{q}_2) \rangle \, Q_{\lambda_1} Q_{\lambda_2} \, I(|\mathbf{k}_1-\mathbf{q}_1|,q_1) I(|\mathbf{k}_2-\mathbf{q}_2|,q_2),
\end{equation}
with kernel \(I(p,q,\tau)\) incorporating the Green's function and scalar transfer functions. In the Gaussian limit, Wick's theorem reduces the four-point function to products of two-point functions, yielding the standard integral over variables \(u = |\mathbf{k}-\mathbf{q}|/k\), \(v = q/k\) \cite{Kohri:2018awv,Espinosa:2018eve}:
\begin{equation}
\Omega_{\rm GW}^{\rm G}(k) = \frac{2}{3} \int_0^\infty dv \int_{|1-v|}^{1+v} du \, \overline{J^2(u,v,u,v)} \frac{\Delta_g^2(vk)}{v^2} \frac{\Delta_g^2(uk)}{u^2},
\end{equation}
where \(\overline{J^2}\) contains the known kernels \(I_A, I_B, I_C\) (see Refs.~\cite{Kohri:2018awv,Adshead:2021hnm,Li:2023qua} for full expressions).

\subsection{Non-Gaussian contributions and scale dependence}

For non-Gaussian \(\zeta\), the four-point function also contains connected contributions\footnote{See the corresponding discussion in Ref.~\cite{Adshead:2021hnm}.}. We first recall the constant local-type ansatz up to quadratic order:
\begin{equation}
\zeta(\mathbf{x}) = \zeta_g(\mathbf{x}) + F_{\rm NL} \bigl( \zeta_g^2(\mathbf{x}) - \langle \zeta_g^2 \rangle \bigr).
\end{equation}
Our convention is $F_{\rm NL}=3f_{\rm NL}/5$ relative to the commonly used CMB convention. In the scale-dependent model below, $\bar f_{\rm NL}$ is the reference value of this same quadratic coefficient, so the constant-weight limit is $F_{\rm NL}=\bar f_{\rm NL}$.
\begingroup
To avoid a convention ambiguity, $f_{\rm NL}$ without a bar refers only to the standard CMB local parameter, $F_{\rm NL}=3f_{\rm NL}/5$ denotes the constant real-space quadratic coefficient, and $\bar f_{\rm NL}$ denotes the reference amplitude multiplying the internal-leg kernel used in this paper. We retain $\bar f_{\rm NL}$ because it is the parameter used in the numerical configurations and figure labels; no equality between $\bar f_{\rm NL}$ and the conventional CMB $f_{\rm NL}$ is implied.
\endgroup
In Fourier space (dropping the zero-mode subtraction for the convolution),
\begin{equation}
\zeta(\mathbf{k}) = \zeta_g(\mathbf{k}) + F_{\rm NL} \int \frac{d^3q}{(2\pi)^{3/2}} \zeta_g(\mathbf{q}) \zeta_g(\mathbf{k}-\mathbf{q}).
\end{equation}

To incorporate scale dependence, we use
\begin{equation}
\zeta(\mathbf{k})=\zeta_g(\mathbf{k})+\bar f_{\rm NL}\int\frac{d^3q}{(2\pi)^{3/2}}f(q)f(\lVert\mathbf{k}-\mathbf q\rVert)\zeta_g(\mathbf q)\zeta_g(\mathbf{k}-\mathbf q).
\label{eq:internal_leg_kernel_main}
\end{equation}
This internal-leg separable kernel is symmetric in its Gaussian input momenta. The results apply to this specified kernel and not to an external-output prescription.
Expanding the four-point function and performing Wick contractions yields seven classes of integrals: one Gaussian term, two disconnected/hybrid/reducible terms at \(\mathcal O(\bar f_{\rm NL}^2)\) and \(\mathcal O(\bar f_{\rm NL}^4)\), and four connected terms (C, Z, planar, nonplanar) \cite{Adshead:2021hnm,Li:2023qua}. The explicit expressions (with variables \(u_i,v_i,s_i,t_i,\phi_{ij}\), etc.) are standard and given in the original constant-\(f_{\rm NL}\) analyses; we generalize them by replacing each scalar power spectrum associated with a weighted Gaussian leg by the rescaled power spectrum\footnote{ For compactness, some terms below are written using weighted spectra. This notation is only a bookkeeping shorthand for particular diagrammatic factors after the kernel has been specified. It should not be interpreted as a universal replacement rule $\mathcal P_g\to f^2\mathcal P_g$ for all correlators. }
\begingroup
We distinguish the dimensionful two-point spectrum from the dimensionless spectrum used in the SIGW integrals. With the symmetric Fourier convention adopted in Eq.~\eqref{eq:internal_leg_kernel_main}, define
\begin{equation}
\left\langle \zeta_g(\mathbf k)\zeta_g(\mathbf k')\right\rangle
=\delta^3(\mathbf k+\mathbf k')P_g(k),
\qquad
\mathcal P_g(k)\equiv\Delta_g^2(k)=\frac{k^3}{2\pi^2}P_g(k).
\label{eq:scalar_power_conventions_main}
\end{equation}
Here $P_g$ is dimensionful, whereas $\mathcal P_g=\Delta_g^2$ is dimensionless. Their weighted counterparts are
\begin{equation}
P_g^S(k)\equiv f^2(k)P_g(k),
\qquad
\mathcal P^S(k)\equiv f^2(k)\mathcal P_g(k).
\label{eq:rescaled_power}
\end{equation}
Unless a dimensionful spectrum $P$ is written explicitly, all subsequent $\mathcal P$ and $\Delta^2$ symbols denote dimensionless spectra.
\endgroup
where the superscript $S$ denotes the scale-weighted Gaussian leg. This notation has a direct diagrammatic meaning. Each $\mathcal P^S(p)$ marks a Gaussian two-point contraction whose momentum $p$ receives two factors of the leg weight, one from each occurrence of that Gaussian mode. At $\mathcal O(\bar f_{\rm NL}^2)$ the merged SIGW expressions therefore contain two weighted spectra and one unweighted spectrum, whereas the $\mathcal O(\bar f_{\rm NL}^4)$ sector contains four weighted spectra. The explicit momentum routings remain in Appendix~\ref{app:scale_dependent_extension}, while this counting explains the enhanced sensitivity of the fourth-order sector to running. The full curvature power spectrum receives a one-loop non-Gaussian correction
\begingroup
The one-loop correction is most compactly written for the dimensionful spectrum. Direct contraction of the two quadratic insertions in Eq.~\eqref{eq:internal_leg_kernel_main} gives
\begin{equation} \label{eq:full_power_loop}
P_\zeta(k)=P_g(k)+2\bar f_{\rm NL}^2
\int\frac{d^3q}{(2\pi)^3}
P_g^S(q)P_g^S(\lVert\mathbf k-\mathbf q\rVert).
\end{equation}
The factor $(2\pi)^{-3}$ follows from the two factors $(2\pi)^{-3/2}$ in the quadratic insertions after one Wick-contraction delta function removes the second loop integration. The corresponding dimensionless physical spectrum is $\mathcal P_\zeta(k)=k^3P_\zeta(k)/(2\pi^2)$; it should not be obtained by inserting dimensionless spectra directly into Eq.~\eqref{eq:full_power_loop} without the associated momentum factors.

All numerical comparisons in this work hold the Gaussian seed spectrum $\mathcal P_g$ fixed while varying the leg weight $f$ and the reference quadratic amplitude $\bar f_{\rm NL}$. Consequently, Eq.~\eqref{eq:full_power_loop} implies that the full physical spectrum $\mathcal P_\zeta$ is generally not fixed across those comparisons. The resulting change in the SIGW signal therefore contains both the altered loop-corrected two-point contribution and the altered connected higher-point contributions generated by the same primordial kernel. A comparison at fixed $\mathcal P_\zeta$ would require retuning $\mathcal P_g$ separately for each running template and constitutes a different matching prescription, which is not adopted here.
\begingroup
Thus, statements in this paper that compare running templates at a common scalar source mean a common Gaussian seed source. They do not claim that the higher-point kernel has been varied while the observable two-point spectrum is held exactly fixed. The present setup isolates the consequences of changing one specified quadratic mapping at fixed seed statistics; a fixed-$\mathcal P_\zeta$ analysis would instead isolate higher-point information more cleanly but would require a separate inverse matching of $P_g$ for every parameter choice.
\endgroup
\endgroup

Substituting into the SIGW integrals and merging connected/disconnected contributions yields compact expressions for the \(\mathcal O(\bar f_{\rm NL}^2)\) and \(\mathcal O(\bar f_{\rm NL}^4)\) corrections to \(\Omega_{\rm GW}\):
\begin{align} \label{GW1}
\Omega_{\rm GW}^{(1)}(k)
&=\bar f_{\rm NL}^{2}\,
\mathcal I_1\!\left[k;\mathcal P_g,\mathcal P^S\right],\\
\Omega_{\rm GW}^{(2)}(k)
&=\bar f_{\rm NL}^{4}\,
\mathcal I_2\!\left[k;\mathcal P^S\right],
\end{align}
where $\mathcal I_1$ denotes the sum of the hybrid, C, and Z integral functionals containing two weighted spectra and one unweighted spectrum, while $\mathcal I_2$ denotes the reducible, planar, and nonplanar integral functionals containing four weighted spectra. Their complete momentum-space expressions, including all numerical prefactors and transfer kernels, are given in Eqs.~\eqref{eq:omega_gw_fnl2} and \eqref{eq:omega_gw_fnl4}.

The full kernels involving the time-integrated transfer functions \(\overline{J(a_i,b_i)J(a_j,b_j)}\) are unchanged, since they encode propagation after horizon re-entry rather than primordial statistics \cite{Ellis:2023oxs}.
These integrals are evaluated numerically using a quasi-Monte Carlo (QMC) integrator built in \texttt{JAX} \cite{jax2018github}. Quasi-Monte Carlo and ordinary Monte Carlo integration are conceptually similar, but QMC uses deterministic low-discrepancy sequences rather than pseudo-random samples. In this work we use Sobol sequences, whose more uniform coverage of the integration domain reduces sampling error for smooth integrands \cite{QMCI-mor}.
The dominant connected terms (C and planar) often rival the disconnected contributions, especially in the infrared, and \(\mathcal O(\bar f_{\rm NL}^4)\) terms become essential when \(A_r \bar f_{\rm NL}^2 \gtrsim \mathcal O(1)\), as relevant for PBH formation. Scale dependence introduces momentum-dependent weighting that distorts the spectral shape in ways absent from constant-\(f_{\rm NL}\) calculations.

{Further technical details, including the diagrammatic decomposition and changes of variables, are given in Appendices~\ref{app:ng_4_pt_diagram}-\ref{app:scale_dependent_extension}.}

\subsection{Definition of the scale-dependent quadratic kernel}
Scale dependence may be attached either to a nonlinear output momentum or to the two Gaussian input legs. These prescriptions are inequivalent. This work adopts exclusively the internal-leg separable kernel in Eq.~\eqref{eq:internal_leg_kernel_main}, which is implemented numerically and in Appendix~\ref{app:scale_dependent_extension}. Every $\mathcal P^S(p)=f^2(p)\mathcal P_g(p)$ represents a weighted Gaussian momentum entering a quadratic insertion. The results are not predictions for an external-output kernel or an arbitrary running local bispectrum.
\medskip

\section{Templates for {Scale-dependent} Non-Gaussianity}
\label{sec:templates}

While a constant quadratic coefficient provides a useful baseline, inflationary dynamics generically motivate scale dependence. In multifield scenarios, models with evolving sound speed, and cases with post-horizon conversion, the effective nonlinear coupling can vary with wavenumber \cite{Byrnes:2010ft,Sefusatti:2009xu,Byrnes:2008zy,Becker:2010nm}.

We adopt
\begin{equation}\label{eq:zeta_running_ansatz}
\zeta(\mathbf{k})=\zeta_g(\mathbf{k})+\bar f_{\rm NL}\int\frac{d^3q}{(2\pi)^{3/2}}f(q)f(\lVert\mathbf{k}-\mathbf q\rVert)\zeta_g(\mathbf q)\zeta_g(\mathbf{k}-\mathbf q),
\end{equation}
with $\mathcal F_{\rm NL}=\bar f_{\rm NL}f(q)f(\lVert\mathbf{k}-\mathbf q\rVert)$. The function $f(p)$ is a Gaussian-leg weight, not a one-argument external-output nonlinearity parameter.
\begingroup
The running index $n_f$ therefore characterizes the scale dependence of a single Gaussian-leg transfer factor. It should not be identified directly with the running index of a conventional one-argument nonlinearity parameter. At tree level,
\begin{align}
B_\zeta(k_1,k_2,k_3)=2\bar f_{\rm NL}\big[&f(k_1)f(k_2)P_g(k_1)P_g(k_2)
+f(k_2)f(k_3)P_g(k_2)P_g(k_3)\nonumber\\
&+f(k_3)f(k_1)P_g(k_3)P_g(k_1)\big],
\label{eq:main_tree_bispectrum_running_interpretation}
\end{align}
up to the momentum-conserving delta function and Fourier normalization. In the equilateral configuration $k_1=k_2=k_3=k$, the effective quadratic amplitude scales as
\begin{equation}
F_{\rm NL}^{\rm eq}(k)\equiv \bar f_{\rm NL}f^2(k)
=\bar f_{\rm NL}\left(\frac{k}{k_\star}\right)^{2n_f}
\label{eq:equilateral_effective_running}
\end{equation}
for the power-law leg weight. Hence an equilateral one-argument running convention, $F_{\rm NL}^{\rm eq}\propto k^{n_{\rm NG}^{\rm eq}}$, would correspond to $n_{\rm NG}^{\rm eq}=2n_f$. This relation is configuration dependent: in a squeezed triangle the scale dependence is distributed separately between the long and short input legs and cannot be represented by the same single running index without specifying which momentum is varied.
\endgroup

\subsection{Dynamical motivation from multifield transfer}
\label{sec:multifield_kernel_motivation}

{The input-leg momentum dependence in Eq.~\eqref{eq:zeta_running_ansatz} arises naturally when scale-dependent linear transfer acts on a Gaussian degree of freedom before local nonlinear conversion. This ordering distinguishes the kernel used here from a prescription in which the scale-dependent coefficient depends only on the final output momentum.}

Consider an adiabatic perturbation and a spectator or entropy perturbation whose evolution is dominated, over the momentum interval of interest, by a common Gaussian seed. On a conversion hypersurface $t_c$, the transferred spectator mode can then be written as
\begin{equation}
s_{\mathbf k}(t_c)=\mathcal A_s\,\mathcal T_s(k)\,\zeta_{g,\mathbf k},
\qquad
\mathcal T_s(k)\equiv\frac{T_s(k)}{T_s(k_\star)}.
\label{eq:main_multifield_transfer}
\end{equation}
Here $T_s(k)$ is the linear entropy-to-conversion transfer function, $\mathcal A_s$ is a momentum-independent normalization, and $\mathcal T_s(k_\star)=1$. Equation~\eqref{eq:main_multifield_transfer} is the rank-one, or fully correlated, limit of a more general multifield transfer matrix. {The spectator need not be local in Fourier space; rather, its scale dependence is generated by linear evolution before nonlinear conversion.}

Suppose that the subsequent conversion is local in real space and begins quadratically at a symmetry point,
\begin{equation}
\zeta(\mathbf x)=\zeta_g(\mathbf x)
+\frac12N_{ss}\left[s^2(\mathbf x)-\langle s^2\rangle\right]
+\cdots.
\label{eq:main_local_conversion}
\end{equation}
Substituting Eq.~\eqref{eq:main_multifield_transfer} into Eq.~\eqref{eq:main_local_conversion} and Fourier transforming gives
\begin{align}
\zeta(\mathbf k)
&=\zeta_g(\mathbf k)
+\bar f_{\rm NL}
\int\frac{d^3q}{(2\pi)^{3/2}}
\mathcal T_s(q)\mathcal T_s(\lVert\mathbf k-\mathbf q\rVert)
\nonumber\\
&\hspace{3.0cm}\times
\zeta_g(\mathbf q)\zeta_g(\mathbf k-\mathbf q)
+\cdots,
\label{eq:main_multifield_internal_kernel}
\end{align}
where
\begin{equation}
\bar f_{\rm NL}=\frac12N_{ss}\mathcal A_s^2.
\label{eq:main_multifield_fbar}
\end{equation}
{The two scale weights therefore have a direct origin: each incoming Gaussian mode is transferred linearly before the two transferred fields undergo local quadratic conversion.} A local product in real space consequently generates the product $\mathcal T_s(q)\mathcal T_s(\lVert\mathbf k-\mathbf q\rVert)$, rather than a single weight depending only on the output momentum $k$.

The two templates used below represent two qualitatively different transfer histories. If the transfer evolves slowly over a finite interval of horizon-exit times,
\begin{equation}
\frac{d\ln T_s}{dN_k}\simeq n_f,
\end{equation}
then quasi-de Sitter evolution, $N_k-N_\star\simeq\ln(k/k_\star)$, gives
\begin{equation}
\mathcal T_s(k)\simeq\left(\frac{k}{k_\star}\right)^{n_f}.
\label{eq:main_multifield_powerlaw_transfer}
\end{equation}
A mixing or conversion efficiency localized in time instead changes only across the corresponding interval of horizon-exit scales and approaches a constant outside that interval. This motivates a bounded smooth transition whose logarithmic running vanishes in the ultraviolet. The tanh profile adopted below is a phenomenological realization of this behavior, not a unique solution of a specific multifield Lagrangian.

The normalization in Eq.~\eqref{eq:main_multifield_transfer} is convenient for deriving the momentum routing and directly matches the pivot-normalized power-law template. For the UV-matched tanh convention used numerically, the momentum-independent normalization is instead chosen so that the ultraviolet plateau agrees with the constant-weight kernel. Equivalently, a constant factor may be absorbed into $\mathcal A_s$ and hence into $\bar f_{\rm NL}$, while the scale-dependent part retains the same input-leg origin. Thus the difference between pivot normalization and UV matching changes the reference amplitude but not the physical routing of the two transfer factors.

{This realization also delineates the domain of validity of the separable ansatz.} An independent Gaussian component of the spectator, imperfect adiabatic-entropy correlation, intrinsic spectator self-interactions, or nonlinear transfer would generate additional auto- and cross-spectra and more general mixed momentum kernels. The present calculation isolates the rank-one, weakly nonlinear limit in which those structures are subleading. Appendix~\ref{app:toy_internal_leg_realization} gives the underlying quadratic action, the detailed $\delta N$ construction, the associated bispectrum, and the limitations of this realization.

{We consider two representative functional forms.}

\subsection{{Power-law running}}
A power law is the natural finite-range resummation of a leading logarithmic running. If a slowly evolving interaction coefficient gives $f(k)=1+n_f\ln(k/k_\star)+\cdots$, exponentiating the leading logarithm yields $f(k)\simeq(k/k_\star)^{n_f}$. Closely related power-law parametrizations have been used for running primordial non-Gaussianity in CMB and large-scale-structure analyses and arise in models with a varying sound speed or evolving multifield transfer~\cite{Sefusatti:2009xu,Byrnes:2010ft,Becker:2010nm}. The power law is therefore motivated as the leading finite-range behavior of an evolving nonlinear coupling. It should be understood as an effective description over the momentum range sampled by the scalar source, not as an arbitrarily high-energy extrapolation or by itself as a UV completion. In the numerical analysis, perturbative control is monitored through the weighted hierarchy parameter $A_r\bar f_{\rm NL}^2f_{\rm eff}^4(k)$.
\begin{equation}
f_{\rm PL}(k) = \left( \frac{k}{k_\star} \right)^{n_f},
\label{eq:flog}
\end{equation}
where \(k_\star\) is a normalization scale (typically corresponding to the SIGW/PBH peak) and \(n_f\) parametrizes the running. This form arises naturally as the small-\(n_f\) limit of a linear logarithmic correction:
\begin{equation}
f(k) \approx 1 + n_f \ln\left( \frac{k}{k_\star} \right) \quad \to \quad \left( \frac{k}{k_\star} \right)^{n_f}.
\end{equation}
Positive (negative) \(n_f\) leads to enhancement (suppression) of non-Gaussianity on small scales. This template has been employed in studies of scale-dependent PNG in the context of the CMB and LSS \cite{Byrnes:2010ft,Sefusatti:2009xu}.

\subsection{UV-matched tanh transition template}
The second template is ultraviolet bounded and UV matched to the constant-weight coupling. We use
\begin{equation}
f_{\rm tanh}(k)=\tanh\!\left[(k/k_\star)^{n_f}\right],
\label{eq:ftanh}
\end{equation}
so that
\begin{equation}
f_{\rm tanh}(k_\star)=\tanh(1),
\qquad
f_{\rm tanh}(k)\longrightarrow 1\quad(k\gg k_\star).
\end{equation}
At low momentum,
\begin{equation}
f_{\rm tanh}(k)\simeq (k/k_\star)^{n_f}\quad(k\ll k_\star).
\end{equation}
Thus the UV-matched tanh weight suppresses weighted legs below and around the pivot and recovers the absolute constant-weight normalization at sufficiently large scalar momentum. Its logarithmic running tends to zero in the ultraviolet, so it tests a finite transition followed by saturation to the constant kernel. The exponent $n_f>0$ controls the sharpness of the transition. The precise tanh profile is a phenomenological smooth interpolation, not the unique prediction of a particular ultraviolet Lagrangian. Unlike the power-law weight, which obeys $f_{\rm PL}(k_\star)=1$, the UV-matched tanh weight has $f_{\rm tanh}(k_\star)=\tanh(1)$. The rescaled spectrum $\mathcal P^S(k)=f^2(k)\mathcal P_g(k)$ is substituted into the non-Gaussian integrals as detailed in Sec.~\ref{sec:formalism}. We use $\bar f_{\rm NL}\sim5{-}20$ and vary $n_f$ to compare persistent power-law running with a bounded transition whose logarithmic slope and absolute weight both approach the constant-kernel values in the ultraviolet.
\begingroup
Accordingly, the raw power-law and tanh spectra are not matched to the same effective quadratic amplitude at the pivot: $F_{\rm NL}^{\rm eq}(k_\star)=\bar f_{\rm NL}$ for the power law but $F_{\rm NL}^{\rm eq}(k_\star)=\bar f_{\rm NL}\tanh^2(1)$ for the UV-matched tanh profile. The comparison is intentionally normalized in the ultraviolet rather than at $k_\star$. A pivot-matched comparison could instead use $f_{\rm tanh}^{\rm piv}(k)=\tanh[(k/k_\star)^{n_f}]/\tanh(1)$, or equivalently rescale $\bar f_{\rm NL}$ by $1/\tanh^2(1)$ at tree level. We do not reinterpret the existing numerical curves as pivot matched. Whenever shape rather than absolute normalization is the object of interest, the normalized statistic $\mathcal S(k)$ in Eq.~\eqref{eq:shape_diagnostic} removes a constant run-to-constant ratio at its normalization point; nevertheless, because different perturbative sectors receive different powers of the leg weight, a change in their relative normalization can still affect the shape of the total spectrum.
\endgroup
\subsection{Analytical estimates and scaling expectations}
\label{sec:analytic_estimates}
Before presenting the numerical spectra, we isolate the analytical scaling relations implied by the separable running prescription. These estimates do not replace the multidimensional QMC calculation; rather, they provide controlled limiting checks and explain why the numerical curves behave as they do. \begingroup Throughout this subsection, $p_i$ denotes a generic scalar momentum entering a convolution, $K_X(k;\{p_i\})$ denotes the complete reduced integrand factor associated with sector $X$ after extracting the displayed scalar spectra, and $d\Pi_X$ denotes the corresponding integration measure. For Gaussian and disconnected contributions this factor can be treated as a non-negative weighting kernel. For connected sectors, however, polarization factors such as $\cos(2\phi_{ij})$ and products of transfer functions make the integrand sign indefinite. The quantities introduced below should therefore be interpreted as signed integral ratios or dominant-support diagnostics, not as probability moments, unless positivity is established for the sector under consideration.\endgroup The weighted scalar spectrum is
\begin{equation}
    \mathcal P^S(p)=f^2(p)\mathcal P_g(p),
\end{equation}
so each scale-weighted scalar leg contributes one factor of \(f^2\).

The first estimate concerns the perturbative hierarchy. Before discussing generic broad sources, it is useful to record an exact consistency result. For the internal-leg kernel and an exact monochromatic spectrum $\mathcal P_g(k)=A_r\delta[\ln(k/k_\star)]$, every weighted Gaussian leg is evaluated at $k_\star$. The three sectors therefore factorize as
\begin{align}
\Omega_{{\rm GW},\delta}^{(0)}(k)&=A_r^2\mathcal K_0(k/k_\star),\\
\Omega_{{\rm GW},\delta}^{(1)}(k)&=\bar f_{\rm NL}^2A_r^3 w_\star^4\mathcal K_1(k/k_\star),\\
\Omega_{{\rm GW},\delta}^{(2)}(k)&=\bar f_{\rm NL}^4A_r^4 w_\star^8\mathcal K_2(k/k_\star),
\label{eq:main_mono_sector_scaling}
\end{align}
where $w_\star\equiv f(k_\star)$ is the dimensionless leg weight evaluated at the pivot and the transfer-shape functions $\mathcal K_i$ are independent of the running template. Thus the pivot-normalized power law gives the constant-kernel result sector by sector, whereas the UV-matched tanh template leaves each sector shape unchanged but multiplies the $\mathcal O(\bar f_{\rm NL}^{2})$ and $\mathcal O(\bar f_{\rm NL}^{4})$ sectors by the constants $\tanh^4(1)$ and $\tanh^8(1)$, respectively. Any shape deformation in the numerical narrow-source curves is consequently a finite-width effect. The final appendix supplies the complete derivation; placing the result here makes its role as a code and interpretation check explicit. {Appendix~\ref{app:exact_monochromatic_limit} provides the derivation and Table~\ref{tab:exact_delta_hierarchy_benchmarks} summarizes the corresponding hierarchy estimates for the benchmark amplitudes.}

If the scalar spectrum has a characteristic amplitude \(A_r\), dimensional power counting gives
\begin{align}
    \Omega_{\rm GW}^{(0)}(k) &\sim A_r^2\,\mathcal K_0(k),\\
    \Omega_{\rm GW}^{(1)}(k) &\sim \bar f_{\rm NL}^{\,2}A_r^3\,\mathcal K_1(k)\,\mathcal M_2(k),\\
    \Omega_{\rm GW}^{(2)}(k) &\sim \bar f_{\rm NL}^{\,4}A_r^4\,\mathcal K_2(k)\,\mathcal M_4(k),
\end{align}
where $\mathcal K_i(k)$ are the constant-$f_{\rm NL}$ shape kernels and the running-dependent integral ratios are schematically
\begin{equation}
    \mathcal M_N(k)\equiv
    \frac{\displaystyle \int d\Pi_N\,K_N(k;\{p_i\})\prod_i \mathcal P_g(p_i)\prod_{j=1}^{N}f^2(p_j)}
    {\displaystyle \int d\Pi_N\,K_N(k;\{p_i\})\prod_i \mathcal P_g(p_i)}.
\end{equation}
\begingroup Here $N=2$ denotes the two doubly weighted spectra in the compact $\mathcal O(\bar f_{\rm NL}^2)$ prescription used in this manuscript, and $N=4$ denotes the four weighted spectra in the $\mathcal O(\bar f_{\rm NL}^4)$ sector. For a sign-indefinite connected contribution, $\mathcal M_N$ can be negative, singular near a cancellation of its denominator, or unrepresentative of a unique momentum scale. In those cases the subsequent $p_{\rm eff}$ formulas are saddle or dominant-support estimates rather than literal averages.   It follows that
\begin{align}
    \frac{\Omega_{\rm GW}^{(1)}}{\Omega_{\rm GW}^{(0)}}
    &\sim A_r\bar f_{\rm NL}^{\,2}\,\mathcal M_2(k),\\
    \frac{\Omega_{\rm GW}^{(2)}}{\Omega_{\rm GW}^{(1)}}
    &\sim A_r\bar f_{\rm NL}^{\,2}\,\frac{\mathcal M_4(k)}{\mathcal M_2(k)}.
\end{align}
Thus the familiar constant-\(f_{\rm NL}\) criterion \(A_r\bar f_{\rm NL}^{\,2}\gtrsim1\) is generalized to the scale-dependent condition
\begin{equation}
    A_r\bar f_{\rm NL}^{\,2} f_{\rm eff}^4(k)\gtrsim 1,
    \qquad
    f_{\rm eff}^4(k)\equiv \mathcal M_2(k),
    \label{eq:analytic_hierarchy_condition}
\end{equation}
up to order-one kernel factors. This explains why the plotted low-amplitude examples remain perturbative, while the same diagrammatic sector becomes essential in PBH-level regimes. It also explains why the fourth-order component is more sensitive to running: it contains more weighted scalar legs.

For the power-law template, \(f(p)=(p/k_\star)^{n_f}\), if a diagram is dominated by an effective scalar momentum \(p_{\rm eff}(k)\), the above moments reduce to
\begin{equation}
    \mathcal M_N(k)\simeq \left(\frac{p_{\rm eff}(k)}{k_\star}\right)^{2N n_f}.
\end{equation}
Consequently,
\begin{equation}
    \frac{\Omega_{\rm GW}^{(2)}}{\Omega_{\rm GW}^{(1)}}
    \propto A_r\bar f_{\rm NL}^{\,2}
    \left(\frac{p_{\rm eff}(k)}{k_\star}\right)^{4n_f},
\end{equation}
so positive running compresses the perturbative hierarchy most strongly on the ultraviolet side of the signal, where the convolution samples momenta larger than \(k_\star\). For the UV-matched tanh template, \(f(p)=\tanh[(p/k_\star)^{n_f}]\), one instead has
\begin{equation}
    f(p)\simeq \left(\frac{p}{k_\star}\right)^{n_f}\quad (p\ll k_\star),
    \qquad
    f(p)\to 1\quad (p\gg k_\star),
\end{equation}
so the running produces a delayed onset and a transition-region suppression but no unbounded ultraviolet tilt. This is precisely the qualitative difference between the red curves in the power-law and tanh panels of the numerical figures.

The second useful estimate is the effective local-slope shift. Suppose that over a finite momentum interval the Gaussian scalar spectrum behaves as
\begin{equation}
    \mathcal P_g(p)\propto p^s.
\end{equation}
For power-law running one obtains
\begin{equation}
    \mathcal P^S(p)=f^2(p)\mathcal P_g(p)
    \propto p^{s+2n_f}.
    \label{eq:effective_slope_shift}
\end{equation}
Therefore every weighted scalar leg shifts the local scalar tilt by \(2n_f\). On a UV tail \(\mathcal P_g\propto p^{-\beta}\), the effective fall-off becomes \(p^{-\beta+2n_f}\), so positive \(n_f\) weakens the UV suppression. On an IR rise \(\mathcal P_g\propto p^{\alpha}\), the effective rise becomes \(p^{\alpha+2n_f}\). For broken and double-broken power laws this gives the interval-by-interval replacements
\begin{equation}
    \alpha_{\rm eff}=\alpha+2n_f,
    \qquad
    \beta_{\rm eff}=\beta-2n_f,
\end{equation}
and analogously for each local slope in the double-broken-power-law template. This slope-renormalization picture explains why the numerical spectra with multi-slope scalar sources show stronger shape modulation than single-bump sources.
\begingroup
The relation $s_{\rm eff}=s+2n_f$ applies to an individual weighted scalar spectrum, not directly to the observable ultraviolet slope of $\Omega_{\rm GW}$. The GW slope follows only after convolution with the radiation-era transfer kernel, the remaining weighted and unweighted scalar spectra, the angular factors, and the allowed momentum domain. We therefore use the scalar-leg shift as a mechanism for, rather than an equality with, the template-dependent deformation of the SIGW ultraviolet tail.
\endgroup

\begingroup
The same local-slope description gives a useful sufficient condition for the existence of weighted scalar moments. Define
\begin{equation}
I_m\equiv\int_0^\infty d\ln p\,f^{2m}(p)\mathcal P_g(p),
\label{eq:weighted_scalar_moment_definition}
\end{equation}
where $m>0$ is the power of the squared leg weight in the moment. For a scalar spectrum with asymptotic behavior
\begin{equation}
\mathcal P_g(p)\propto p^\alpha\quad(p\to0),
\qquad
\mathcal P_g(p)\propto p^{-\beta}\quad(p\to\infty),
\end{equation}
where $\alpha$ and $\beta$ are positive infrared and ultraviolet slope parameters, respectively, the power-law template gives
\begin{equation}
f_{\rm PL}^{2m}(p)\mathcal P_g(p)\propto
\begin{cases}
p^{\alpha+2mn_f}, & p\to0,\\
p^{-\beta+2mn_f}, & p\to\infty.
\end{cases}
\end{equation}
The logarithmic moment $I_m$ is therefore finite if
\begin{equation}
\alpha+2mn_f>0,
\qquad
\beta-2mn_f>0,
\label{eq:powerlaw_moment_integrability}
\end{equation}
or equivalently $-\alpha/(2m)<n_f<\beta/(2m)$. For the UV-matched tanh template with $n_f>0$, the infrared condition remains $\alpha+2mn_f>0$, whereas its ultraviolet saturation, $f_{\rm tanh}\to1$, reduces the ultraviolet condition to $\beta>0$. Equations~\eqref{eq:powerlaw_moment_integrability} are sufficient conditions for the isolated scalar moments in Eq.~\eqref{eq:weighted_scalar_moment_definition}; they are not necessary and sufficient conditions for every complete SIGW diagram, whose convergence also depends on transfer kernels, integration measures, triangle constraints, and correlations among internal momenta. Nevertheless, violation of these inequalities provides an analytical warning that a weighted sector may become sensitive to an infrared or ultraviolet integration boundary.
\endgroup

For a log-normal source the same statement can be made more explicitly by a saddle estimate. Writing
\begin{equation}
    x\equiv \ln(p/k_\star),
    \qquad
    \mathcal P_g(p)=\frac{A_r}{\sqrt{2\pi}\sigma}
    \exp\left[-\frac{x^2}{2\sigma^2}\right],
\end{equation}
the weighted spectrum under power-law running is
\begin{equation}
    \mathcal P^S(p)
    \propto \exp\left(2n_f x-\frac{x^2}{2\sigma^2}\right)
    =\exp\left[-\frac{(x-2n_f\sigma^2)^2}{2\sigma^2}\right]
    \exp\left(2n_f^2\sigma^2\right).
    \label{eq:lognormal_saddle}
\end{equation}
Thus the running does not broaden the weighted scalar spectrum in this simple scalar-leg estimate, but shifts its effective support to
\begin{equation}
    p_{\rm eff}=k_\star \exp(2n_f\sigma^2),
    \label{eq:lognormal_effective_peak}
\end{equation}
and enhances the weighted scalar amplitude by \(\exp(2n_f^2\sigma^2)\). For the benchmark \(\sigma=0.8\) and \(n_f=0.5\), this gives \(2n_f\sigma^2=0.64\), hence \(p_{\rm eff}\simeq 1.90\,k_\star\). This analytical estimate anticipates the numerical observation that the power-law running log-normal case preferentially lifts the high-frequency side of the SIGW peak.

\begingroup
The log-normal calculation can be generalized to an arbitrary weighted moment without further approximation. For $m>0$, define the normalized moment
\begin{equation}
\left\langle f_{\rm PL}^{2m}\right\rangle_g
\equiv
\frac{1}{A_r}\int_{-\infty}^{\infty}d\ln p\,
 f_{\rm PL}^{2m}(p)\mathcal P_g(p),
\end{equation}
where the subscript $g$ indicates averaging with the Gaussian scalar spectrum and $A_r=\int d\ln p\,\mathcal P_g(p)$. Since $f_{\rm PL}^{2m}=\exp(2mn_fx)$, completing the square gives
\begin{align}
2mn_fx-\frac{x^2}{2\sigma^2}
&=-\frac{(x-2mn_f\sigma^2)^2}{2\sigma^2}
+2m^2n_f^2\sigma^2,
\label{eq:lognormal_general_square}\\
\left\langle f_{\rm PL}^{2m}\right\rangle_g
&=\exp\!\left(2m^2n_f^2\sigma^2\right).
\label{eq:lognormal_general_moment}
\end{align}
The probability density proportional to $f_{\rm PL}^{2m}\mathcal P_g$ remains Gaussian in $x=\ln(p/k_\star)$, with
\begin{equation}
\langle x\rangle_m=2mn_f\sigma^2,
\qquad
{\rm Var}_m(x)=\sigma^2,
\qquad
p_{{\rm eff},m}=k_\star\exp(2mn_f\sigma^2),
\label{eq:lognormal_general_support}
\end{equation}
where $\langle\cdot\rangle_m$ and ${\rm Var}_m$ denote the mean and variance with respect to this normalized weighted density. The quadratic dependence on $m$ in Eq.~\eqref{eq:lognormal_general_moment} shows that repeated weighting of the same scalar momentum amplifies finite-width sensitivity more rapidly than a linear leg-counting estimate would suggest. In a complete SIGW diagram, however, the weighted momenta are generally distinct and correlated through the transfer kernels and triangle constraints. Consequently, $m$ in Eqs.~\eqref{eq:lognormal_general_moment} and \eqref{eq:lognormal_general_support} characterizes an isolated repeated-momentum moment and should not be identified automatically with the number of independent weighted spectra in a diagram.
\endgroup

{A third estimate quantifies the displacement of peaks and shoulders.} Let \(\Omega_0(k)\) denote a constant-\(f_{\rm NL}\) contribution and let running multiply it by a smooth effective weight \(W(k)\), so that \(\Omega(k)\simeq W(k)\Omega_0(k)\). If \(k_0\) is the peak of \(\Omega_0\), then expanding the peak condition \(d\ln\Omega/d\ln k=0\) around \(k_0\) gives
\begin{equation}
    \Delta\ln k_{\rm pk}
    \equiv \ln\left(\frac{k_{\rm pk}}{k_0}\right)
    \simeq
    -\frac{\left. d\ln W/d\ln k\right|_{k_0}}
    {\left. d^2\ln\Omega_0/d(\ln k)^2\right|_{k_0}}.
    \label{eq:peak_shift_estimate}
\end{equation}
The denominator is negative at a maximum. Therefore a positive power-law running weight, for which \(d\ln W/d\ln k>0\), shifts the apparent peak or shoulder toward larger \(k\). 

\begingroup
For later use, it is helpful to rewrite Eq.~\eqref{eq:peak_shift_estimate} in terms of two dimensionless local diagnostics. For a diagrammatic sector $X$, define the curvature of the constant-weight spectrum at its unperturbed peak by
\begin{equation}
\kappa_X\equiv-
\left.\frac{d^2\ln\Omega_{X,{\rm const}}}{d(\ln k)^2}
\right|_{k=k_{{\rm pk},X}^{(0)}}>0,
\end{equation}
where $k_{{\rm pk},X}^{(0)}$ is the peak position of $\Omega_{X,{\rm const}}$. Let $p_{{\rm eff},X}(k)$ denote the effective scalar momentum controlling the running weight in that sector and define its local response to the external tensor momentum by
\begin{equation}
\gamma_X\equiv
\left.\frac{d\ln p_{{\rm eff},X}}{d\ln k}
\right|_{k=k_{{\rm pk},X}^{(0)}}.
\end{equation}
If sector $X$ contains $N_X$ weighted scalar spectra and is locally approximated by
\begin{equation}
W_X(k)\propto
\left[\frac{p_{{\rm eff},X}(k)}{k_\star}\right]^{2N_Xn_f},
\end{equation}
then Eq.~\eqref{eq:peak_shift_estimate} becomes
\begin{equation}
\Delta\ln k_{{\rm pk},X}^{({\rm an})}
\simeq\frac{2N_X\gamma_Xn_f}{\kappa_X}.
\label{eq:sector_peak_shift_curvature_form}
\end{equation}
Thus the peak displacement grows with the number $N_X$ of weighted spectra, the running index $n_f$, and the rate $\gamma_X$ at which the dominant internal momentum moves with the external scale, while a sharply curved peak with large $\kappa_X$ is more resistant to displacement. Equation~\eqref{eq:sector_peak_shift_curvature_form} is a local analytical estimate rather than a numerical prediction: both $\gamma_X$ and $\kappa_X$ depend on the scalar source, diagrammatic sector, and transfer kernel and must be extracted from the corresponding convolution or spectrum before a numerical value is assigned.
\endgroup

For the UV-matched tanh template,
\begin{equation}
    \frac{d\ln f_{\rm tanh}}{d\ln k}
    =n_f\,\frac{x\,\mathrm{sech}^2 x}{\tanh x},
    \qquad x=\left(\frac{k}{k_\star}\right)^{n_f},
\end{equation}
which is largest around the transition and tends to zero at high momentum. Hence UV-matched tanh running produces localized shoulder-like distortions rather than a persistent UV tilt.

\begingroup
This saturation can be quantified exactly. Introduce the dimensionless transition variable
\begin{equation}
\chi(p)\equiv\left(\frac{p}{k_\star}\right)^{n_f},
\qquad
f_{\rm tanh}(p)=\tanh\chi(p).
\end{equation}
Using $d\chi/d\ln p=n_f\chi$ and ${\rm sech}^2\chi/\tanh\chi=2/\sinh(2\chi)$ gives
\begin{equation}
\frac{d\ln f_{\rm tanh}}{d\ln p}
=n_f\frac{2\chi}{\sinh(2\chi)}.
\label{eq:tanh_exact_log_slope}
\end{equation}
At the pivot, $p=k_\star$ and $\chi=1$, so
\begin{equation}
\left.\frac{d\ln f_{\rm tanh}}{d\ln p}\right|_{p=k_\star}
=\frac{2}{\sinh 2}\,n_f\simeq0.5514\,n_f.
\end{equation}
The asymptotic behavior is
\begin{equation}
\frac{d\ln f_{\rm tanh}}{d\ln p}\simeq
\begin{cases}
n_f, & p\ll k_\star,\\
4n_f\chi e^{-2\chi}, & p\gg k_\star,
\end{cases}
\end{equation}
which proves that the low-momentum logarithmic slope approaches the power-law value while the ultraviolet running vanishes exponentially. The response to the running index itself is
\begin{equation}
\frac{\partial\ln f_{\rm tanh}(p)}{\partial n_f}
=\ln\!\left(\frac{p}{k_\star}\right)
\frac{2\chi}{\sinh(2\chi)}.
\label{eq:tanh_nf_response}
\end{equation}
Here the logarithm records whether the sampled scalar momentum lies below or above the pivot, while the factor $2\chi/\sinh(2\chi)$ suppresses sensitivity once the UV-matched template reaches its plateau. Equations~\eqref{eq:tanh_exact_log_slope}-\eqref{eq:tanh_nf_response} provide the analytical counterpart of the transition-localized deformations in the tanh panels.
\endgroup

{The deep-infrared behavior follows from phase space and the causal structure of the induced tensor source.} For compact scalar spectra and \(k\ll k_\star\), the scalar momenta in the convolution remain near the support of \(\mathcal P_g\), so the running weights are approximately independent of the external tensor momentum \(k\). Each diagram therefore has the form
\begin{equation}
    \Omega_{\rm GW}^{(X)}(k\ll k_\star)
    \simeq C_X[f] \, k^3
    \left[1+\alpha_X[f]\ln(k_c/k)+\cdots\right],
    \label{eq:ir_running_estimate}
\end{equation}
where \(C_X[f]\) and \(\alpha_X[f]\) are weighted moments of the running template and \(k_c\) is a scale of order the scalar peak. The leading \(k^3\) scaling is therefore unchanged by smooth running, while the coefficient and logarithmic correction are diagram dependent. This explains why the numerical spectra preserve the expected IR behavior even when the peak and UV side are visibly distorted.

The preservation of the leading infrared power in Eq.~\eqref{eq:ir_running_estimate} is conditional rather than universal. The argument assumes that the scalar source has compact support, or decreases sufficiently rapidly away from its characteristic scale, so that the weighted convolution moments are finite and remain dominated by scalar momenta of order the source scale as the external tensor momentum tends to zero. It also assumes that $f(p)$ is smooth over the scalar support and that $f^2(p)\mathcal P_g(p)$ does not introduce a new infrared or ultraviolet non-integrability. Under these conditions the leg weights modify the coefficients of the infrared expansion, but do not supply an additional leading power of the external momentum $k$.

The order of limits is important for a regulated narrow source. We first take $k/k_\star\to0$ at fixed nonzero source width and only then consider the distributional zero-width limit. When the weighted scalar moments remain finite, the internal momenta stay localized near the source support and the factors $f(p_i)$ are independent of the external momentum at leading order. The phase-space and radiation-era transfer kernels then determine the leading $k^3$ behavior, while the running enters through diagram-dependent normalization factors and logarithmic corrections.

These assumptions can fail for scalar spectra with extended infrared support, sufficiently slowly decaying ultraviolet tails, or running weights that make the weighted spectrum non-integrable. If $\mathcal P_g(p)\propto p^\alpha$ in the infrared, the power-law weight gives $\mathcal P^S(p)\propto p^{\alpha+2n_f}$. Sufficiently negative $n_f$ can therefore enhance soft weighted legs, shift the dominant momentum region of a convolution, or make a weighted moment infrared sensitive. In that regime the leading external-$k$ scaling must be derived from the full diagrammatic integral rather than inferred from Eq.~\eqref{eq:ir_running_estimate}. For the UV-matched tanh template with $n_f>0$, $f_{\rm tanh}(p)\sim(p/k_\star)^{n_f}$ at small $p$ and tends to unity at large $p$; it is UV bounded and suppresses sufficiently soft weighted legs, but this boundedness alone does not guarantee infrared localization of an extended scalar source.

Accordingly, the form in Eq.~\eqref{eq:ir_running_estimate} is claimed here only for compact or sufficiently rapidly decaying scalar sources with finite weighted moments. We do not extrapolate it to arbitrary extended spectra, singular leg weights, or parameter choices for which a convolution is dominated by an infrared or ultraviolet boundary.

{These estimates motivate the following shape-only diagnostic:}
\begin{equation}
    \mathcal S(k)
    \equiv
    \frac{\Omega_{\rm GW}^{\rm run}(k)/\Omega_{\rm GW}^{\rm const}(k)}
    {\Omega_{\rm GW}^{\rm run}(k_{\rm pk})/\Omega_{\rm GW}^{\rm const}(k_{\rm pk})}.
    \label{eq:shape_diagnostic}
\end{equation}
If \(\mathcal S(k)=1\), running acts only as an amplitude rescaling. Deviations from unity isolate genuine spectral morphology. In the local-slope approximation, power-law running gives \(\mathcal S(k)\sim[p_{\rm eff}(k)/p_{\rm eff}(k_{\rm pk})]^{2Nn_f}\), with \(N=2\) or \(4\) for the second- and fourth-order sectors, respectively. This explains why the higher-order contributions are often the most sensitive probes of running.

Figures~\ref{fig:shape_statistic_primary_sources} and \ref{fig:shape_statistic_additional_sources} evaluate this diagnostic for representative localized, finite-width, and multi-slope scalar sources. The power-law weight produces a persistent frequency-dependent departure from $\mathcal S=1$, reflecting the continued redistribution of internal momentum support. The UV-matched tanh weight instead produces its largest departure around the transition and tends back toward amplitude-only behavior as the weighted scalar momenta enter the ultraviolet plateau. For the narrow regularized source, the departure from unity is a finite-width effect; in the exact delta limit the power-law result satisfies $\mathcal S_{\rm PL}^{\delta}=1$, while the UV-matched tanh template can alter the normalized total only through its different constant suppressions of the non-Gaussian sectors.

\begin{figure}
    \centering
    \includegraphics[width=0.92\linewidth]{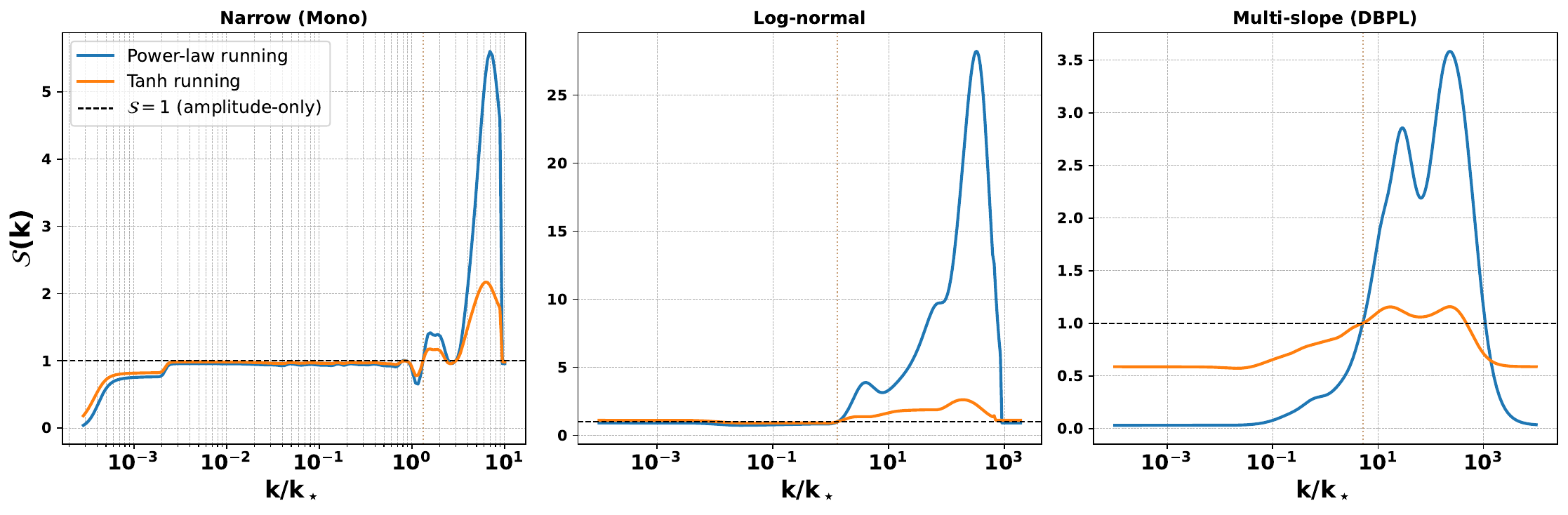}
    \caption{{\it Normalized shape statistic $\mathcal S(k)$, defined in Eq.~\eqref{eq:shape_diagnostic}, for the narrow regularized source (left), finite-width log-normal source (middle), and multi-slope double-broken-power-law source (right). The horizontal axis is $k/k_\star$. Solid blue and solid orange curves show the power-law and UV-matched tanh leg weights, respectively, while the black dashed line $\mathcal S=1$ denotes an amplitude-only deformation with no change of spectral morphology. Departures from unity isolate the scale-dependent redistribution of power after normalization at the peak. The narrow-source result is a finite-width effect and must be interpreted using Eq.~\eqref{eq:narrow_source_regulator_results}.}}
    \label{fig:shape_statistic_primary_sources}
\end{figure}
\begin{figure}
    \centering
    \includegraphics[width=0.92\linewidth]{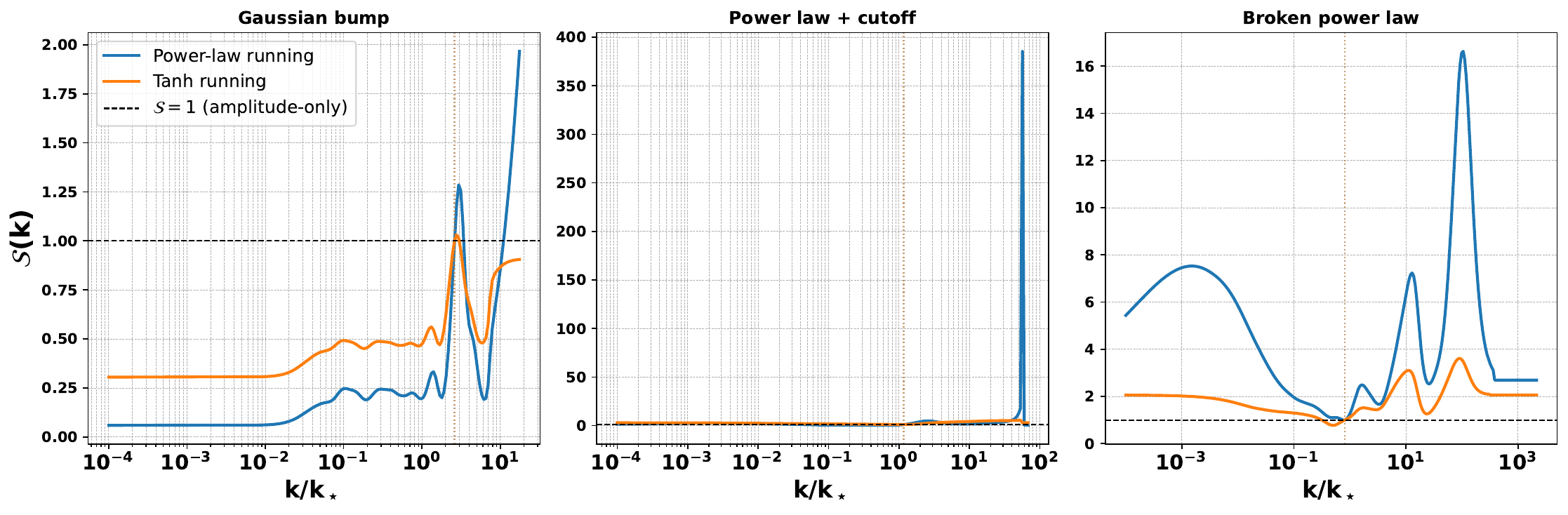}
    \caption{{\it Normalized shape statistic $\mathcal S(k)$ for the Gaussian scalar bump (left), power-law source with exponential cutoff (middle), and broken-power-law source (right). Solid blue and solid orange curves correspond to the power-law and UV-matched tanh leg weights; the black dashed line marks the amplitude-only limit $\mathcal S=1$. The unequal frequency dependence of the two running templates demonstrates persistent redistribution for the power-law weight and a transition-localized deformation with saturated logarithmic running for the UV-matched tanh weight.}}
    \label{fig:shape_statistic_additional_sources}
\end{figure}

For clarity, the peak displacement and the fixed-point shape ratios can be reported analytically only as estimates unless they are extracted from the numerical spectra. For a sector $X$ whose running result is locally $\Omega_{X,\rm run}(k)\simeq W_X(k)\Omega_{X,\rm const}(k)$, the analytical displacement is
\begin{equation}
\Delta\ln k_{{\rm pk},X}^{({\rm an})}
\simeq-
\left.\frac{d\ln W_X/d\ln k}
{d^2\ln\Omega_{X,\rm const}/d(\ln k)^2}
\right|_{k=k_{{\rm pk},X}^{(0)}}.
\label{eq:analytical_peak_displacement}
\end{equation}
A measured displacement should instead be obtained from
\begin{equation}
\Delta\ln k_{\rm pk}^{({\rm num})}
=\ln\!\left[
\frac{\operatorname*{arg\,max}_k\Omega_{\rm GW}^{\rm run}(k)}
{\operatorname*{arg\,max}_k\Omega_{\rm GW}^{\rm const}(k)}
\right],
\label{eq:numerical_peak_displacement_definition}
\end{equation}
and no numerical value is assigned here.

Within the effective-momentum approximation, the sector-level power-law result at the requested reference points is
\begin{equation}
\mathcal S_X^{({\rm an})}(xk_\star)
\simeq
\left[
\frac{p_{{\rm eff},X}(xk_\star)}
{p_{{\rm eff},X}(k_{{\rm pk},X})}
\right]^{2N_Xn_f},
\qquad x\in\{0.3,1,3\},
\label{eq:shape_statistic_fixed_points_powerlaw}
\end{equation}
where $N_X=2$ for $\Omega_{\rm GW}^{(1)}$ and $N_X=4$ for $\Omega_{\rm GW}^{(2)}$. For the UV-matched tanh template,
\begin{equation}
\mathcal S_{X,\tanh}^{({\rm an})}(xk_\star)
\simeq
\left\{
\frac{\tanh\!\left[(p_{{\rm eff},X}(xk_\star)/k_\star)^{n_f}\right]}
{\tanh\!\left[(p_{{\rm eff},X}(k_{{\rm pk},X})/k_\star)^{n_f}\right]}
\right\}^{2N_X},
\qquad x\in\{0.3,1,3\}.
\label{eq:shape_statistic_fixed_points_tanh}
\end{equation}
Numerical values are not quoted because $p_{{\rm eff},X}(k)$ is source-, diagram-, and transfer-kernel-dependent. Assuming $p_{\rm eff}\propto k$ without verifying the convolution support would add an uncontrolled approximation. For the total spectrum, the Gaussian, $\mathcal O(\bar f_{\rm NL}^2)$, and $\mathcal O(\bar f_{\rm NL}^4)$ sectors carry different effective weights and may combine with algebraic signs. The total peak displacement and the values $\mathcal S(0.3k_\star)$, $\mathcal S(k_\star)$, and $\mathcal S(3k_\star)$ are therefore numerical observables unless a single sector dominates throughout the comparison interval.

\subsection{Analytical-numerical comparison and physical implications}
\label{sec:analytic_numeric_implications}

The estimates above provide a useful way to interpret the numerical spectra without relying only on visual inspection. \begingroup In particular, Eqs.~\eqref{eq:lognormal_general_moment}, \eqref{eq:tanh_exact_log_slope}, \eqref{eq:powerlaw_moment_integrability}, and \eqref{eq:sector_peak_shift_curvature_form} separate four effects that otherwise appear simultaneously in the spectra: finite-width moment enhancement, ultraviolet saturation, boundary sensitivity, and peak displacement.\endgroup A convenient diagnostic is the fractional hierarchy parameter
\begin{equation}
    \epsilon_{\rm NG}(k)\equiv A_r\bar f_{\rm NL}^{\,2} f_{\rm eff}^4(k),
    \label{eq:epsilon_ng_definition}
\end{equation}
where \(f_{\rm eff}\) is defined by Eq.~\eqref{eq:analytic_hierarchy_condition}. When \(\epsilon_{\rm NG}\ll1\), the numerical spectrum should show a clear ordering \(\Omega_{\rm GW}^{(0)}\gg \Omega_{\rm GW}^{(1)}\gg \Omega_{\rm GW}^{(2)}\), modulo kernel enhancements in narrow regions of support. When \(\epsilon_{\rm NG}\sim1\), the second- and fourth-order sectors should become comparable and the perturbative hierarchy is compressed. This is exactly the trend seen in the benchmark plots: the low-amplitude examples are mainly shape diagnostics, while the parameter region relevant for PBH production is where the fourth-order sector must be retained for perturbative consistency rather than treated as a negligible correction.

The narrow-source benchmark must be interpreted using the internal-leg kernel. For an exact delta source all weighted Gaussian legs satisfy $p=k_\star$, so {both a pivot-normalized power law and a UV-matched tanh weight leave the normalized sector shapes unchanged.} {The numerical narrow source is the finite-width log-normal regulator in Eq.~\eqref{eq:narrow_source_regulator_results}; its leg weights vary across the support.} The displayed deformations are finite-width corrections that vanish as the width tends to zero, consistently with the final appendix.
For the monochromatic and Gaussian scalar bump benchmarks with \(A_r\sim10^{-5}\) and \(\bar f_{\rm NL}=5\), the constant-weight estimate gives \(A_r\bar f_{\rm NL}^2\simeq2.5\times10^{-4}\). Even allowing for an order-one or few-times enhancement from \(f_{\rm eff}^4\), Eq.~\eqref{eq:epsilon_ng_definition} predicts that \(\Omega_{\rm GW}^{(2)}\) should remain subleading. The numerical decompositions are consistent with this expectation: the fourth-order curves are visible in selected shoulders or post-peak regions, but the total spectrum is mainly controlled by the Gaussian and \(\mathcal O(\bar f_{\rm NL}^2)\) pieces. This agreement is important because it shows that the plotted distortions are not artifacts of an uncontrolled fourth-order dominance.

For the log-normal benchmark the comparison is more quantitative. The analytical saddle estimate gives
\begin{equation}
    p_{\rm eff}=k_\star e^{2n_f\sigma^2}.
\end{equation}
{For \(\sigma=0.8\) and \(n_f=0.5\), this yields \(p_{\rm eff}\simeq1.90\,k_\star\).} In other words, the running-weighted scalar legs preferentially probe momenta almost a factor of two above the nominal scalar peak. The numerical log-normal plots display precisely this effect as an enhancement of the high-frequency flank rather than a uniform rescaling. Moreover, because the induced tensor frequency is set by a convolution of two scalar momenta, this scalar-side shift does not translate into a one-to-one displacement of the GW peak; instead, it broadens the peak and lifts the UV shoulder. This explains why the numerical spectra show both peak deformation and high-frequency enhancement.

For the broken- and double-broken-power-law benchmarks, the analytical local-slope rule
\begin{equation}
    s_{\rm eff}=s+2n_f
\end{equation}
explains the stronger morphology dependence. On the UV side of a broken power law, \(s=-\beta\), so the weighted slope becomes \(-\beta+2n_f\). Thus, positive power-law running systematically reduces the steepness of the weighted UV tail. In the numerical spectra this appears as an uplift of the high-frequency side and as a redistribution of power between the peak, shoulder, and tail. The effect becomes more pronounced for multi-slope spectra because different intervals have different local values of \(s\), so the same running template modifies several parts of the convolution support at once. This is why the double-broken-power-law examples show richer shape changes than the Gaussian scalar bump example, even when the overall amplitude scale is similar.

The UV-matched tanh template provides a useful control case. {Since $f_{\rm tanh}\to1$ at high momentum, the analytical estimates predict recovery of the constant-kernel normalization and no persistent change in the UV slope relative to the constant-$f_{\rm NL}$ case.} The numerical tanh spectra follow this expectation: the largest differences occur around the transition region, where \(d\ln f_{\rm tanh}/d\ln k\) is nonzero, {while the deep UV approaches the constant-kernel shape and normalization with vanishing logarithmic running.} This contrast strengthens the physical interpretation of the power-law running results, because it shows that the UV uplift is tied to the power-law growth of the running kernel rather than to generic non-Gaussian mode coupling.

The physical implication is that running non-Gaussianity acts as a momentum-dependent filter on the scalar modes that source SIGWs. Gaussian SIGWs mainly measure the convolution of the scalar power spectrum with the radiation-era transfer kernel. In contrast, the non-Gaussian pieces measure weighted convolutions in which different internal scalar momenta are amplified or suppressed depending on \(f(p)\). Therefore the GW spectrum is sensitive not only to the amplitude of the primordial scalar perturbations but also to how the non-Gaussian kernel distributes weight across the scalar support. This is the origin of the peak shifts, shoulders, and UV-tail modifications in the numerical results.

These analytical-numerical comparisons also clarify the experimental relevance. PTA observations probe the nHz band, where many SIGW explanations sample the infrared or near-peak side of a scalar source. For the localized source templates and finite weighted moments considered here, Eq.~\eqref{eq:ir_running_estimate} shows that smooth running preserves the leading \(k^3\) infrared scaling. PTA measurements of the low-frequency slope alone may not uniquely determine the running template. However, deviations from a pure amplitude rescaling, quantified by \(\mathcal S(k)\), can still appear if the PTA band overlaps the transition or shoulder region. In that case, a joint fit to the spectral slope, curvature, and turnover frequency would be more informative than a single-amplitude comparison.

For LISA-like mHz observations, the impact can be stronger because the detector band can overlap the peak and UV shoulder of the induced signal. The peak-shift estimate in Eq.~\eqref{eq:peak_shift_estimate} predicts that positive power-law running moves the effective shoulder toward higher frequency and broadens the signal. Numerically this is the region where power-law running most visibly separates from the scale-independent case. Therefore, a LISA-band stochastic background with a measurable asymmetric shoulder or a frequency-dependent excess relative to a constant-\(f_{\rm NL}\) template would be a natural target for the shape diagnostic \(\mathcal S(k)\).

For higher-frequency or broader-band experiments such as DECIGO, BBO, and third-generation ground-based detectors, the UV-tail estimate is particularly relevant. If a scalar source peaks below the most sensitive part of a detector band, positive power-law running can lift the high-frequency tail and improve detectability in a band that would otherwise probe only the decaying side of the signal. Conversely, a UV-matched tanh transition does not generate the same persistent UV uplift, so a non-detection in a high-frequency band combined with a lower-frequency detection could help distinguish saturating running from power-law running. A multi-band comparison between PTA, LISA, and higher-frequency detectors would therefore directly test the analytical slope-shift mechanism in Eq.~\eqref{eq:effective_slope_shift}.

In practice, these implications suggest three useful observables for future forecasts. First, one should fit not only the peak amplitude but also the normalized shape ratio \(\mathcal S(k)\). Second, one should track the apparent peak or shoulder displacement \(\Delta\ln k_{\rm pk}\) defined in Eq.~\eqref{eq:peak_shift_estimate}. Third, one should compare the measured UV slope with the expectation \(s_{\rm eff}=s+2n_f\) for power-law running. These observables are less degenerate with an overall change in \(A_r\) than the total amplitude alone, and therefore provide a cleaner route to separating running primordial non-Gaussianity from a constant-\(f_{\rm NL}\) model or from a different scalar power-spectrum shape.

\section{Numerical results for representative scalar sources}
\label{sec:results}
{In this section, we present numerical results for the SIGW spectrum} \(\Omega_{\rm GW}(f)\) sourced by the scale-dependent internal-leg separable kernel. {For a self-contained comparison, we treat all scalar-source benchmarks in a single section:} a narrow (near-monochromatic) peak, a finite-width log-normal bump, a USR-inspired spectrum, {a Gaussian scalar bump}, a power-law spectrum with exponential cutoff, a broken power-law spectrum, and a double-broken-power-law spectrum. This broader set of templates allows us to test how generic the effects of running non-Gaussianity are.

For each scalar template we compute the Gaussian contribution \(\Omega_{\rm GW}^{(0)}\), the \(\mathcal O(\bar f_{\rm NL}^2)\) correction \(\Omega_{\rm GW}^{(1)}\), and the \(\mathcal O(\bar f_{\rm NL}^4)\) correction \(\Omega_{\rm GW}^{(2)}\). We compare the scale-independent case, \(f(k)=1\), with the two running templates introduced in Sec.~\ref{sec:templates},
\begin{align}
 f_{\rm PL}(k) &= \left( \frac{k}{k_\star} \right)^{n_f}, \\
  f_{\rm tanh}(k) &= \tanh\!\left[(k/k_\star)^{n_f}\right],
\end{align}
and use the corresponding rescaled power spectrum \(\mathcal P^S(k)=f^2(k)\mathcal P_g(k)\) consistently for the scale-weighted Gaussian legs in the non-Gaussian integrals. This is precisely the internal-leg separable prescription of Eq.~\eqref{eq:zeta_running_ansatz}; it is not an external-output prescription. Appendix~\ref{app:scale_dependent_extension} gives the weighted and unweighted factors explicitly. Throughout this section, the plotted spectra compare the scale-independent result with the power-law and UV-matched tanh running cases using the same decomposition into Gaussian, reducible, and connected pieces as in Sec.~\ref{sec:formalism}. All multidimensional integrals are evaluated numerically with a quasi-Monte Carlo integrator.
{
The sectors are combined with their algebraic signs. In logarithmic order-by-order figures, the orange, green, and red curves show $|\Omega_{\rm GW}^{(0)}|$, $|\Omega_{\rm GW}^{(1)}|$, and $|\Omega_{\rm GW}^{(2)}|$, whereas the blue curve shows
\begin{equation}
|\Omega_{\rm GW}^{\rm tot}|=|\Omega_{\rm GW}^{(0)}+\Omega_{\rm GW}^{(1)}+\Omega_{\rm GW}^{(2)}|.
\label{eq:signed_total_plotting_convention}
\end{equation}
Cancellations can therefore place the total below an individual displayed sector magnitude. All totals and ratios are formed from signed quantities before an absolute value is taken for logarithmic display.
}

When translating comoving wavenumber into present-day detector frequency, we use the standard mapping \footnote{The radiation-era late-time result is converted to the present-day spectrum by \[ \Omega_{\rm GW,0}(k) = \Omega_{r,0} \left(\frac{g_{*,0}}{g_{*,c}}\right)^{1/3} \Omega_{\rm GW,c}(k), \] where \(g_{*,c}\) is evaluated at horizon re-entry of the source scale. Unless stated otherwise, detector comparisons refer to \(\Omega_{\rm GW,0}\). }
\begin{equation}
 f \simeq 1.5\times 10^{-15}\,\left(\frac{k}{\mathrm{Mpc}^{-1}}\right)\,\mathrm{Hz},
 \label{eq:k_to_f_conversion}
\end{equation}
so that any quoted \(f_\star\) should be understood as the present-day frequency associated with the comoving pivot scale \(k_\star\), rather than the naive flat-space relation \(k_\star = 2\pi f_\star\). The benchmark parameters are chosen so that the Gaussian SIGW peak lies in the LISA or PTA bands. The amplitudes shown in many figures are deliberately perturbative and are used to diagnose spectral-shape effects; quantitative PBH statements require larger scalar amplitudes and should be associated with the regime $A_r\bar f_{\rm NL}^2\gtrsim1$. Our main goal is to identify, template by template, how the shape of the scalar spectrum controls the relative importance of \(\mathcal O(\bar f_{\rm NL}^2)\) and \(\mathcal O(\bar f_{\rm NL}^4)\)\footnote{ Our use of the phrase \(\mathcal O(\bar f_{\rm NL}^4)\) refers only to insertions of the primordial non-Gaussian statistics into the second-order tensor source. It should not be interpreted as a complete fourth-order cosmological perturbation-theory calculation, nor as an all-order treatment of the nonlinear curvature field. }, and how these orders respond differently to power-law and UV-matched tanh running of the non-Gaussian kernel.

The benchmark set spans narrow, finite-width, asymmetric, and multi-slope scalar sources, allowing the two running templates to be compared across qualitatively different momentum supports.

\subsection{{Narrow regularized scalar spectrum}}
{
A sharply peaked scalar spectrum provides the cleanest illustration of non-Gaussian effects. The exact distributional benchmark is
\begin{equation}
\Delta_{\mathcal R,\delta}^2(k)=A_{\mathcal R}\delta\!\left[\ln\left(\frac{k}{k_\star}\right)\right].
\label{eq:exact_delta_source_results}
\end{equation}
For numerical evaluation, we replace it by the finite-width log-normal regulator
\begin{equation}
\Delta_{\mathcal R,\sigma_\delta}^2(k)=\frac{A_{\mathcal R}}{\sqrt{2\pi}\sigma_\delta}
\exp\!\left[-\frac{\ln^2(k/k_\star)}{2\sigma_\delta^2}\right],
\qquad
\sigma_\delta=0.8.
\label{eq:narrow_source_regulator_results}
\end{equation}
The regulator satisfies $\int d\ln k\,\Delta_{\mathcal R,\sigma_\delta}^2=A_{\mathcal R}$ and recovers the exact delta source as $\sigma_\delta\to0$. The displayed narrow-source curves use Eq.~\eqref{eq:narrow_source_regulator_results}.
}
With $A_{\mathcal R}=1\times10^{-3}$, $\sigma_\delta=0.2$, and $k_\star$ chosen so that the present-day peak lies in the mHz or nHz band, the Gaussian spectrum approaches the known monochromatic result as $\sigma_\delta\to0$~\cite{Kohri:2018awv}.

Figures~\ref{fig:mono_all_orders} and \ref{fig:mono_final} display the order-by-order decomposition and the impact of running non-Gaussianity on the total spectrum. The connected terms (especially C and planar) contribute comparably to the disconnected ones and peak at slightly different wavenumbers, leading to a richer spectral structure. For \(A_r \bar f_{\rm NL}^2 \gtrsim 1\), \(\mathcal O(\bar f_{\rm NL}^4)\) terms become non-negligible, consistent with earlier constant-\(f_{\rm NL}\) analyses \cite{Adshead:2021hnm,Li:2023qua} \footnote{The numerical ``monochromatic'' source is a narrow finite-width log-normal regularization. The exact delta-source limit is derived in the final appendix.}

For the finite-width source used numerically, scale-dependent leg weights produce visible shape distortions: the power-law template with positive $n_f$ enhances the high-frequency flank and broadens the peak, {while the UV-matched tanh weight suppresses the sub-pivot and pivot regions and approaches the constant-weight normalization in the ultraviolet.} These deformations vanish in the exact delta-source limit for a pivot-normalized power law. In the infrared, both templates recover the expected \(k^3\) scaling with mild logarithmic corrections, in agreement with known IR subtleties \cite{Yuan:2019wwo,Domenech:2025bvr}.

\begin{figure}[H]
\centering
\includegraphics[width=0.48\textwidth]{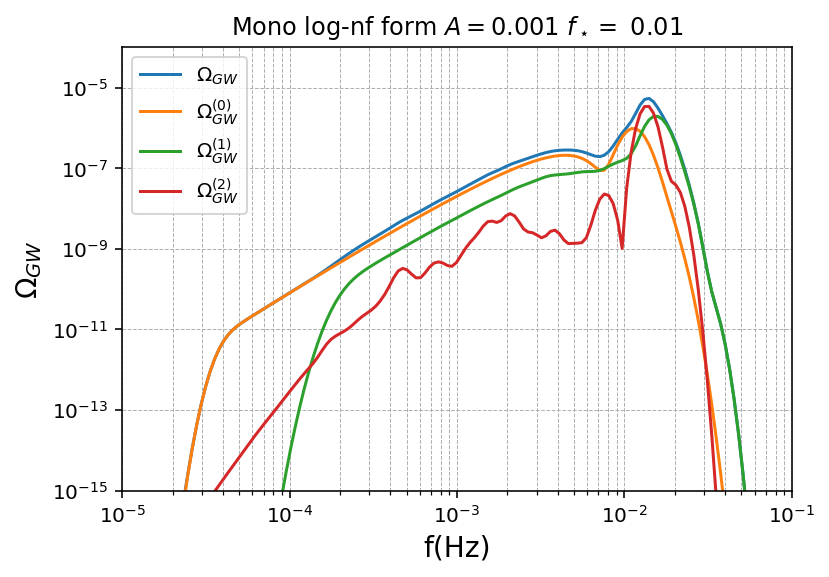}
\includegraphics[width=0.48\textwidth]{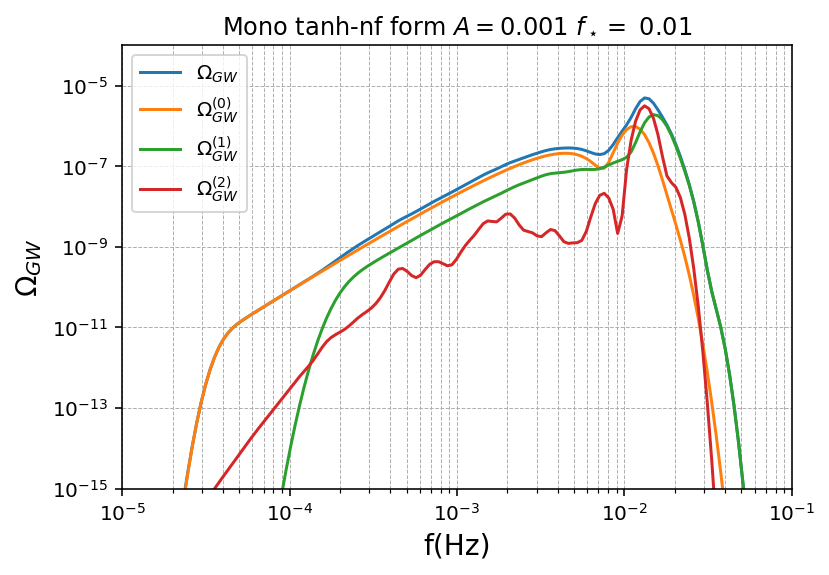}
\caption{{\it Contributions to $\Omega_{\rm GW}$ for the narrow regularized scalar source with $A_{\mathcal R}=10^{-3}$ and $f_\star=10^{-2}\,{\rm Hz}$. Left: power-law running. Right: UV-matched tanh running. Both panels use $\bar f_{\rm NL}=15$ and $n_f=0.8$. Orange, green, and red show $|\Omega_{\rm GW}^{(0)}|$, $|\Omega_{\rm GW}^{(1)}|$, and $|\Omega_{\rm GW}^{(2)}|$, while blue shows the magnitude of their signed sum. Algebraic cancellations can place the blue total below an individual sector magnitude, as specified in Eq.~\eqref{eq:signed_total_plotting_convention}.}}
\label{fig:mono_all_orders}
\end{figure}

\begin{figure}[H]
\centering
\includegraphics[width=0.48\textwidth]{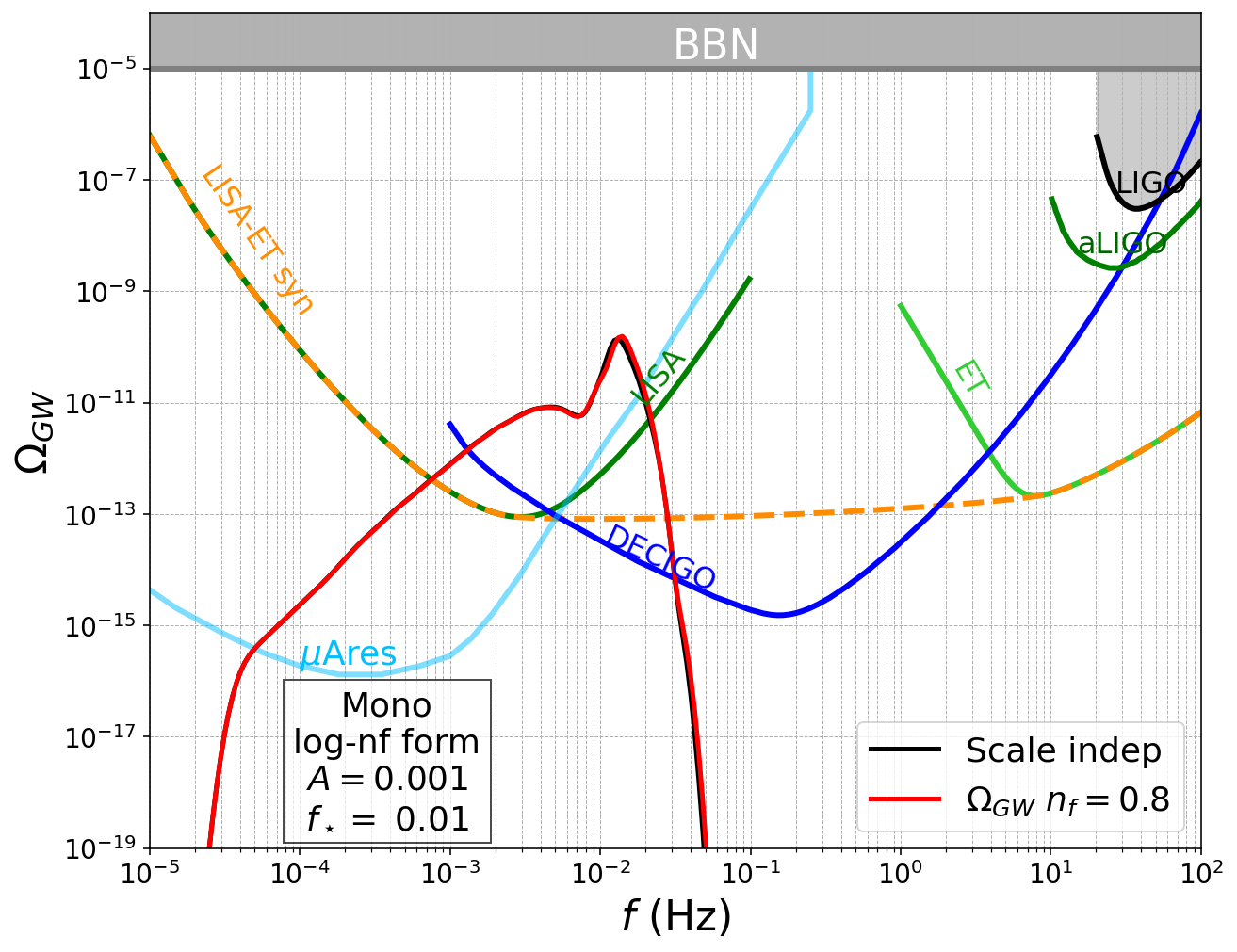}
\includegraphics[width=0.48\textwidth]{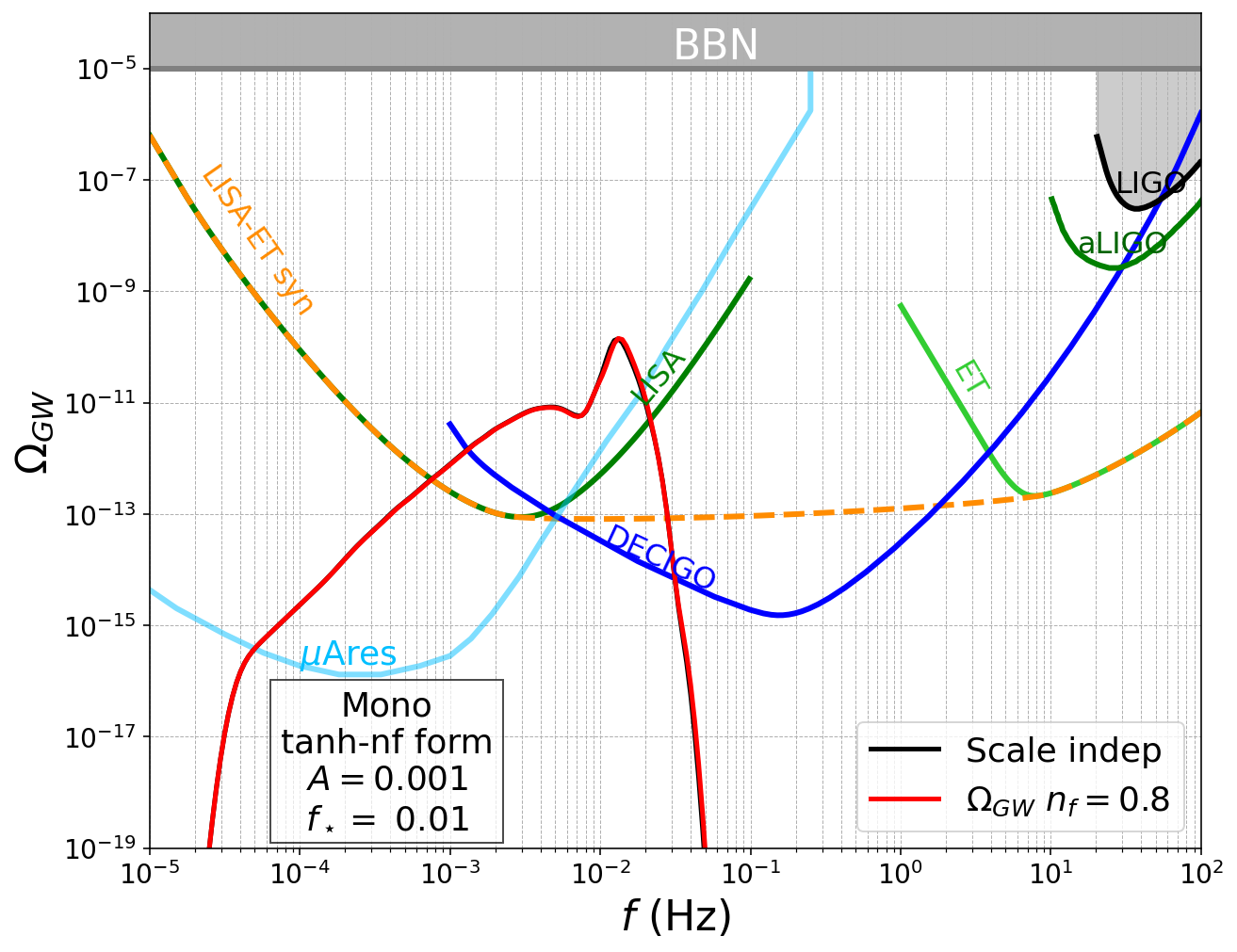}
\caption{\it Total $\Omega_{\rm GW}$ comparing the constant-weight (black) and scale-dependent (red) cases for the same narrow finite-width source. Power-law running enhances the high-$k$ flank, whereas the tanh weight delays the non-Gaussian onset and saturates at high scalar momentum. LISA and PTA sensitivity curves are indicated. The gray BBN/CMB and LVK regions are visual guides: the corresponding bounds apply to $\int d\ln f\,\Omega_{\rm GW,0}(f)$ and therefore depend on the bandwidth and full spectral shape.}
\label{fig:mono_final}
\end{figure}

\subsection{Log-normal scalar spectrum}

A finite-width log-normal spectrum is a realistic benchmark for many inflationary mechanisms:
\begin{equation}
\Delta_{\mathcal R}^2(k) = \frac{A_r}{\sqrt{2\pi}\sigma} \exp\left( -\frac{\ln^2(k/k_\star)}{2\sigma^2} \right),
\end{equation}
normalized such that \(\int d\ln k\,\Delta_{\mathcal R}^2(k) = A_r\). We take \(\sigma = 0.8\) and \(A_r = 1\times10^{-3}\).

The finite width smooths the sharp multi-peak structure seen in the monochromatic case while preserving a localized signal near the peak. Nevertheless, the scale-dependent leg weight produces visible template-level distortions (Figs.~\ref{fig:lognormal_all} and \ref{fig:lognormal_final}). The power-law template systematically enhances the high-frequency side of the peak by up to \(\mathcal O(10)\) for \(n_f \approx 0.5{-}1\), while the UV-matched tanh template leads to a suppression through the transition region and then saturates toward the scale-independent weighting at sufficiently large \(k\). {For the localized log-normal source, the infrared retains a leading $k^3$-type behavior multiplied by slowly varying logarithmic corrections, with connected contributions remaining important.}

\begin{figure}[H]
\centering
\includegraphics[width=0.48\textwidth]{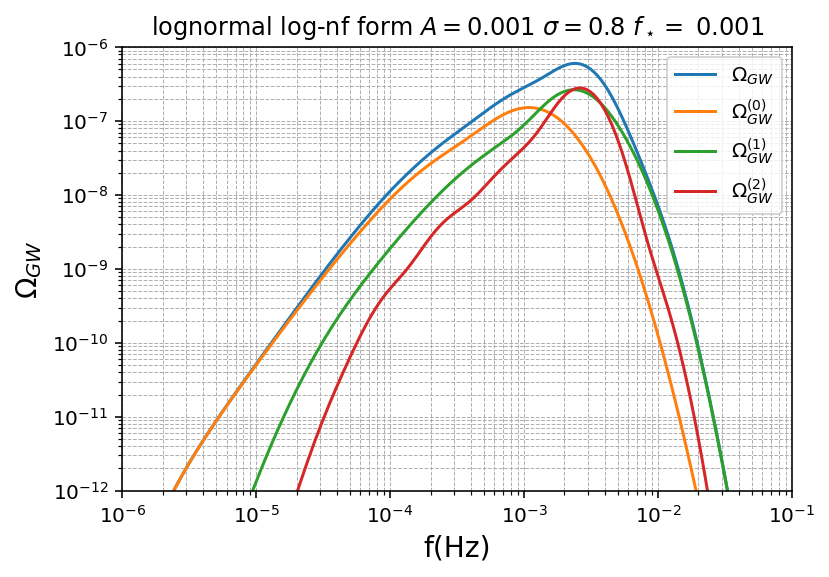}
\includegraphics[width=0.48\textwidth]{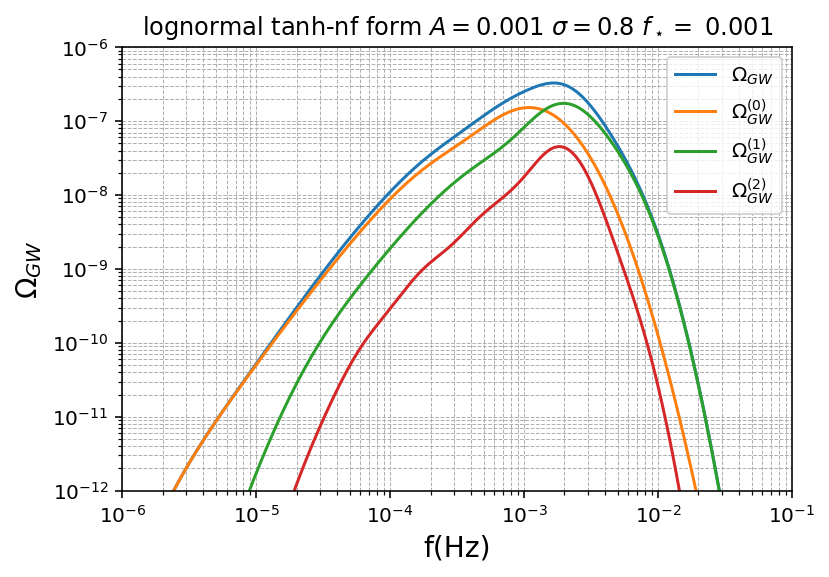}
\caption{\it Decomposition for log-normal source with \(A_r=10^{-3}\), \(\sigma=0.8\), \(f_\star=10^{-3}\) Hz. Scale-dependent running modifies both amplitude and width of the main peak.}
\label{fig:lognormal_all}
\end{figure}

\begin{figure}[H]
\centering
\includegraphics[width=0.48\textwidth]{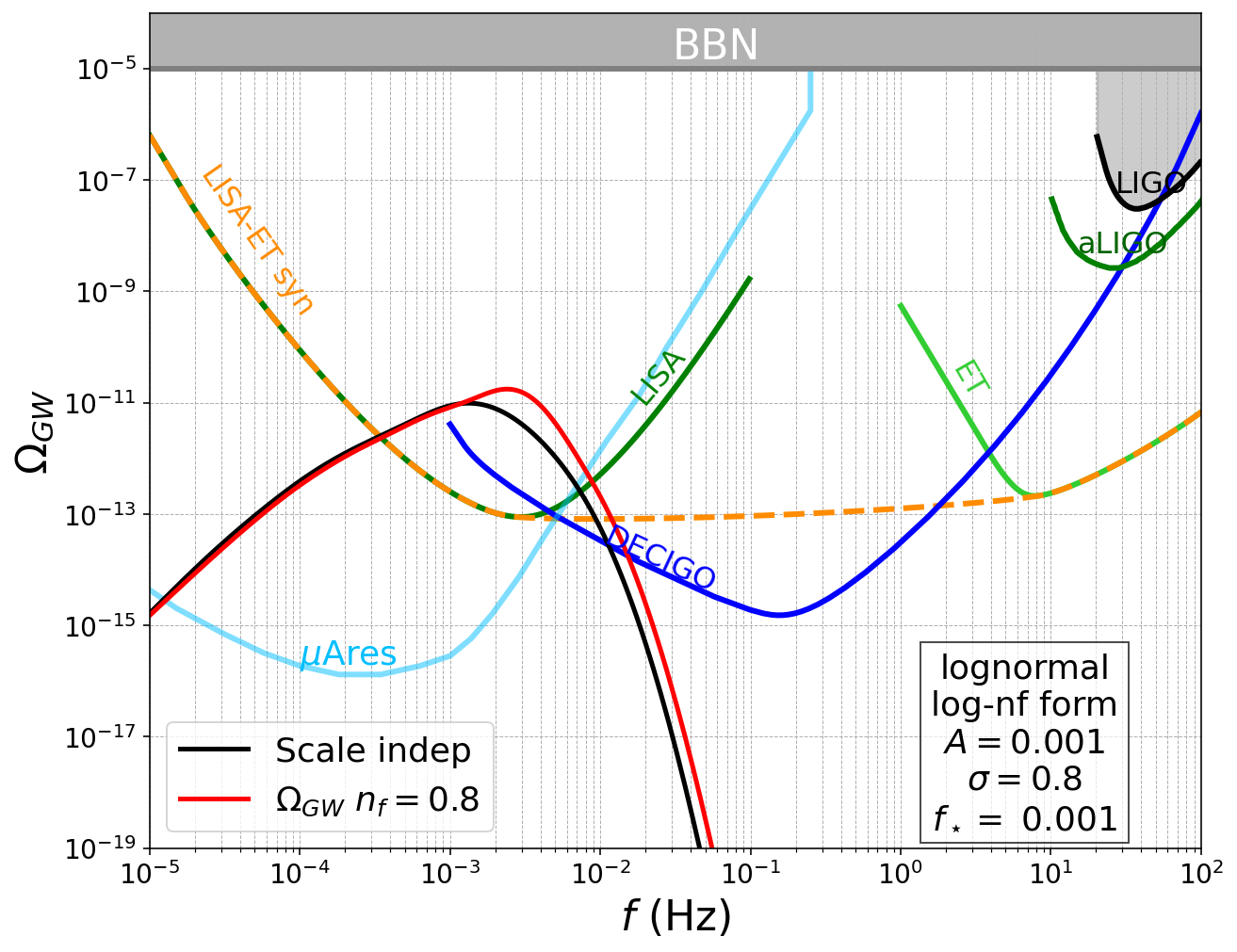}
\includegraphics[width=0.48\textwidth]{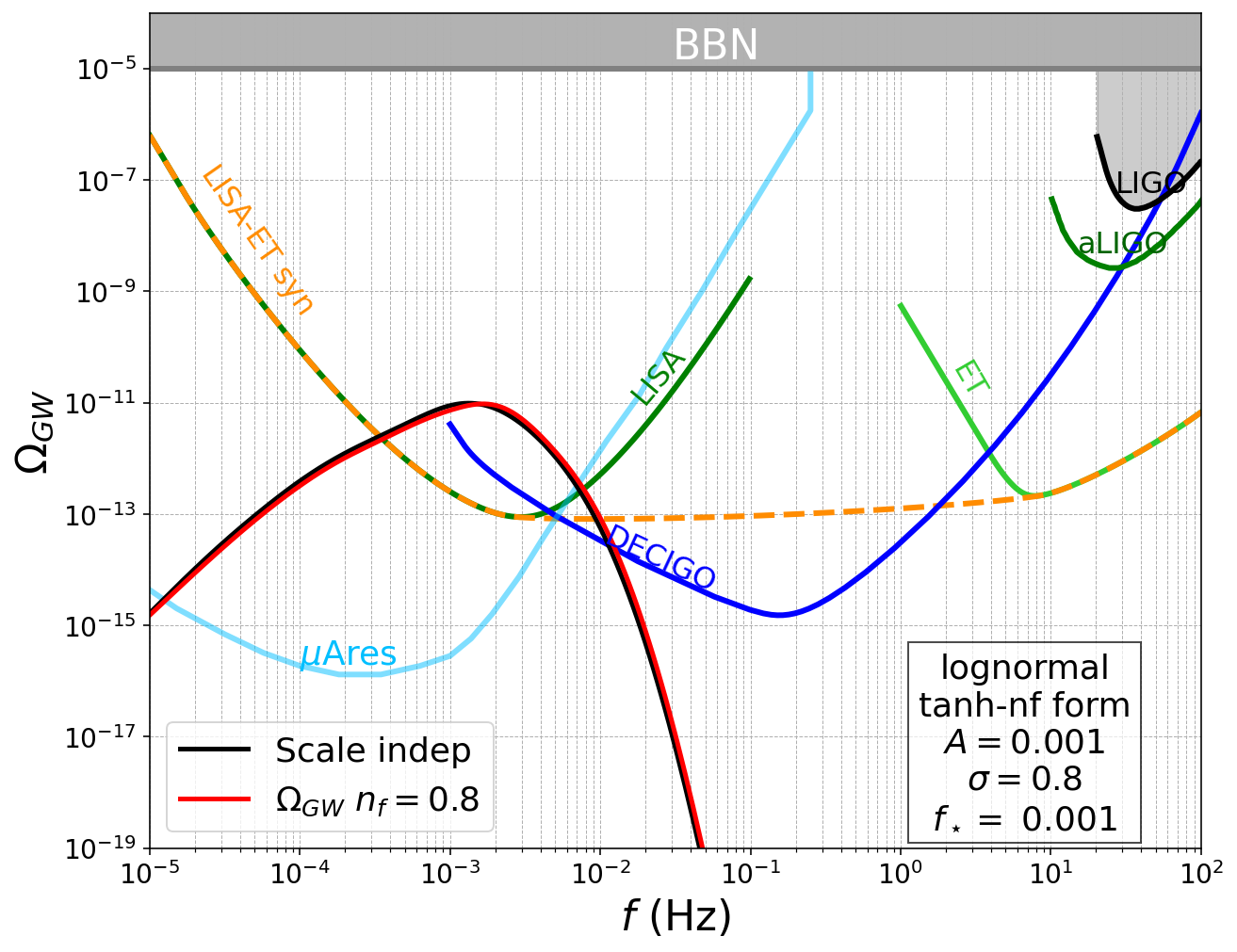}
\caption{\it Total spectrum for the log-normal source. Power-law running (left) enhances the high-frequency flank, {whereas UV-matched tanh running (right) suppresses the sub-pivot and pivot regions and approaches the constant-weight normalization above the transition.} These changes are not equivalent to a uniform rescaling of the constant-weight result.}
\label{fig:lognormal_final}
\end{figure}

\subsection{USR-inspired scalar spectrum}

Ultra-slow-roll phases during inflation naturally produce peaked spectra with a characteristic \(k^4\) rise followed by a decline. We use the regularized Pad\'e approximant of the USR template \cite{Tasinato:2020vdk,Tasinato:2023ukp}:
\begin{equation}
\Delta_{\mathcal R}^2(k) = A_s \, \Pi_{\rm new}(k/k_1),
\end{equation}
\begin{equation}
    \Pi_{\rm new}(x)=1-\Pi_0\frac{28x^2(6255-2578x^2)}{15(8757+1645x^2+153x^4)}+\Pi_0^2\frac{700x^4}{9(175+35x^2+4x^4)}
\end{equation}
with parameters yielding a peak at \(k_\star\). In the displayed example, we use \(A_s = 1 \times 10^{-11}\), $n_f=0.5$, and $\Pi_0=1000$ which should be interpreted as a shape-normalized perturbative test case rather than as a PBH-producing amplitude. A broken power-law form \(\Delta_{\mathcal R}^2(k) \propto (k/k_\star)^\alpha / [1 + (k/k_\star)^{\alpha+\beta}]\) provides a similar phenomenology.

In this class of spectra, the same type of non-Gaussian corrections can be particularly relevant for PBH formation once the scalar amplitude is raised to the PBH-relevant regime. {Scale-dependent running modifies the high-\(k\) tail and can shift the effective peak position, as illustrated in Fig.~\ref{fig:usr_final}, with potential consequences for SIGW detectability and the conditional PBH scaling discussed in Sec.~\ref{sec:phenom}.}

\begin{figure}[H]
\centering
\includegraphics[width=0.48\textwidth]{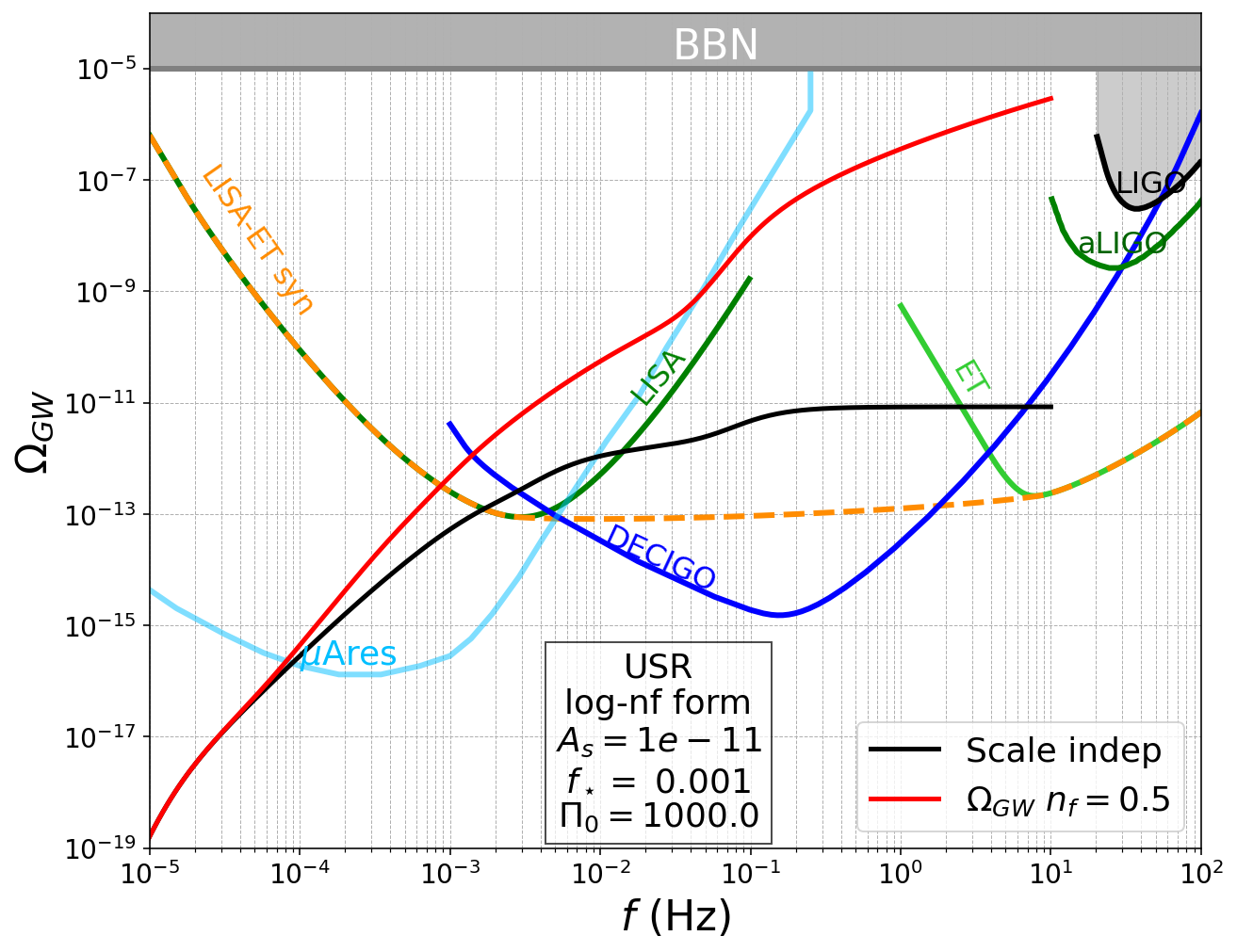}
\includegraphics[width=0.48\textwidth]{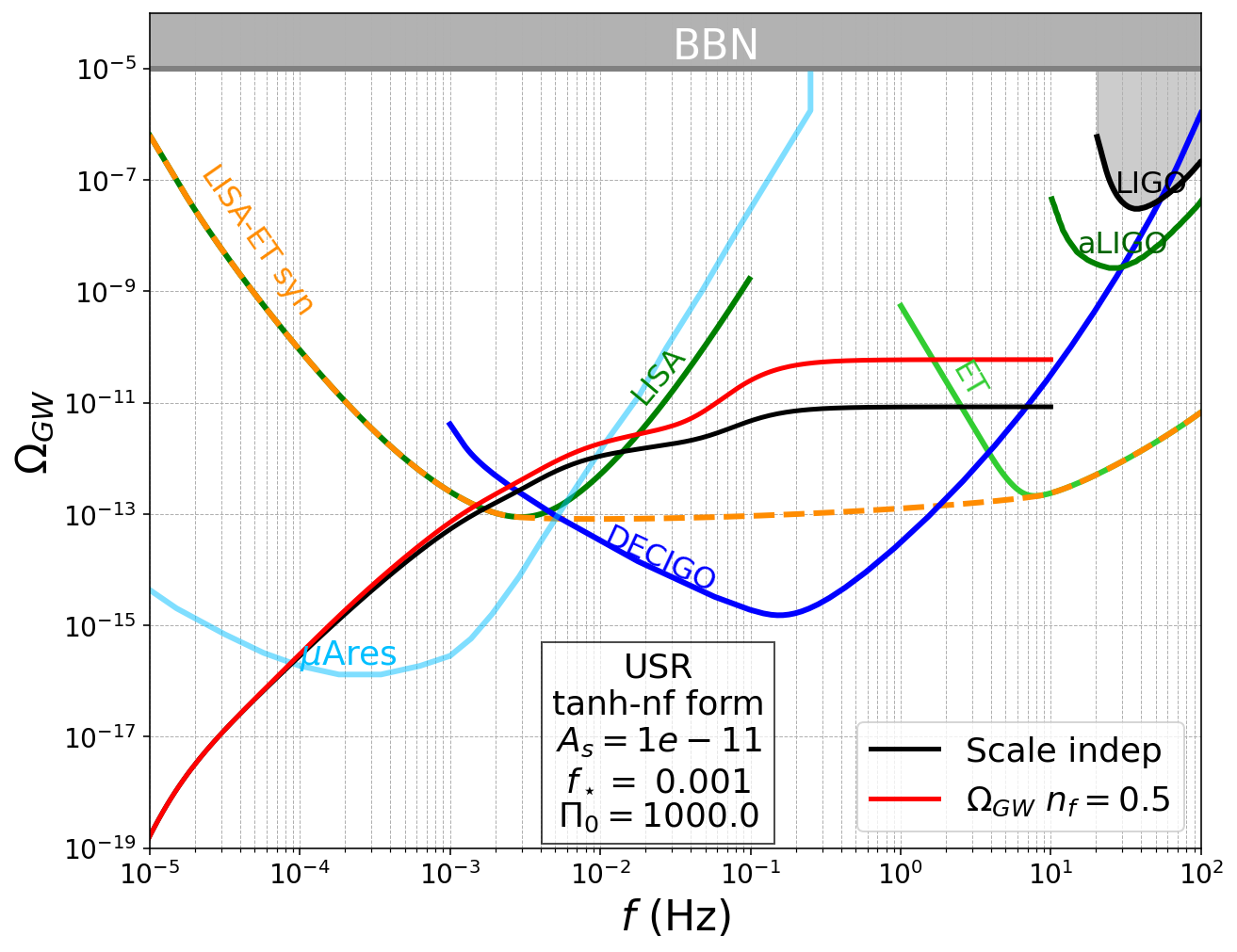}
\caption{\it Total \(\Omega_{\rm GW}\) for USR source. Power-law running produces enhancement on one side of the peak; UV-matched tanh running yields suppression. Red curves include full scale dependence.}
\label{fig:usr_final}
\end{figure}

{Across these benchmarks, the scale-dependent separable kernel primarily modifies spectral shape rather than producing a simple amplitude rescaling.} This produces a characteristic template-level signature that can be tested against constant-kernel and source-shape alternatives in a dedicated inference analysis. Quantitative results depend on the sign and magnitude of \(n_f\) as well as the location of \(k_\star\) relative to detector bands.

\subsection{Additional scalar power spectrum templates}
\label{sec:extra_templates}
To make the phenomenological survey fully self-contained, we collect here the additional scalar-source benchmarks that complement the monochromatic, log-normal, and USR-inspired examples discussed above. These extra templates broaden the coverage of the analysis and show that the qualitative conclusions of the paper are not tied to a single source morphology. In every case we use the same computational framework described in Secs.~\ref{sec:formalism} and \ref{sec:templates}: we evaluate the Gaussian contribution \(\Omega_{\rm GW}^{(0)}\), the \(\mathcal O(\bar f_{\rm NL}^2)\) correction \(\Omega_{\rm GW}^{(1)}\), and the \(\mathcal O(\bar f_{\rm NL}^4)\) correction \(\Omega_{\rm GW}^{(2)}\), and we compare the scale-independent case with both power-law running and UV-matched tanh running of the {internal-leg separable kernel}.

The purpose of this extended benchmark set is twofold. First, it documents the explicit scalar-spectrum choices used in the numerical scans. Second, it demonstrates that the central conclusions of the paper, especially the appearance of scale-dependent distortions in the shape of \(\Omega_{\rm GW}(f)\) and the benchmark-dependent competition between \(\mathcal O(\bar f_{\rm NL}^2)\) and \(\mathcal O(\bar f_{\rm NL}^4)\), persist across compact bumps, broad asymmetric sources, and multi-slope spectra. Throughout this subsection we retain the same decomposition into Gaussian, reducible, and connected non-Gaussian pieces used in the rest of the paper, and whenever a frequency label is attached to a benchmark it is understood using the present-day conversion in Eq.~\eqref{eq:k_to_f_conversion}.

\subsection{Gaussian scalar bump}
\label{sec:gaussian_bump_extra}

A useful benchmark interpolating between a very narrow source and a smooth broad bump is a Gaussian profile in linear wavenumber,
\begin{equation}
\Delta_{\mathcal R}^2(k)
= \left( \frac{k}{k_\star} \right)^3
\frac{A_r}{\sqrt{2\pi\,(\sigma/k_\star)^2}}
\exp\!\left[-\frac{(k-k_\star)^2}{2\sigma^2}\right],
\label{eq:gaussian_bump_extra}
\end{equation}
where $A_r$ is the amplitude parameter used in the numerical benchmark. For this finite-width profile it is not identical to $\int d\ln k\,\Delta_{\mathcal R}^2(k)$ unless an additional width-dependent normalization is included. The prefactor $(k/k_\star)^3$ guarantees the correct low-$k$ behaviour of the scalar source and makes the profile particularly convenient for studying how localized bumps in linear momentum are mapped into induced-gravitational-wave spectra.

For this benchmark we use $A_r=10^{-3}$, a dimensionless width parameter $\sigma/k_\star\simeq 0.8$, and peak frequencies in the mHz and nHz windows in order to illustrate both LISA-like and PTA-like responses. The corresponding scalar power spectrum is shown in Fig.~\ref{fig:app_gaussian_ps}. The resulting decomposition of the SIGW signal into $\mathcal O(\bar f_{\rm NL}^0)$, $\mathcal O(\bar f_{\rm NL}^2)$, and $\mathcal O(\bar f_{\rm NL}^4)$ pieces is displayed in Fig.~\ref{fig:app_gaussian_orders}, while the total spectra are shown in Figs.~\ref{fig:app_gaussian_total_lisa} and \ref{fig:app_gaussian_total_pta}.

The main feature of this benchmark is that the scalar source is sufficiently smooth to wash out the sharp multi-peak structure familiar from the monochromatic limit, while still remaining localized enough for shape distortions induced by the running leg weight $f(k)$ to be clearly visible. In the power-law running case, the high-frequency side of the SIGW peak is selectively enhanced, and the peak acquires a mild asymmetry. The enhancement is most visible near and slightly above the peak frequency, where the scale-dependent weighting of the non-Gaussian convolution favors the ultraviolet side of the scalar support. By contrast, the UV-matched tanh template primarily delays the onset of the non-Gaussian contribution through the transition region and then saturates, so the resulting spectrum is closer to the scale-independent result at high momenta than a power-law running template but can remain visibly depleted over the finite ultraviolet range shown in the plots. Relative to the log-normal case discussed previously, the Gaussian scalar bump source produces a slightly sharper transition between the peak region and the ultraviolet fall-off, and hence isolates more cleanly the difference between enhancement-type and saturation-type running.

This benchmark also provides a useful cross-check of the numerical stability of the integrals at intermediate width. If the source is too narrow, Monte Carlo convergence degrades because the support becomes highly localized in momentum space; if it is too broad, the induced spectrum begins to resemble the scale-independent broad-plateau limit. The Gaussian scalar bump sits between these two extremes and confirms the main conclusion of this work; namely that running non-Gaussianity affects the spectral shape more strongly than a simple amplitude rescaling.

\begin{figure}[H]
    \centering
    \includegraphics[width=0.56\textwidth]{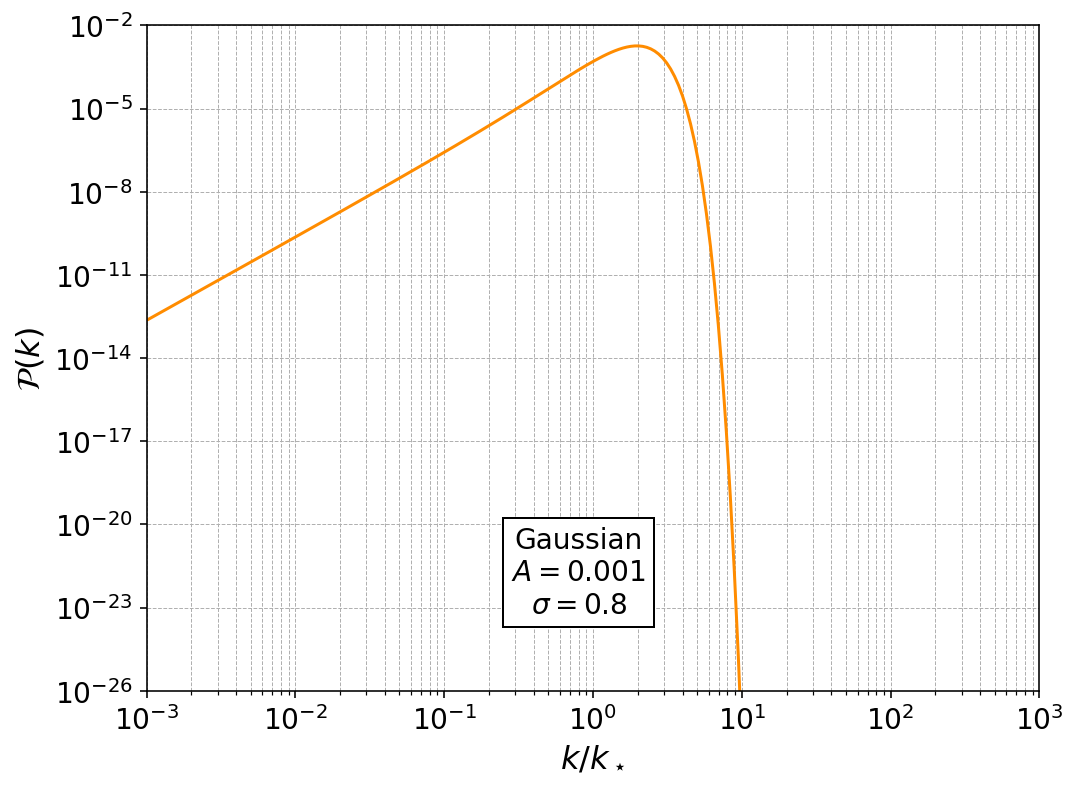}
    \caption{\it Gaussian scalar bump power spectrum used in the additional in-text benchmark, defined in Eq.~\eqref{eq:gaussian_bump_extra}.}
    \label{fig:app_gaussian_ps}
\end{figure}

\begin{figure}[H]
    \centering
    \begin{subfigure}{0.48\textwidth}
        \includegraphics[width=\textwidth]{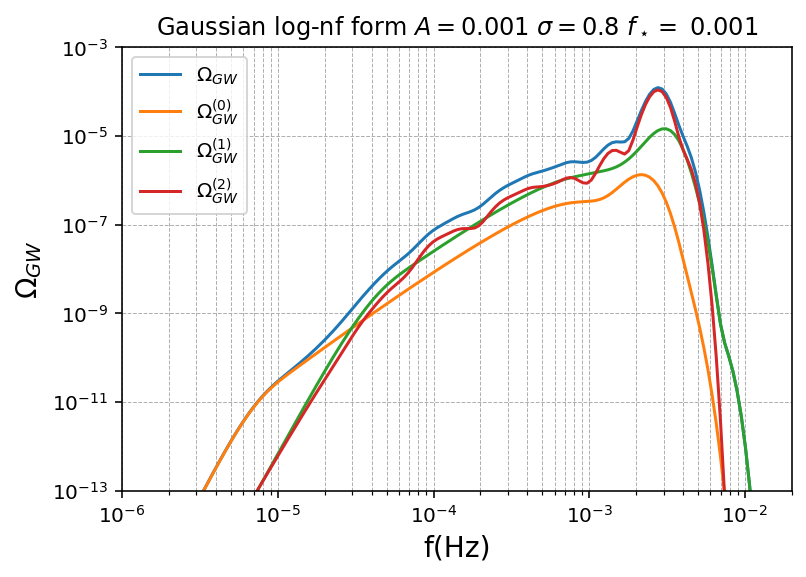}
    \end{subfigure}
    \hfill
    \begin{subfigure}{0.48\textwidth}
        \includegraphics[width=\textwidth]{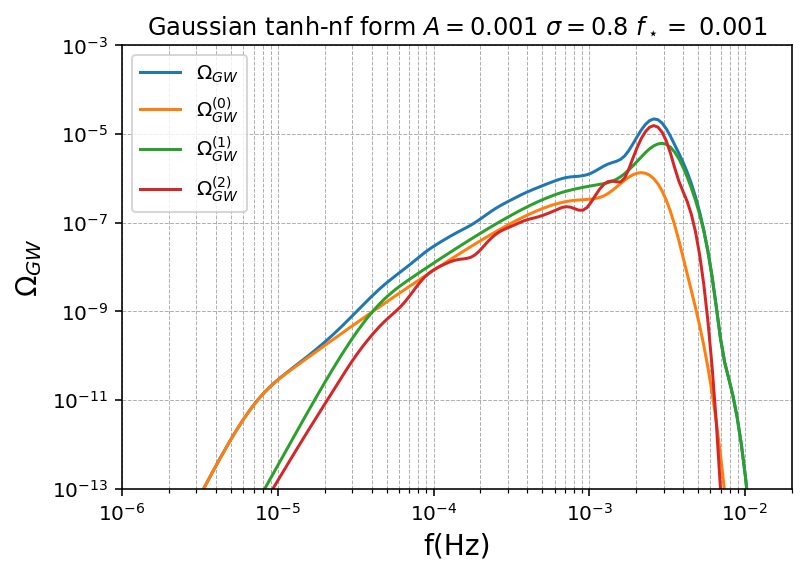}
    \end{subfigure}
    \caption{\it Order-by-order SIGW decomposition for the Gaussian scalar bump. Left: power-law weight. Right: tanh weight. Orange, green, red, and blue denote the Gaussian, $\mathcal O(\bar f_{\rm NL}^2)$, $\mathcal O(\bar f_{\rm NL}^4)$, and total spectra, respectively.}
    \label{fig:app_gaussian_orders}
\end{figure}

\begin{figure}[H]
    \centering
    \begin{subfigure}{0.48\textwidth}
        \includegraphics[width=\textwidth]{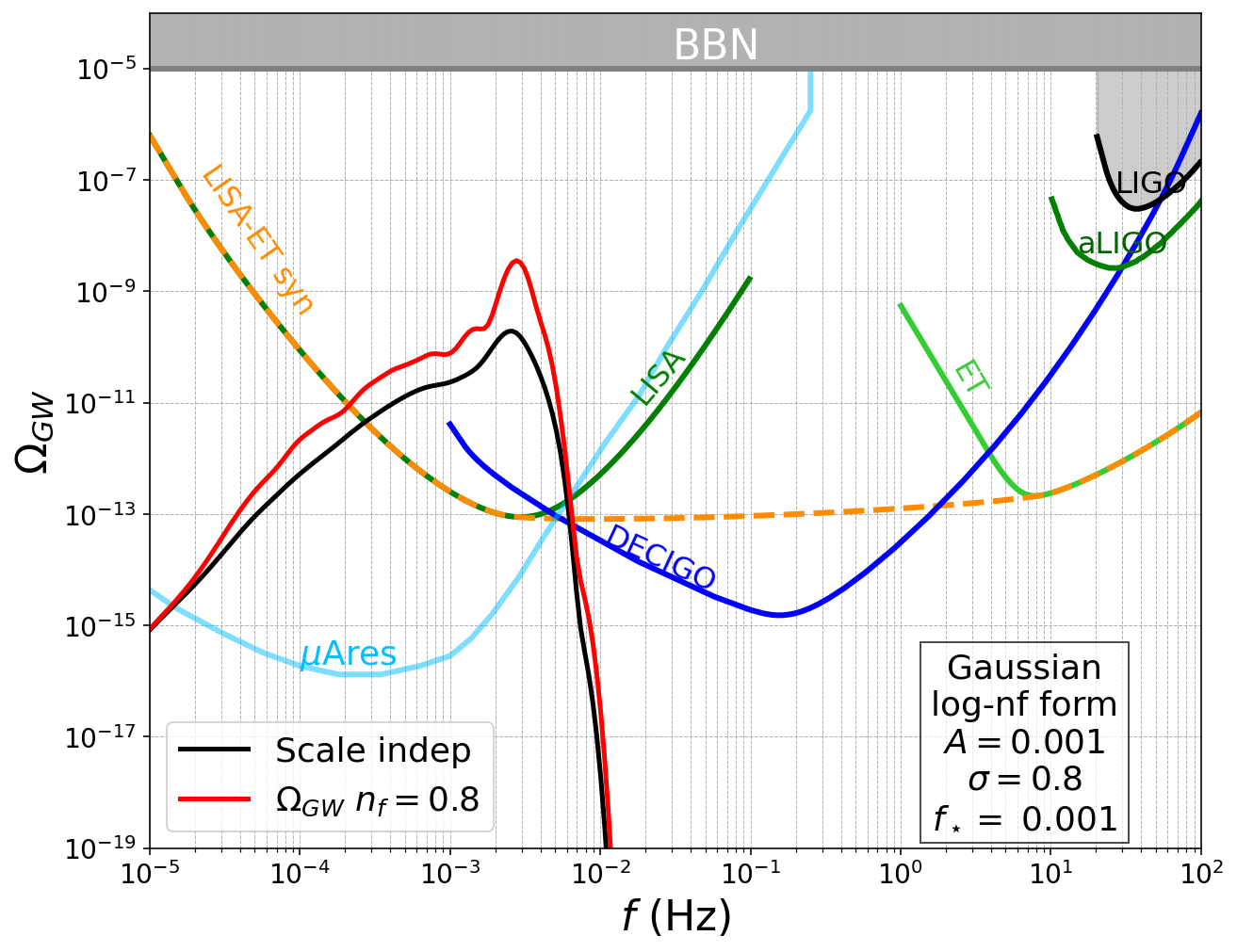}
    \end{subfigure}
    \hfill
    \begin{subfigure}{0.48\textwidth}
        \includegraphics[width=\textwidth]{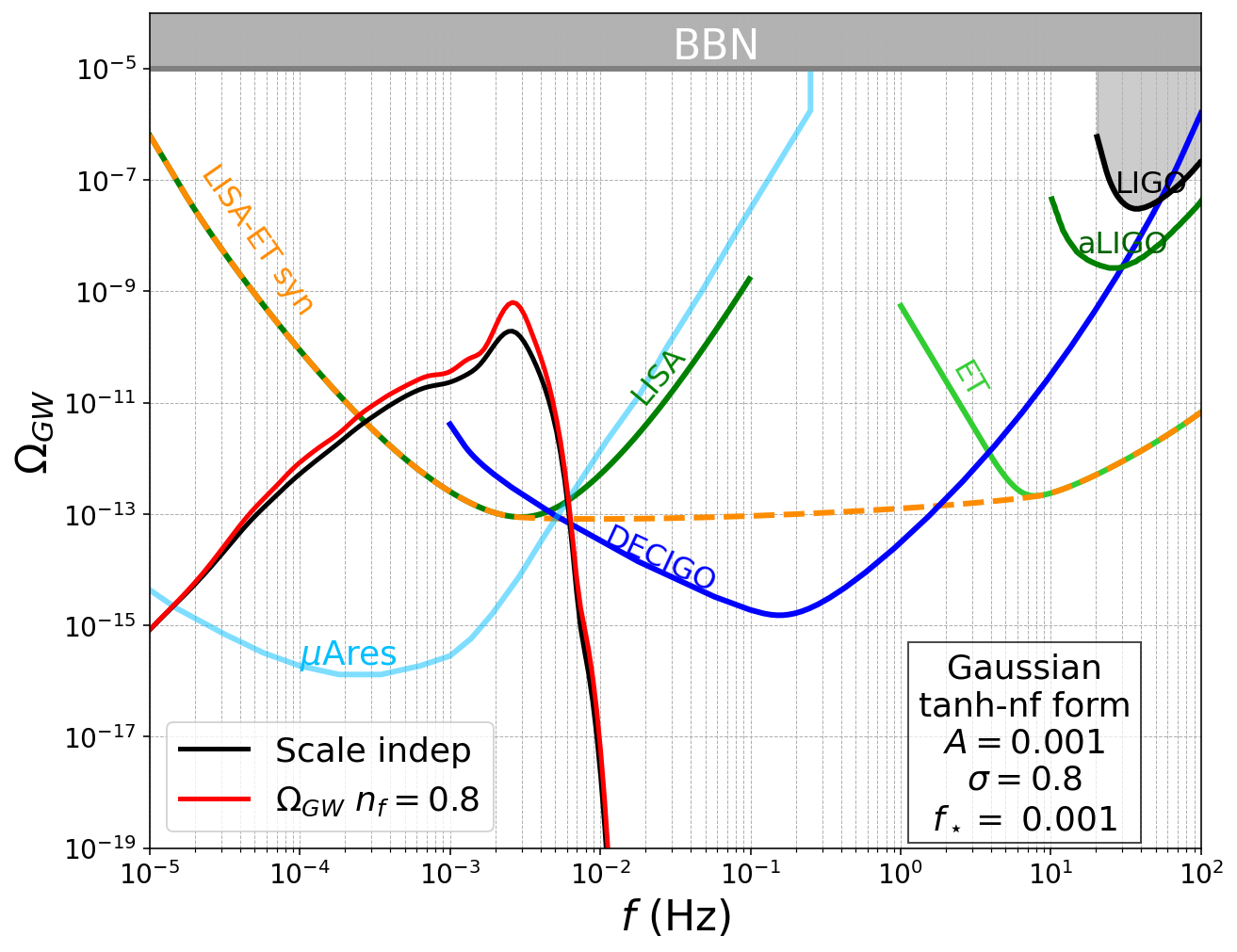}
    \end{subfigure}
    \caption{\it Total induced GW spectrum for the Gaussian scalar bump source with a peak placed in the LISA band. Black: scale-independent non-Gaussianity. Red: scale-dependent running.}
    \label{fig:app_gaussian_total_lisa}
\end{figure}

\begin{figure}[H]
    \centering
    \begin{subfigure}{0.48\textwidth}
        \includegraphics[width=\textwidth]{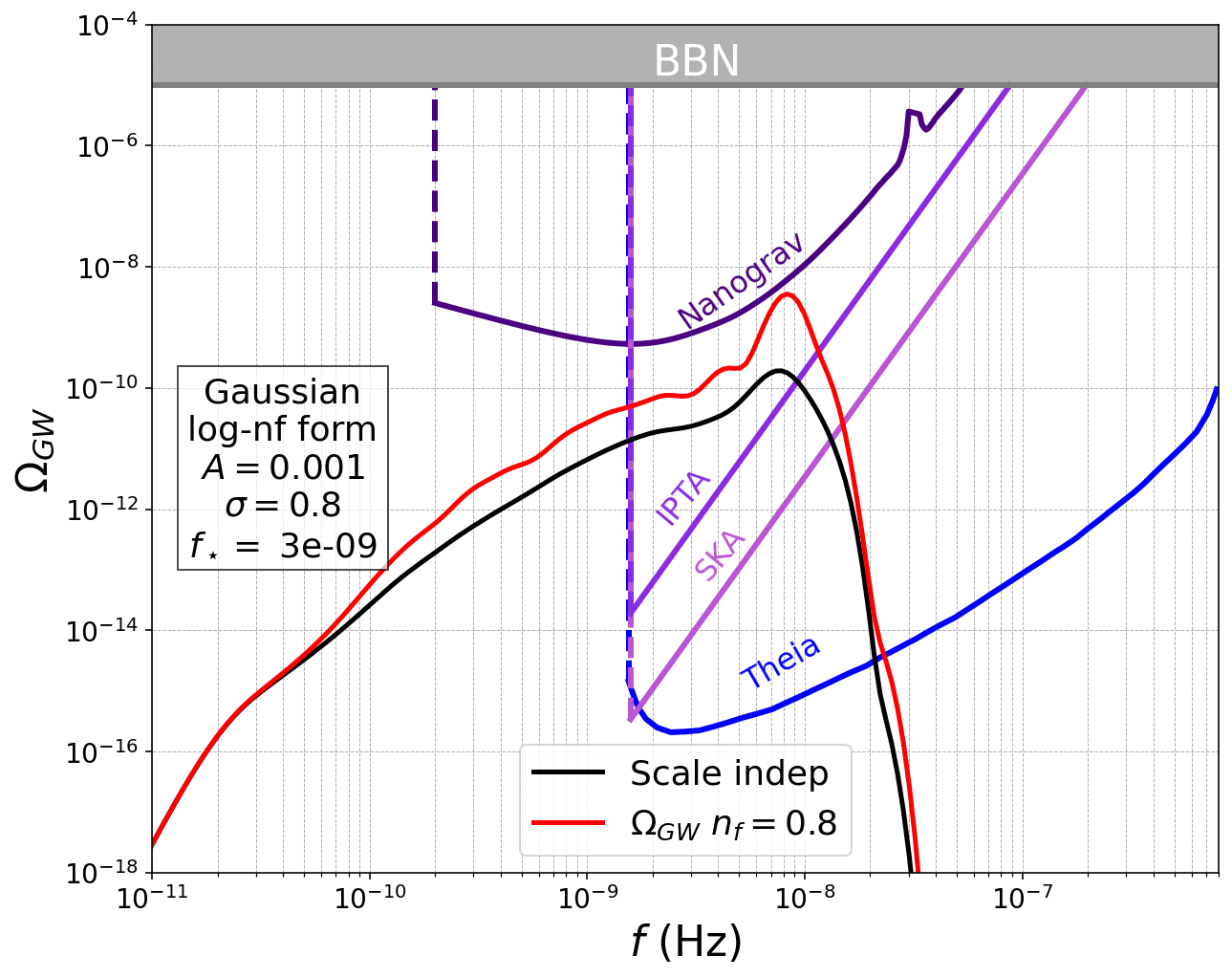}
    \end{subfigure}
    \hfill
    \begin{subfigure}{0.48\textwidth}
        \includegraphics[width=\textwidth]{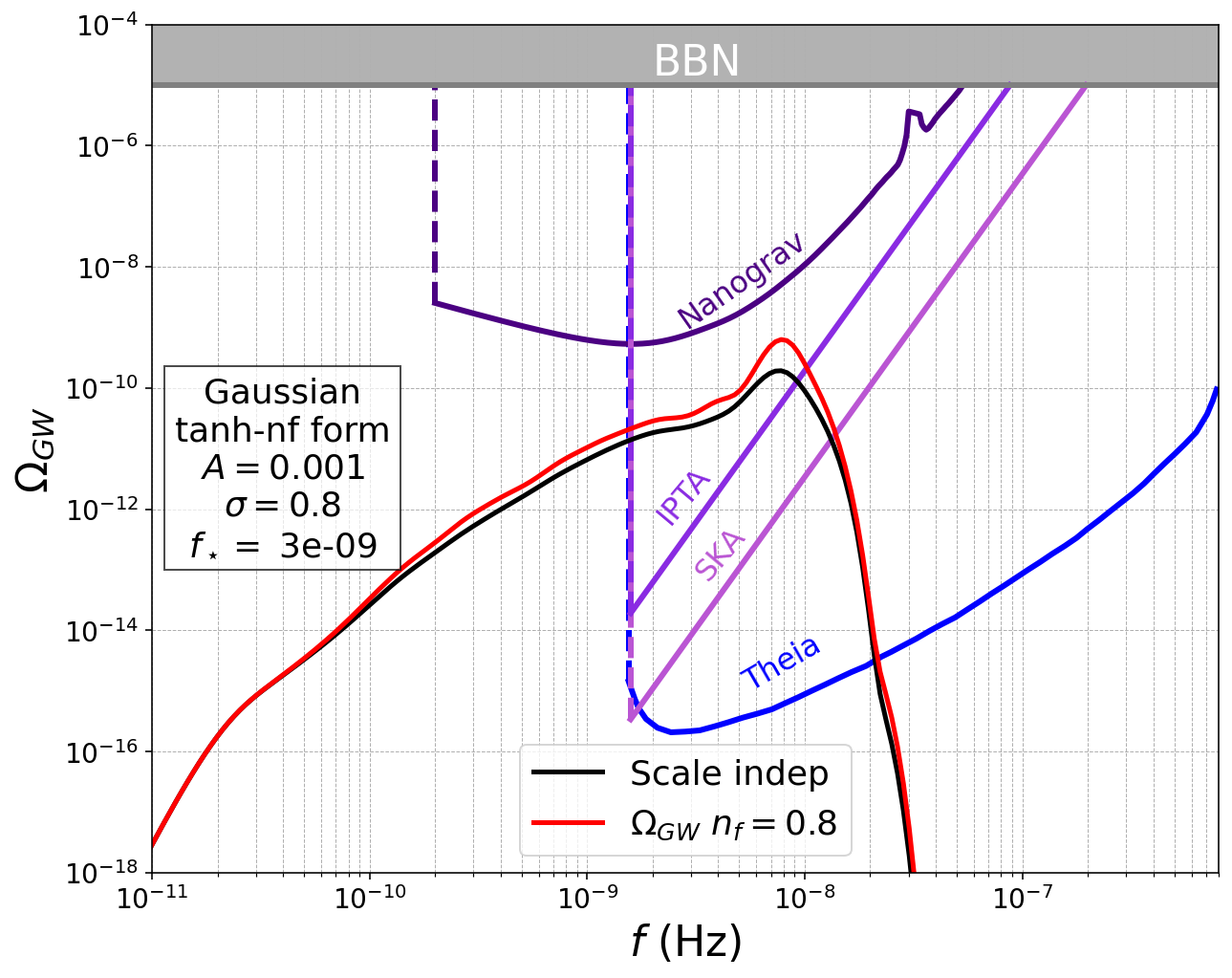}
    \end{subfigure}
    \caption{\it Same as Fig.~\ref{fig:app_gaussian_total_lisa}, but with the peak tuned to the PTA band.}
    \label{fig:app_gaussian_total_pta}
\end{figure}

\subsection{Power-law spectrum with exponential cutoff}
\label{sec:powerlaw_cutoff_extra}

A second benchmark is the power-law source with an exponential cutoff,
\begin{equation}
\Delta_{\mathcal R}^2(k)
= A_r\left(\frac{k}{k_\star}\right)^{\alpha}
\exp\!\left[-\alpha\left(\frac{k}{k_\star}-1\right)\right],
\label{eq:powerlaw_exp_extra}
\end{equation}
which peaks at $k=k_\star$ with amplitude $A_r$ for any value of the slope parameter $\alpha$. This form is useful when one wants a source that rises as a pure power law in the infrared but is exponentially suppressed at high momentum. Such spectra arise naturally as smooth approximations to peaked inflationary spectra; see, for example, the phenomenological peaked-spectrum discussions above, in which the growth toward the maximum is dominated by an approximately constant local tilt over a finite momentum interval.

For this benchmark we use $A_r=5\times10^{-4}$ and $\alpha=1$, with the peak frequency chosen once in the mHz region and once in the PTA window. The corresponding scalar profile is shown in Fig.~\ref{fig:app_pl_ps}. Unlike the Gaussian scalar bump or log-normal cases, the infrared side of the source here is much broader in logarithmic momentum, and this has a clear impact on the induced signal. Because a sizable portion of the convolution support already lies at $k<k_\star$, the Gaussian and non-Gaussian pieces remain more similar in shape than in sharply localized templates and develop comparatively broad peaks.

The decomposition in Fig.~\ref{fig:app_pl_orders} shows that the $\mathcal O(\bar f_{\rm NL}^2)$ and $\mathcal O(\bar f_{\rm NL}^4)$ contributions remain relevant but do not generate as striking a multi-feature structure as in the monochromatic benchmark. Instead, the most visible imprint of scale-dependent running is a redistribution of power between the shoulder of the peak and the ultraviolet cutoff region. In the power-law running case, the red scale-dependent curve is lifted relative to the scale-independent one over a broad range of $k \gtrsim k_\star$, while the UV-matched tanh template causes a smooth transition-region depletion and then approaches a saturated high-momentum weighting. In both cases the net effect on the total spectrum is of order unity rather than an order-of-magnitude reshaping, making this benchmark particularly useful for illustrating that the detectability of running non-Gaussianity depends not only on the amplitude of the primordial bispectrum but also on the detailed morphology of the scalar source.

Another notable feature is the behaviour around the main peak: both running templates can induce a mild dip or flattening relative to the scale-independent curve before the ultraviolet tail separates. This is a consequence of the competition between enhanced/suppressed non-Gaussian weighting and the exponential damping already built into the scalar source. Therefore, for exponentially cut-off spectra, the distinction between shape-induced peak broadening and ultraviolet-tail modification becomes important when translating results into detector-level forecasts. {Figures~\ref{fig:app_pl_total_lisa} and \ref{fig:app_pl_total_pta} show that this morphology is preserved when the characteristic scale is moved between the LISA and PTA bands.}

\begin{figure}[H]
    \centering
    \includegraphics[width=0.60\textwidth]{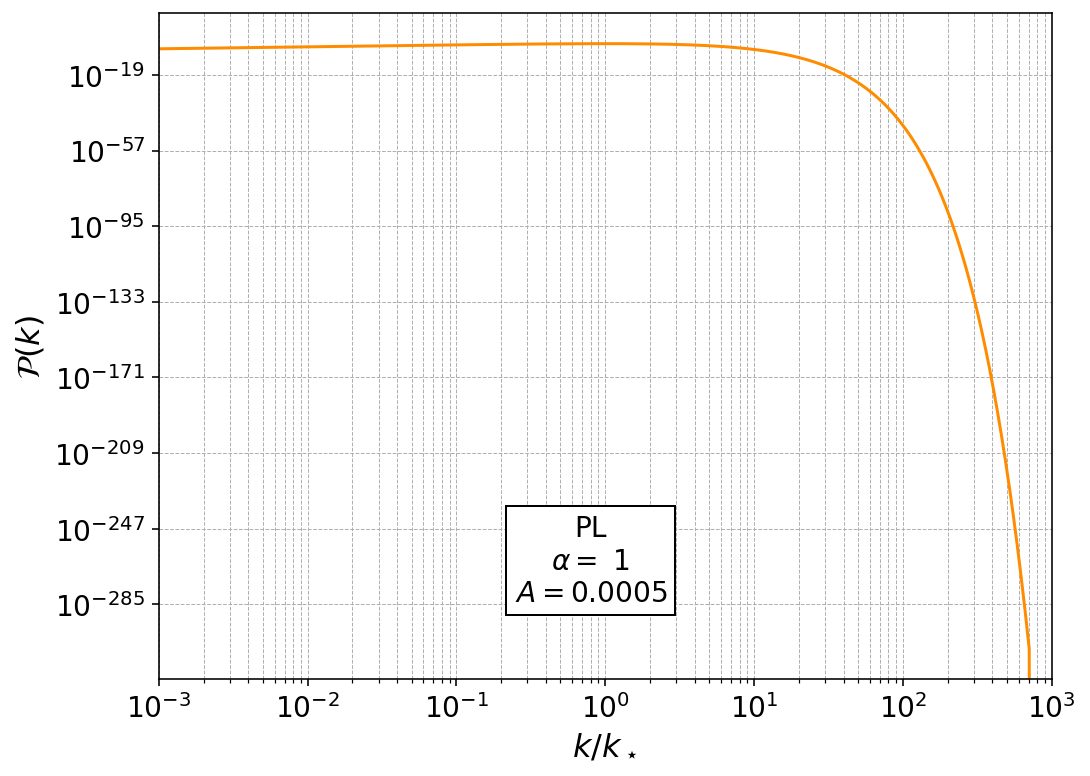}
    \caption{\it Power-law scalar spectrum with exponential cutoff, Eq.~\eqref{eq:powerlaw_exp_extra}.}
    \label{fig:app_pl_ps}
\end{figure}

\begin{figure}[H]
    \centering
    \begin{subfigure}{0.48\textwidth}
        \includegraphics[width=\textwidth]{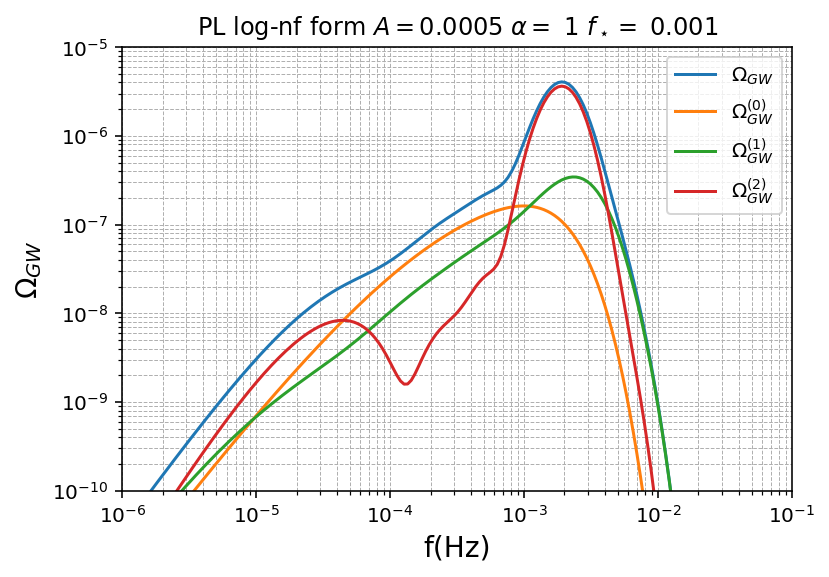}
    \end{subfigure}
    \hfill
    \begin{subfigure}{0.48\textwidth}
        \includegraphics[width=\textwidth]{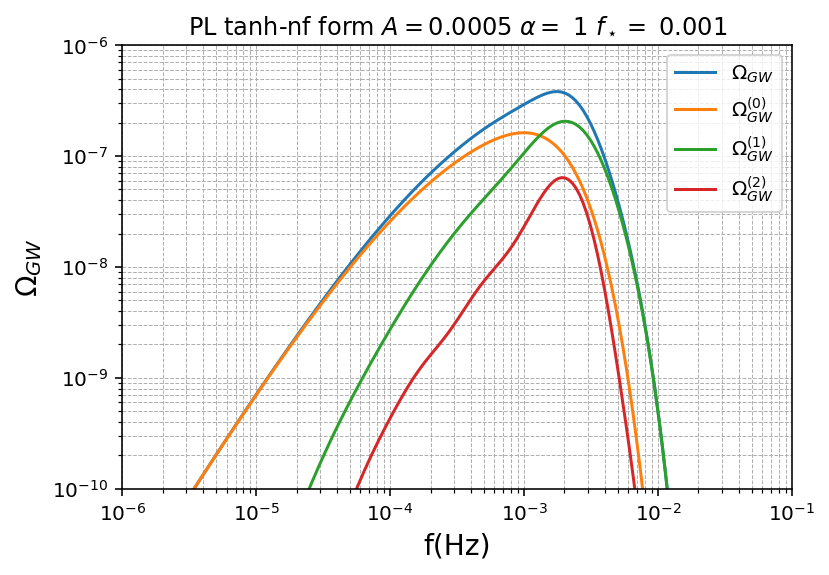}
    \end{subfigure}
    \caption{\it Order-by-order decomposition of the induced GW spectrum for the power-law-with-cutoff source.}
    \label{fig:app_pl_orders}
\end{figure}

\begin{figure}[H]
    \centering
    \begin{subfigure}{0.48\textwidth}
        \includegraphics[width=\textwidth]{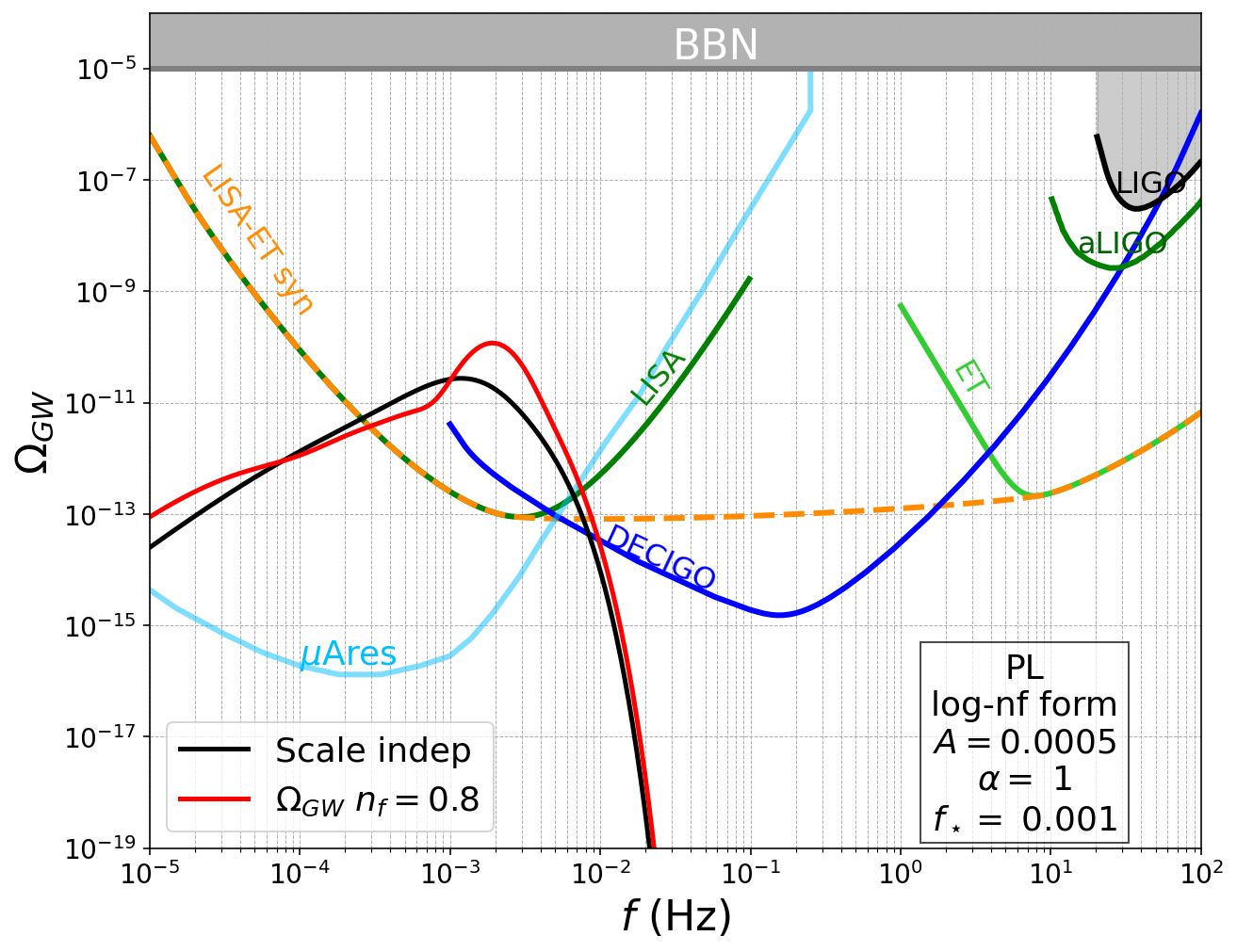}
    \end{subfigure}
    \hfill
    \begin{subfigure}{0.48\textwidth}
        \includegraphics[width=\textwidth]{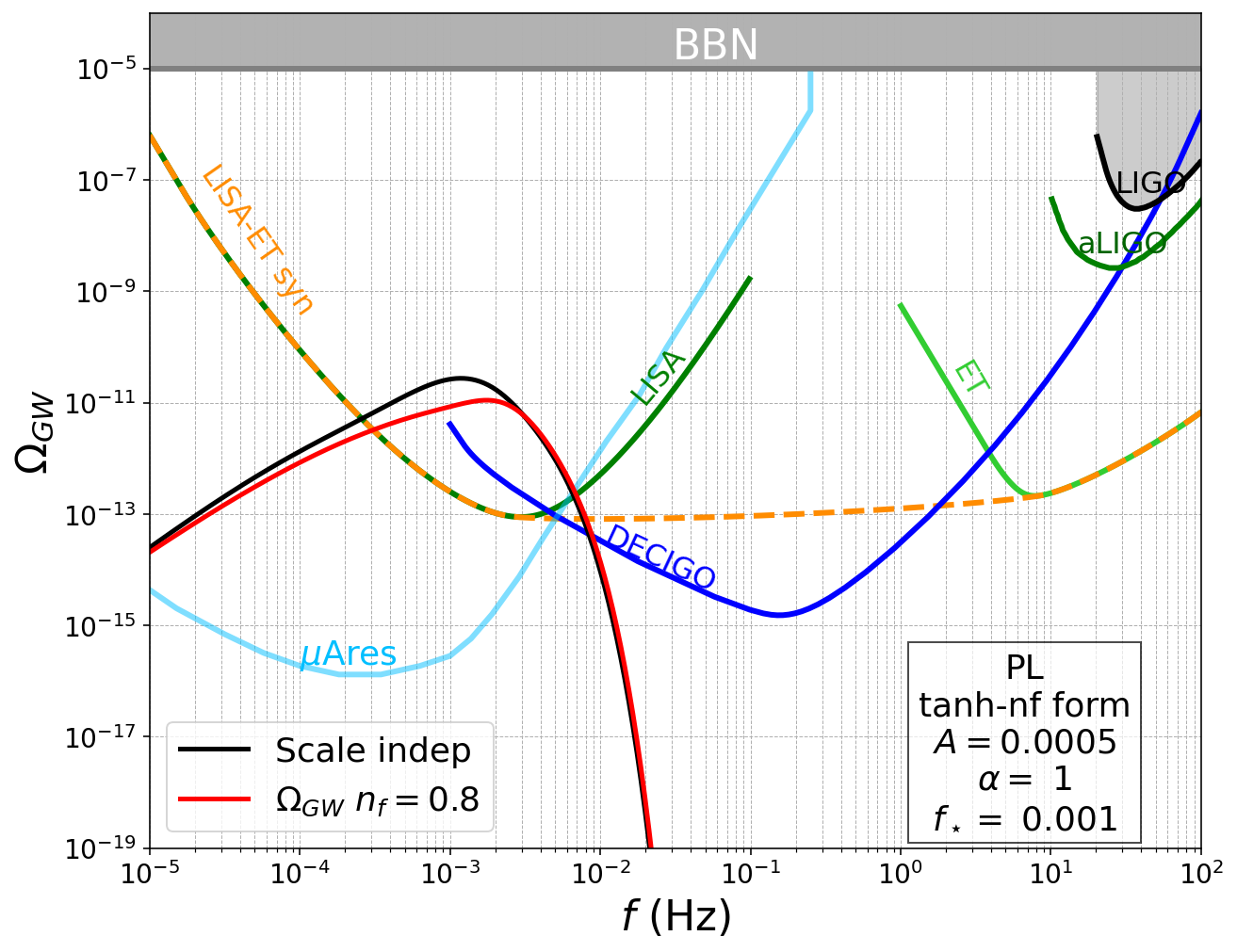}
    \end{subfigure}
    \caption{\it Total induced GW spectrum for the power-law-with-cutoff source with the peak placed in the LISA band.}
    \label{fig:app_pl_total_lisa}
\end{figure}

\begin{figure}[H]
    \centering
    \begin{subfigure}{0.48\textwidth}
        \includegraphics[width=\textwidth]{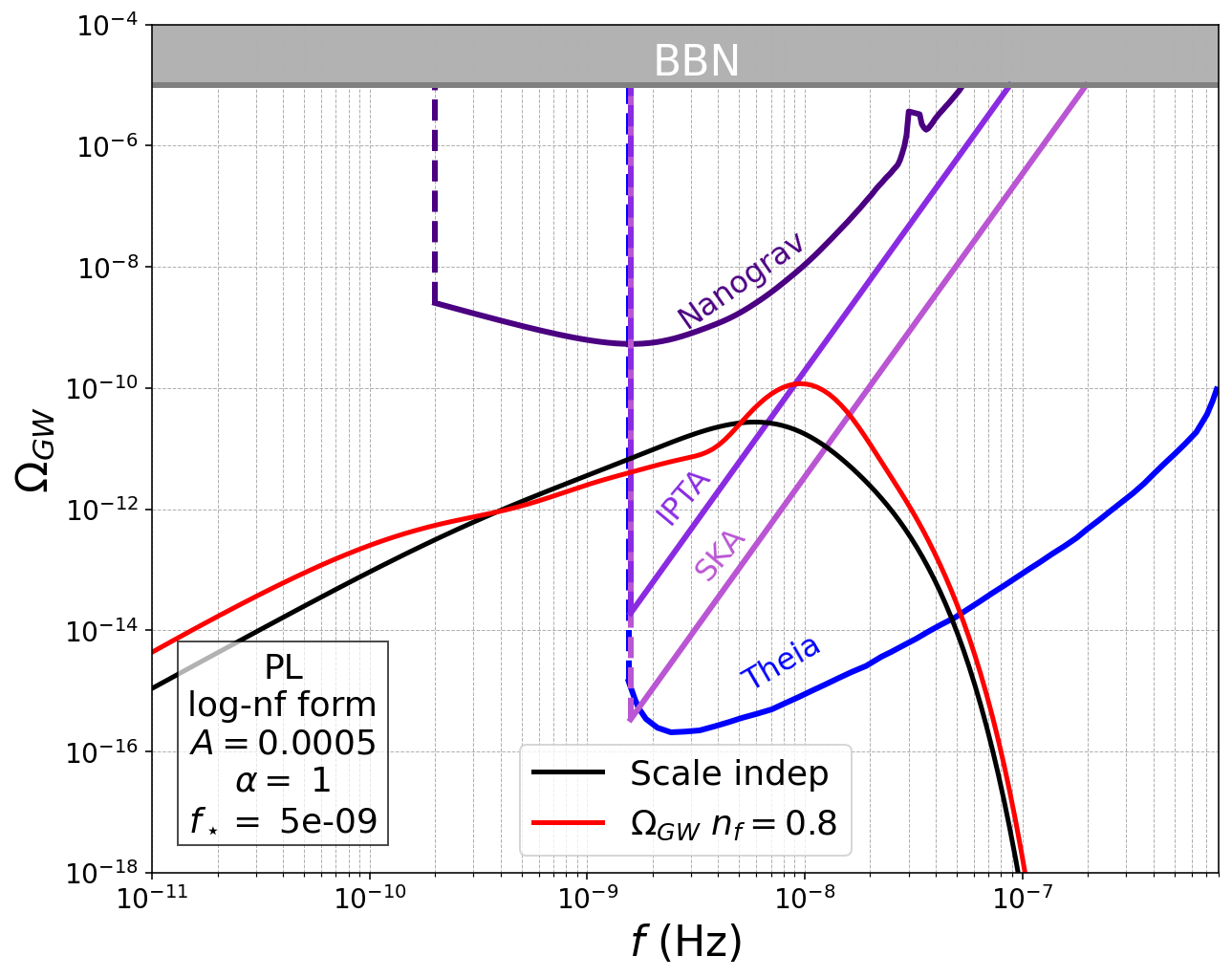}
    \end{subfigure}
    \hfill
    \begin{subfigure}{0.48\textwidth}
        \includegraphics[width=\textwidth]{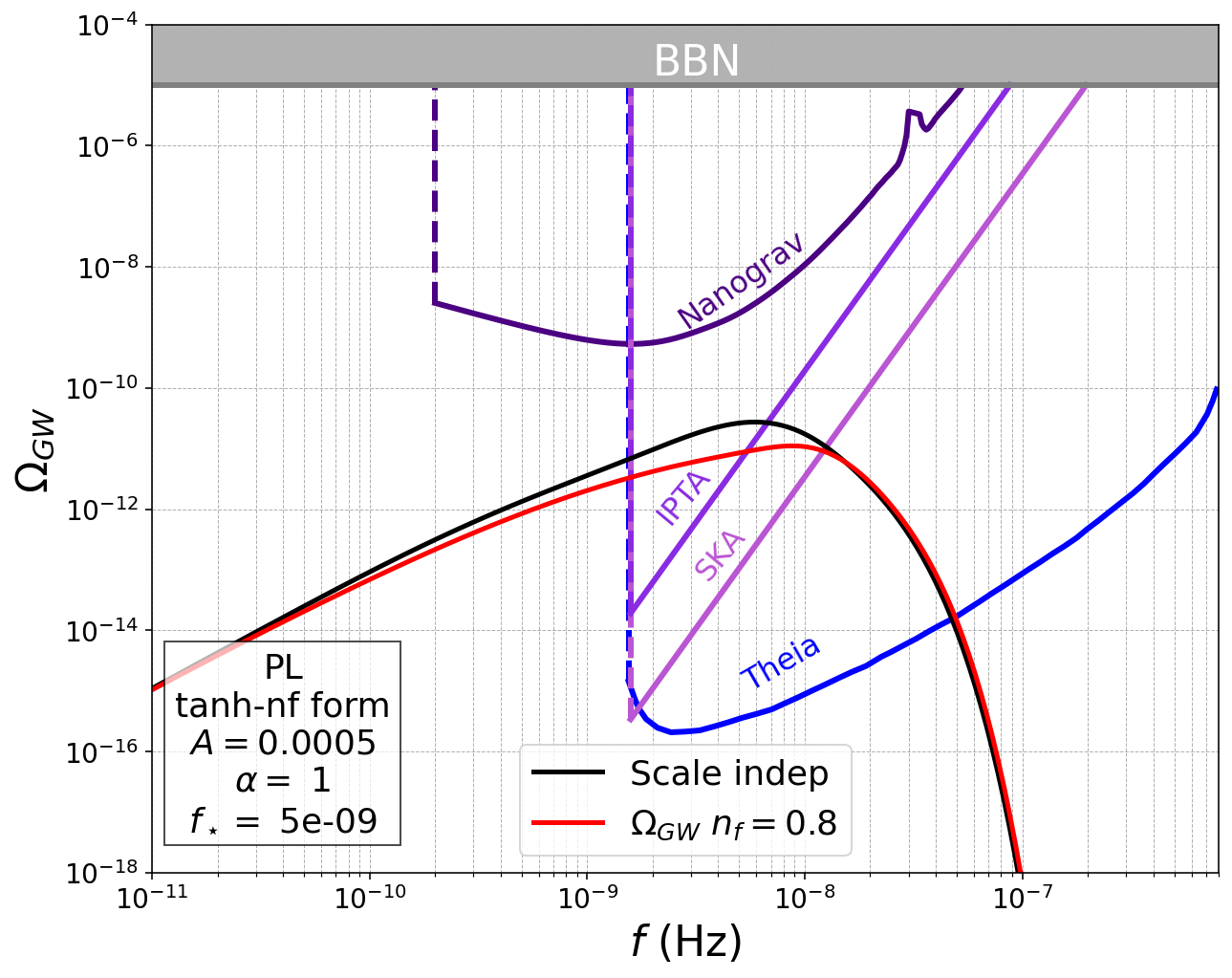}
    \end{subfigure}
    \caption{\it Same as Fig.~\ref{fig:app_pl_total_lisa}, but with the peak tuned to the PTA band.}
    \label{fig:app_pl_total_pta}
\end{figure}

\subsection{Broken power-law spectrum}
\label{sec:broken_power_law_extra}

To model a source with independent infrared and ultraviolet slopes, one may use the broken-power-law parameterization, which we treat here as a phenomenological peaked-spectrum template
\begin{equation}
\Delta_{\mathcal R}^2(k)
= \frac{A_r}{(k/k_\star)^{-\alpha} + (k/k_\star)^{\beta}},
\label{eq:bpl_extra}
\end{equation}
which scales as $k^{\alpha}$ for $k\ll k_\star$ and as $k^{-\beta}$ for $k\gg k_\star$. Here $k_\star$ is the break scale: for $\alpha\neq\beta$ the maximum is displaced from $k_\star$, and the value at the break is $A_r/2$. This is a standard phenomenological form for peaked inflationary power spectra and is especially convenient when one wishes to study how the asymmetry of the source around its peak affects the resulting SIGW signal. In the numerical examples we use $(\alpha,\beta)=(1,5)$, $A_r=1\times10^{-3}$ which corresponds to a relatively gentle infrared rise and a steep ultraviolet decay.

The scalar spectrum is shown in Fig.~\ref{fig:app_bpl_ps}, and the decomposed SIGW contributions are displayed in Fig.~\ref{fig:app_bpl_orders}. Compared with the previous templates, the broken-power-law source produces a visibly asymmetric induced spectrum already in the absence of running non-Gaussianity, since the convolution samples very different amounts of support on the two sides of the peak. The effect of power-law running is then to further accentuate this asymmetry: the left panel of Fig.~\ref{fig:app_bpl_total_lisa} shows that the ultraviolet side of the spectrum is enhanced while the region immediately around the peak can become moderately suppressed relative to the scale-independent case. In other words, the running does not simply ``tilt'' the entire spectrum upward; rather, it redistributes power in a scale-dependent way that depends on how the convolution support overlaps with the steep UV fall-off of Eq.~\eqref{eq:bpl_extra}.

The UV-matched tanh template behaves differently. Because it saturates to a constant at high momentum, it tends to leave the deep ultraviolet closer to the constant-$f_{\rm NL}$ case while suppressing part of the transition region around the peak. Hence the tanh-driven deformation is generally milder and smoother than the power-law case, with the largest differences occurring below or around $k_\star$ rather than far into the ultraviolet tail. This benchmark is therefore particularly useful for disentangling two qualitatively distinct effects of running PNG: enhancement-like reshaping of the UV side versus saturation-like suppression around the transition scale.

The broken-power-law source also makes clear that the importance of non-Gaussian corrections is sensitive to the amount of infrared support. Since the source extends over a wide range of $k<k_\star$, the connected terms contribute over a broad domain in momentum space instead of concentrating only near the peak. The resulting total spectrum in the PTA-tuned configuration, shown in Fig.~\ref{fig:app_bpl_total_pta}, displays both a low-frequency enhancement and a peak-adjacent suppression for the power-law template, whereas the UV-matched tanh template predominantly suppresses the signal. These {behaviors} remain qualitatively stable under shifts of the peak scale from the LISA band to the PTA band.

\begin{figure}[H]
    \centering
    \includegraphics[width=0.60\textwidth]{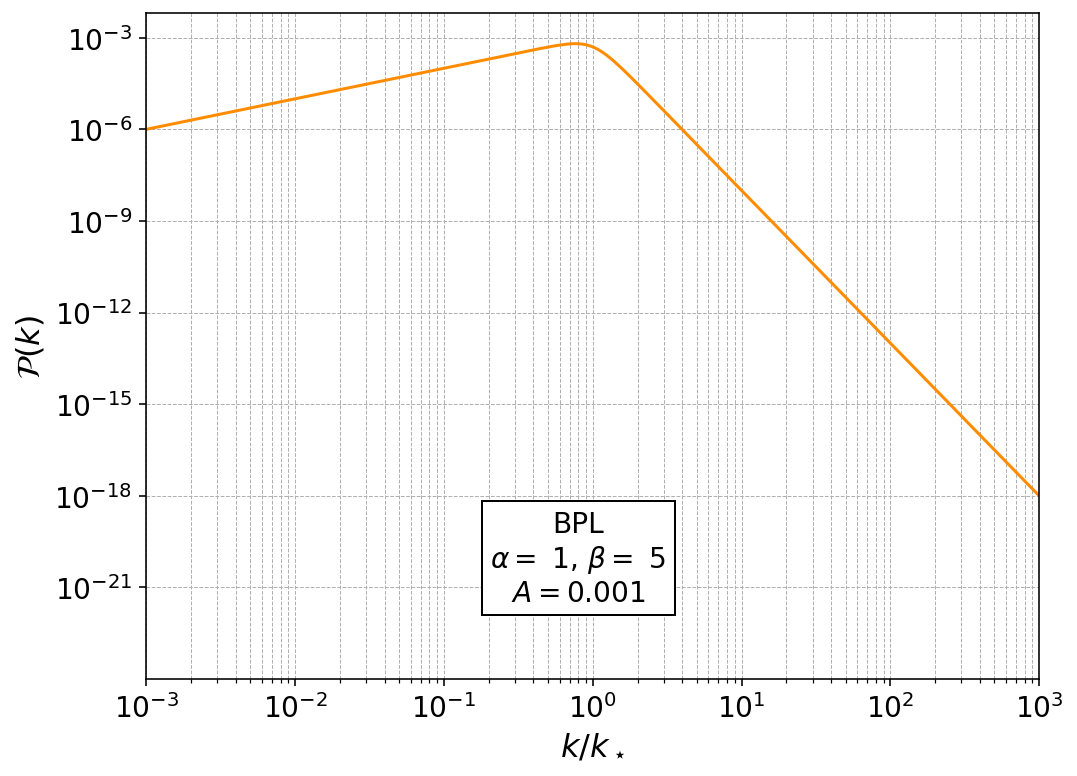}
    \caption{\it Broken-power-law scalar spectrum used for the additional in-text benchmark, Eq.~\eqref{eq:bpl_extra}.}
    \label{fig:app_bpl_ps}
\end{figure}

\begin{figure}[H]
    \centering
    \begin{subfigure}{0.48\textwidth}
        \includegraphics[width=\textwidth]{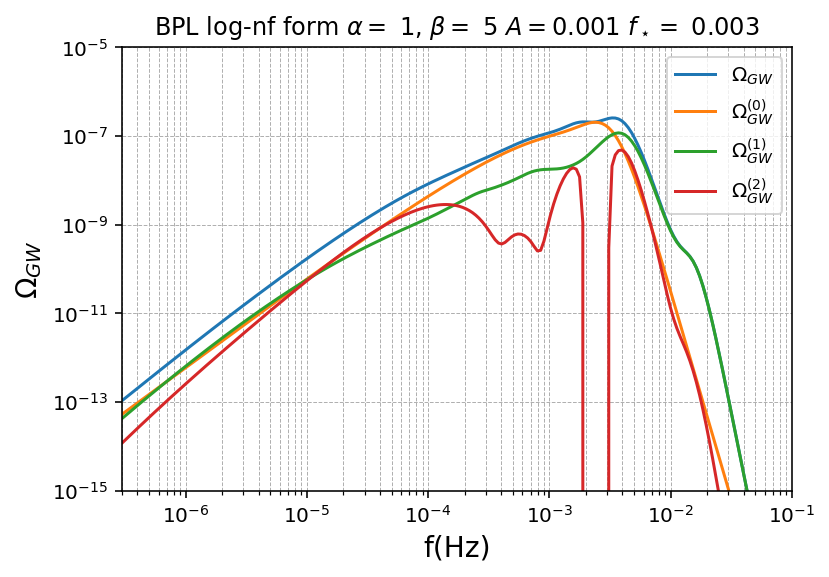}
    \end{subfigure}
    \hfill
    \begin{subfigure}{0.48\textwidth}
        \includegraphics[width=\textwidth]{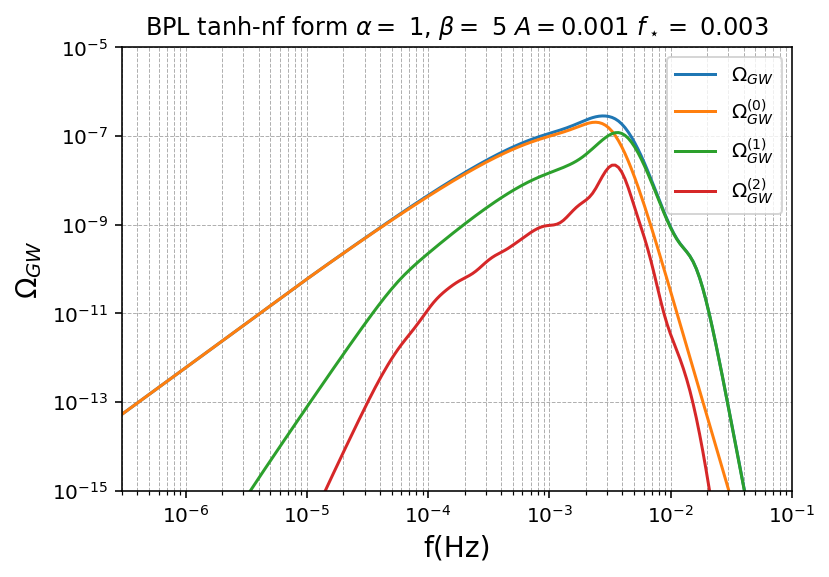}
    \end{subfigure}
    \caption{\it Order-by-order decomposition of the induced GW spectrum for the broken-power-law source.}
    \label{fig:app_bpl_orders}
\end{figure}

\begin{figure}[H]
    \centering
    \begin{subfigure}{0.48\textwidth}
        \includegraphics[width=\textwidth]{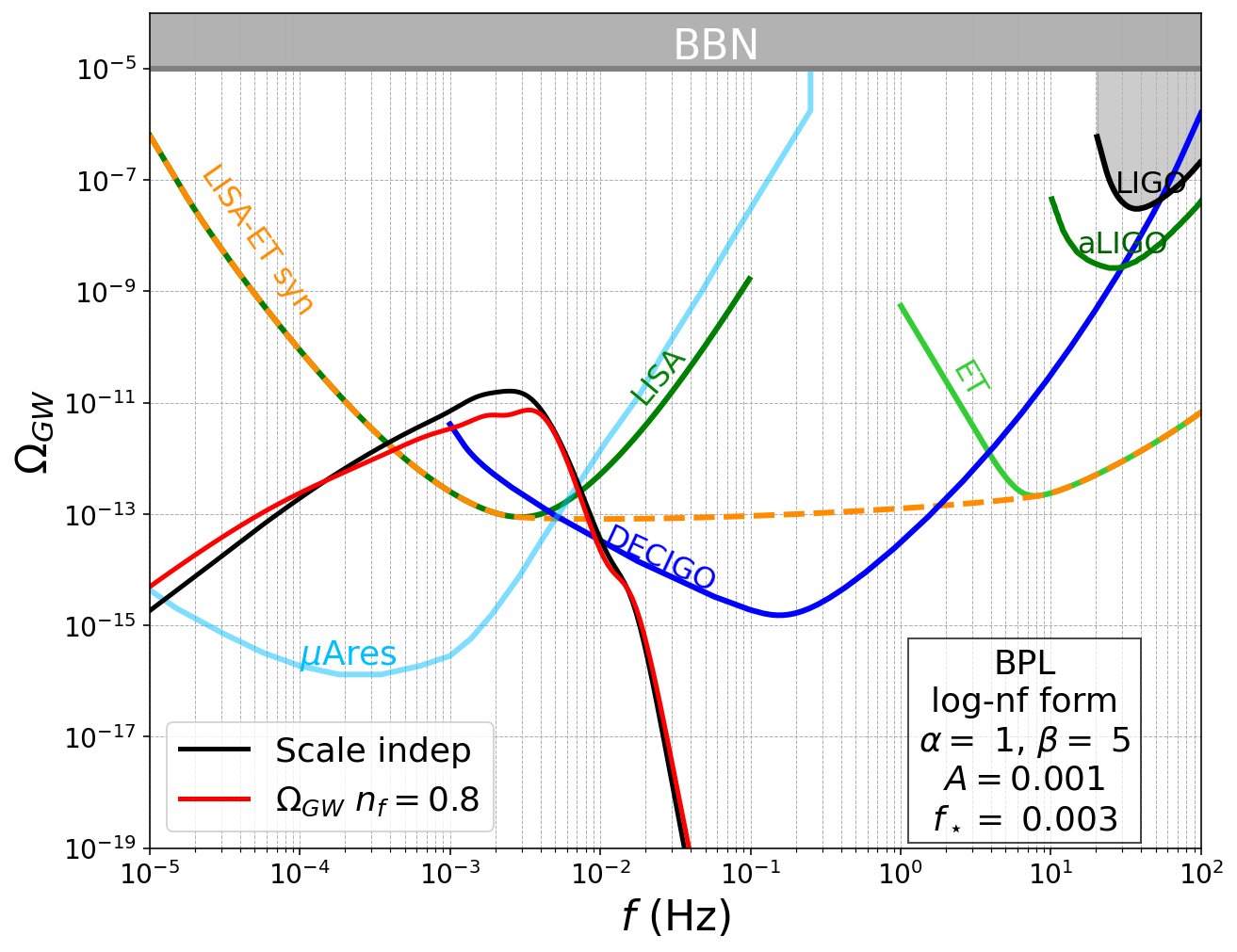}
    \end{subfigure}
    \hfill
    \begin{subfigure}{0.48\textwidth}
        \includegraphics[width=\textwidth]{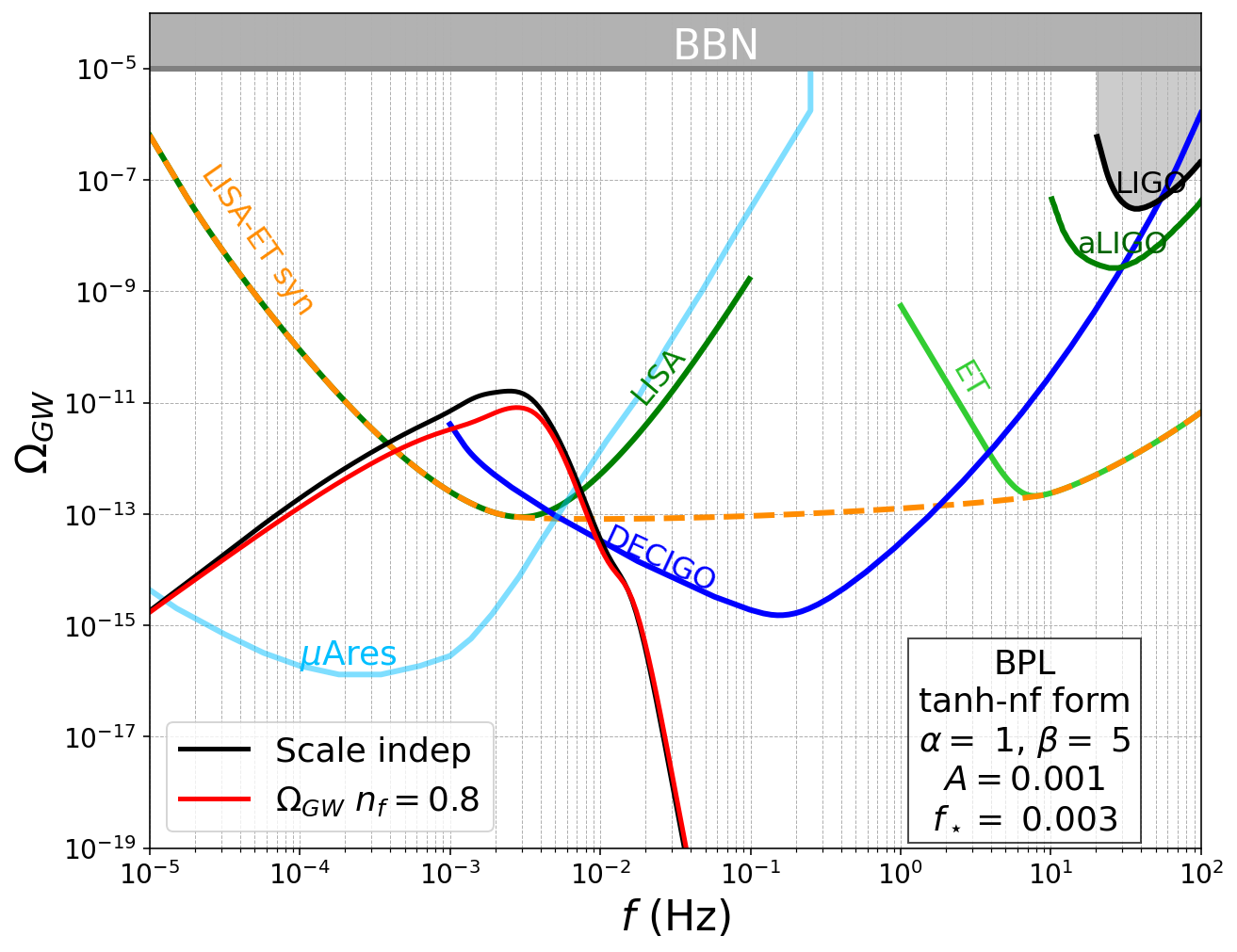}
    \end{subfigure}
    \caption{\it Total induced GW spectrum for the broken-power-law source with the peak placed in the LISA band.}
    \label{fig:app_bpl_total_lisa}
\end{figure}

\begin{figure}[H]
    \centering
    \begin{subfigure}{0.48\textwidth}
        \includegraphics[width=\textwidth]{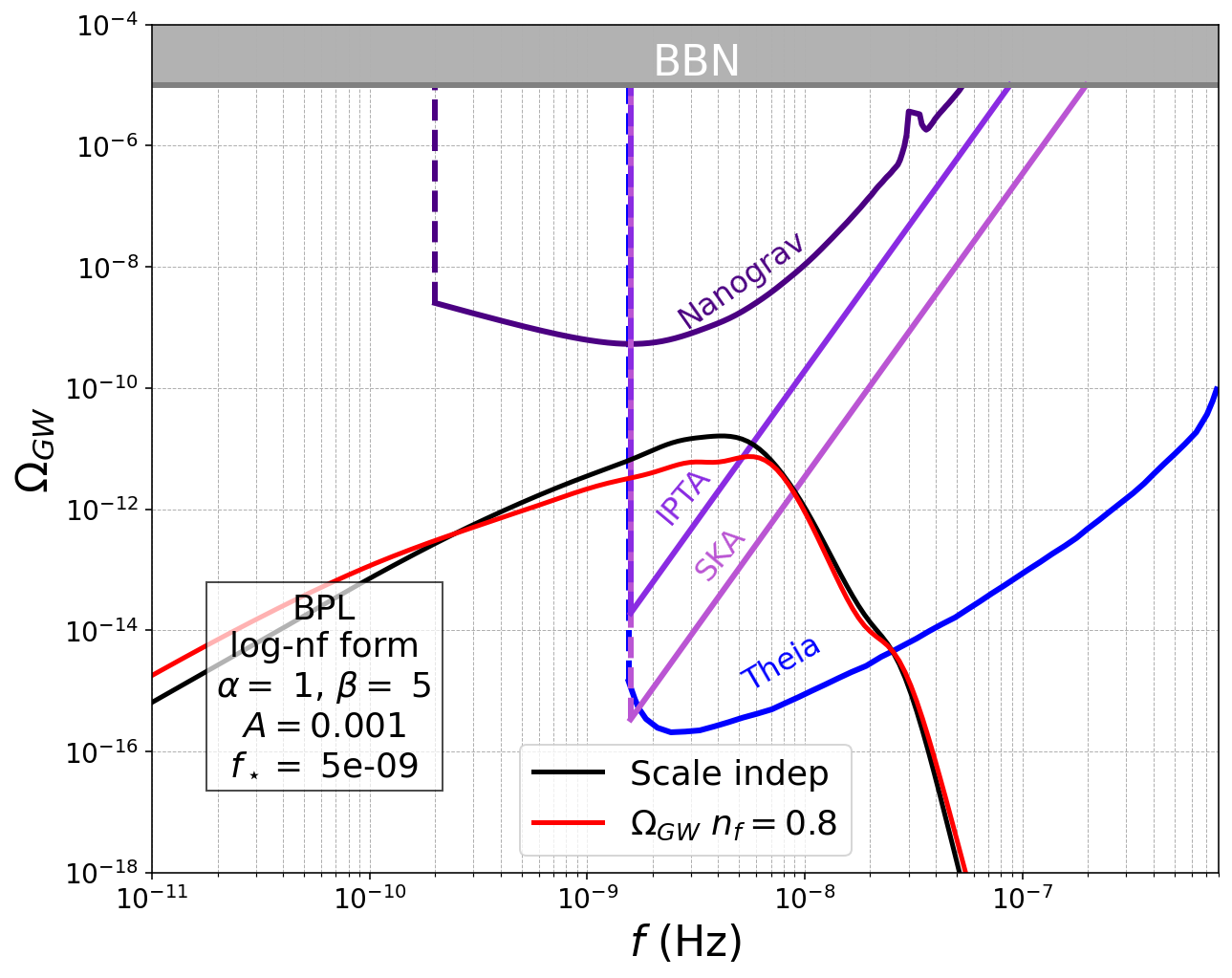}
    \end{subfigure}
    \hfill
    \begin{subfigure}{0.48\textwidth}
        \includegraphics[width=\textwidth]{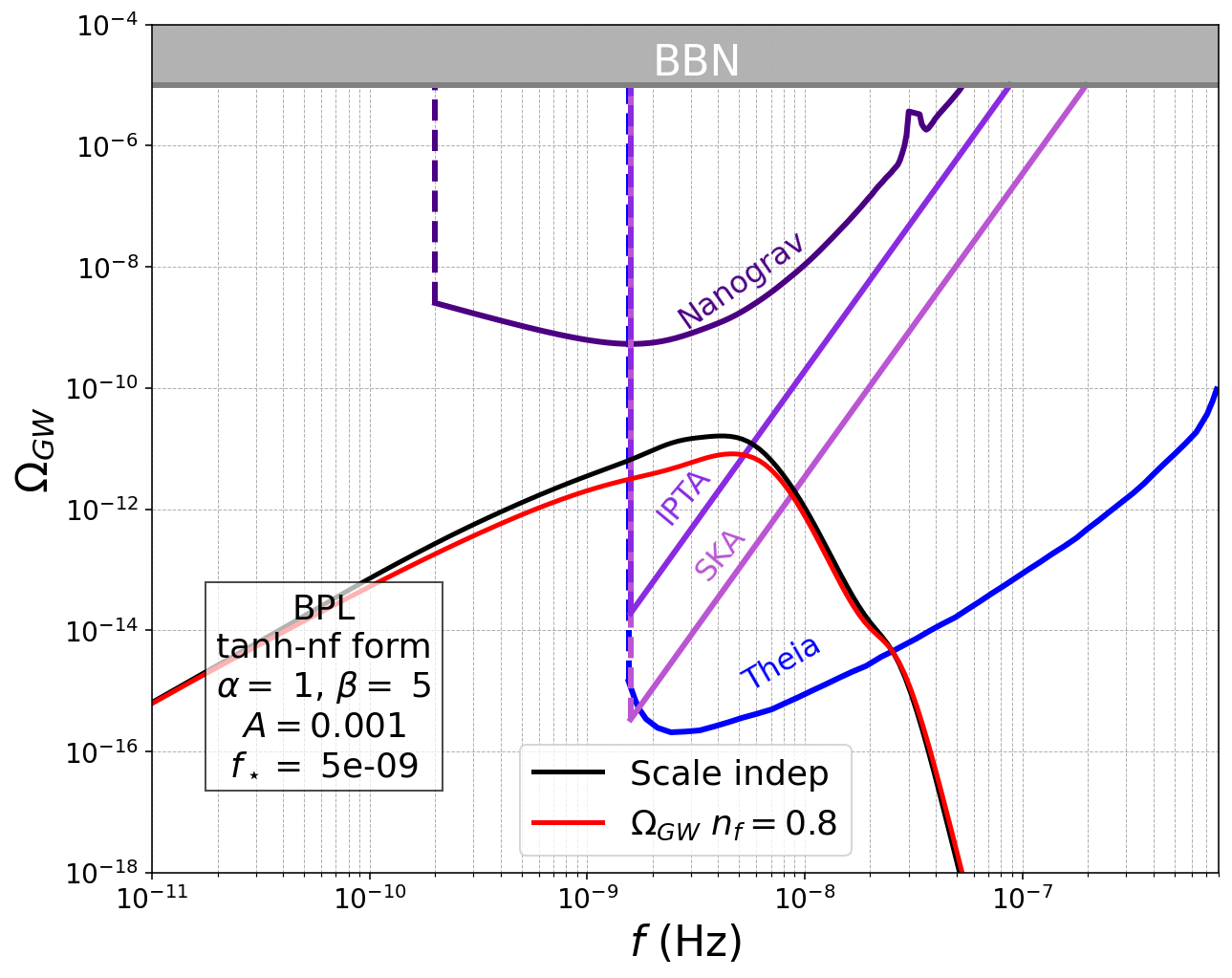}
    \end{subfigure}
    \caption{\it Same as Fig.~\ref{fig:app_bpl_total_lisa}, but with the peak tuned to PTA frequencies.}
    \label{fig:app_bpl_total_pta}
\end{figure}

\subsection{Double broken power-law spectrum}
\label{sec:dbpl_extra}

Finally, we consider a broader class of spectra in which the source interpolates between three distinct scaling regimes. We explored this both through an intermediate ``sum of two broken power laws'' representation and through an explicit three-segment parameterization motivated by the shell-like spectral shapes studied in first-order phase-transition analyses. The numerical results are described by the three-segment form
\begin{align}
\Delta_{\mathcal R}^2(k) &= \frac{A_r}{\left(k/k_l\right)^{-n_l} + \left(k/k_l\right)^{-n_m} + \left(k_h/k_l\right)^{-n_m}\left(k/k_h\right)^{-n_h}}\nonumber\\
&\propto
\begin{cases}
(k/k_l)^{n_l}, & k\ll k_l,\\
(k/k_l)^{n_m}, & k_l \ll k \ll k_h,\\
(k_h/k_l)^{n_m}(k/k_h)^{n_h}, & k_h\ll k,
\end{cases}
\label{eq:dbpl_extra}
\end{align}
with $n_l>n_m>n_h$ and $k_l<k_h$. Here $k_l$ and $k_h$ are comoving break scales. Present-day detector frequencies are obtained from these wavenumbers using Eq.~\eqref{eq:k_to_f_conversion}. We consider two parameter sets: $(n_l,n_m,n_h)=(2,0,-3)$ and $(4,-1,-4)$, with a hierarchy $k_h\simeq k_l/\xi_{\rm shell}$ and $\xi_{\rm shell}=0.1$. These two choices span a moderately broad spectrum and a more sharply varying one, respectively.

The resulting scalar profiles are shown in Fig.~\ref{fig:app_dbpl_ps}. The first parameter set produces a comparatively mild plateau-like intermediate regime, while the second yields a stronger separation between the low-wavenumber rise, the intermediate shoulder, and the ultraviolet descent. This structure is reflected in the induced signal. Because the convolution integrals sample multiple wavenumber intervals with different effective tilts, the non-Gaussian corrections do not simply modify one peak; instead, they reshape the relative weight of the low-wavenumber shoulder, the central plateau/peak region, and the high-wavenumber tail.

This is clearly visible in Figs.~\ref{fig:app_dbpl_orders_a} and \ref{fig:app_dbpl_orders_b}, where the decomposition of the SIGW signal is displayed for the two parameter choices. In the $(2,0,-3)$ case, power-law running enhances the high-frequency side and slightly lifts the intermediate region, whereas the UV-matched tanh template leaves the spectrum much closer to the scale-independent case, with only a mild suppression around the transition and peak scales. In the steeper $(4,-1,-4)$ case, the power-law template again produces the largest deformation, but now the enhancement is focused more strongly on the ultraviolet tail because the underlying scalar source already decays rapidly at high frequency. The UV-matched tanh template continues to act mainly as a saturating profile with a delayed onset, reducing the non-Gaussian contribution near the peak or transition range and producing only a modest change elsewhere.

This benchmark therefore illustrates an important message of the full analysis: the visibility of running non-Gaussianity depends strongly on how many effective slopes the scalar source contains. For smooth single-peak templates, the main observable is usually a deformation of the ultraviolet tail or a mild peak shift. For multi-slope templates such as Eq.~\eqref{eq:dbpl_extra}, the same running can instead alter the balance between several frequency intervals at once. Consequently, if a future stochastic-background signal exhibits more than one characteristic scale or an extended shoulder, the shape information contained in the SIGW spectrum may become particularly powerful for disentangling running PNG from alternative explanations. {The LISA- and PTA-tuned totals in Figs.~\ref{fig:app_dbpl_total_lisa_a}, \ref{fig:app_dbpl_total_pta_a}, \ref{fig:app_dbpl_total_lisa_b}, and \ref{fig:app_dbpl_total_pta_b} demonstrate that the relative shoulder and tail deformations persist under a shift of the overall frequency scale.}

\begin{figure}[H]
    \centering
    \begin{subfigure}{0.48\textwidth}
        \includegraphics[width=\textwidth]{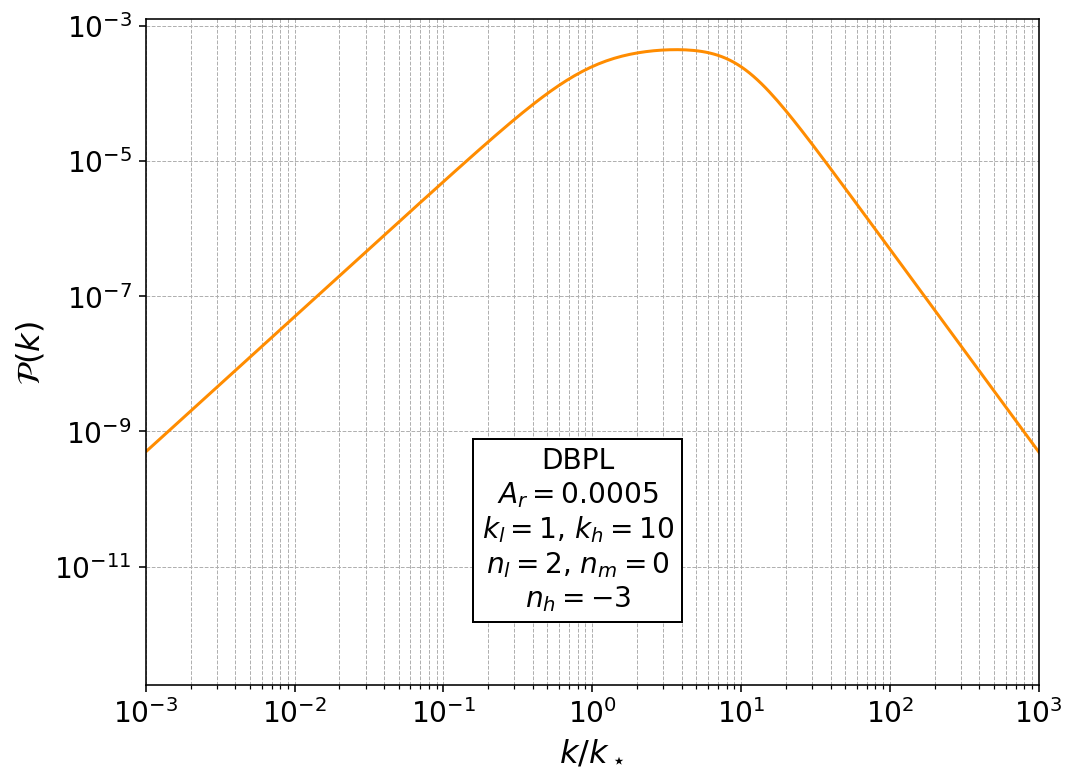}
    \end{subfigure}
    \hfill
    \begin{subfigure}{0.48\textwidth}
        \includegraphics[width=\textwidth]{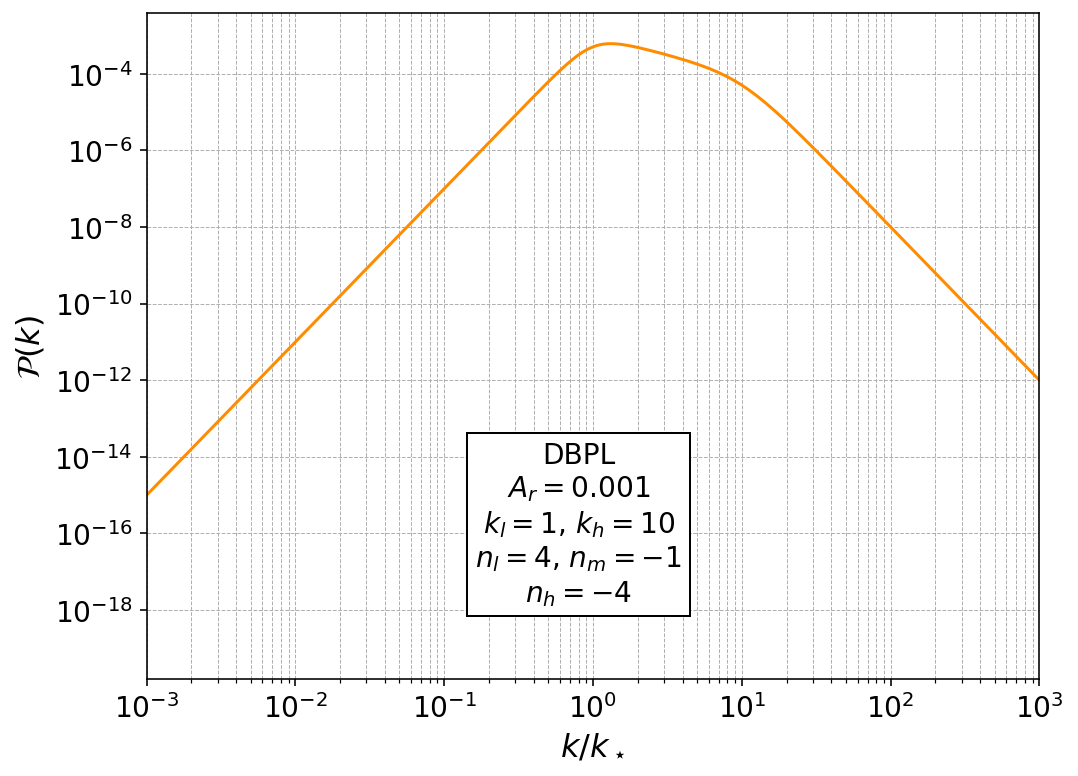}
    \end{subfigure}
    \caption{\it Double-broken-power-law scalar spectra. Left: $(n_l,n_m,n_h)=(2,0,-3)$ $A_r=5\times10^{-4}$. Right: $(n_l,n_m,n_h)=(4,-1,-4)$ $A_r=1\times10^{-3}$.}
    \label{fig:app_dbpl_ps}
\end{figure}

\begin{figure}[H]
    \centering
    \begin{subfigure}{0.48\textwidth}
        \includegraphics[width=\textwidth]{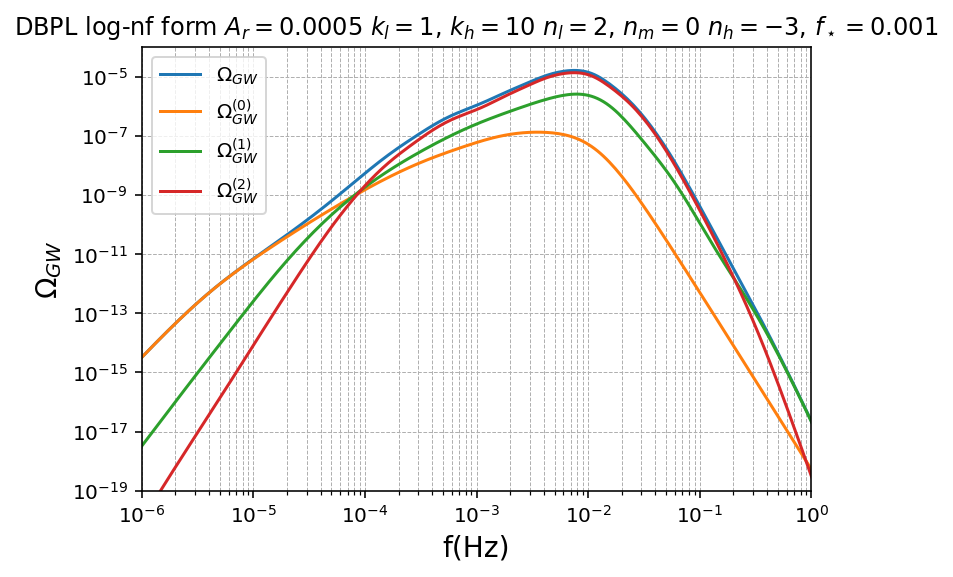}
    \end{subfigure}
    \hfill
    \begin{subfigure}{0.48\textwidth}
        \includegraphics[width=\textwidth]{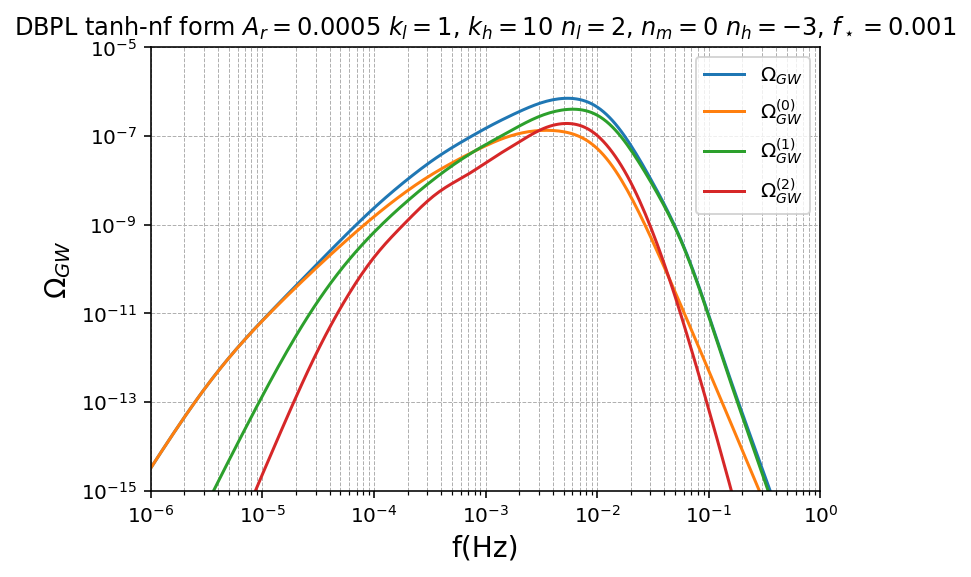}
    \end{subfigure}
    \caption{\it Order-by-order decomposition of the induced GW spectrum for the double-broken-power-law source with $(n_l,n_m,n_h)=(2,0,-3)$, $A_r=5\times10^{-4}$.}
    \label{fig:app_dbpl_orders_a}
\end{figure}

\begin{figure}[H]
    \centering
    \begin{subfigure}{0.48\textwidth}
        \includegraphics[width=\textwidth]{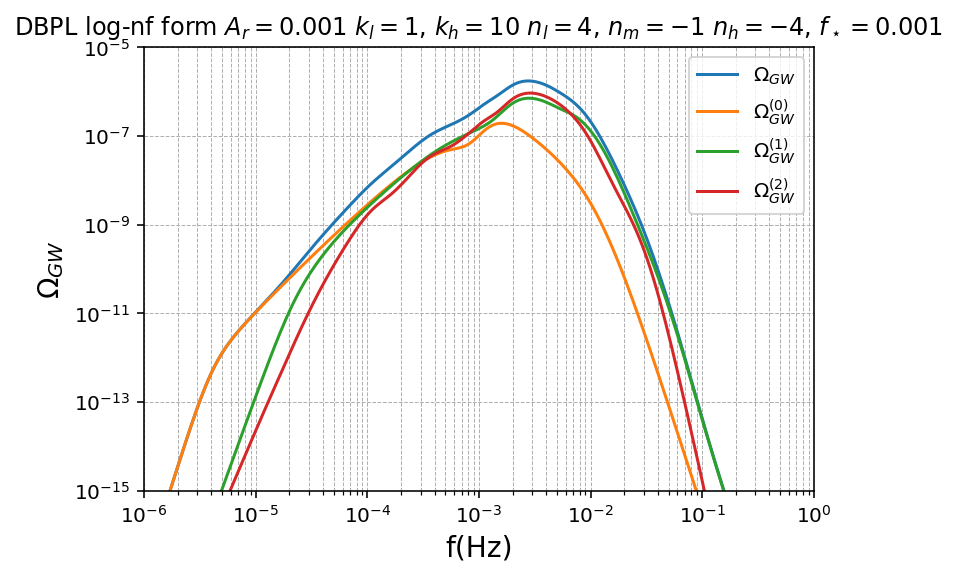}
    \end{subfigure}
    \hfill
    \begin{subfigure}{0.48\textwidth}
        \includegraphics[width=\textwidth]{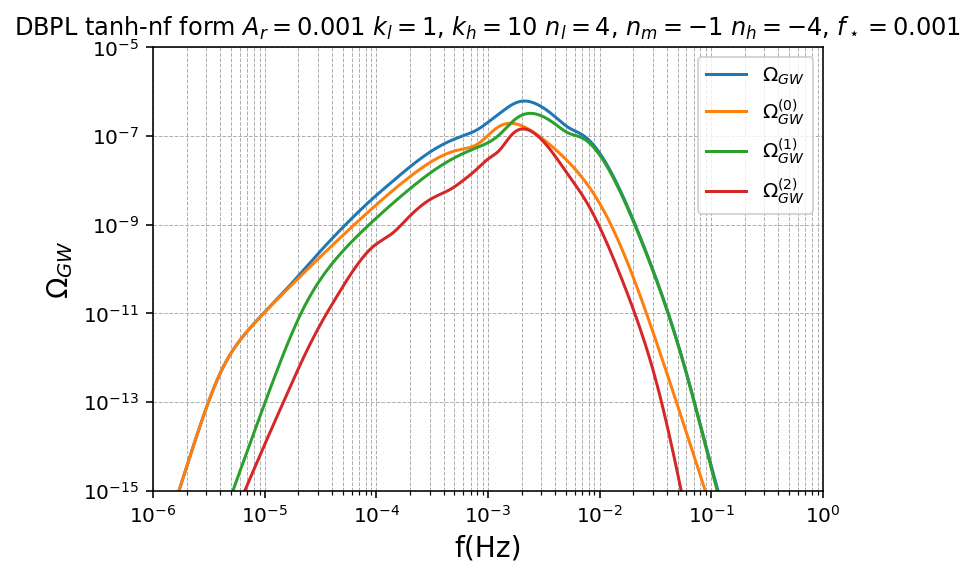}
    \end{subfigure}
    \caption{\it Order-by-order decomposition of the induced GW spectrum for the double-broken-power-law source with $(n_l,n_m,n_h)=(4,-1,-4)$, $A_r=1\times10^{-3}$.}
    \label{fig:app_dbpl_orders_b}
\end{figure}

\begin{figure}[H]
    \centering
    \begin{subfigure}{0.48\textwidth}
        \includegraphics[width=\textwidth]{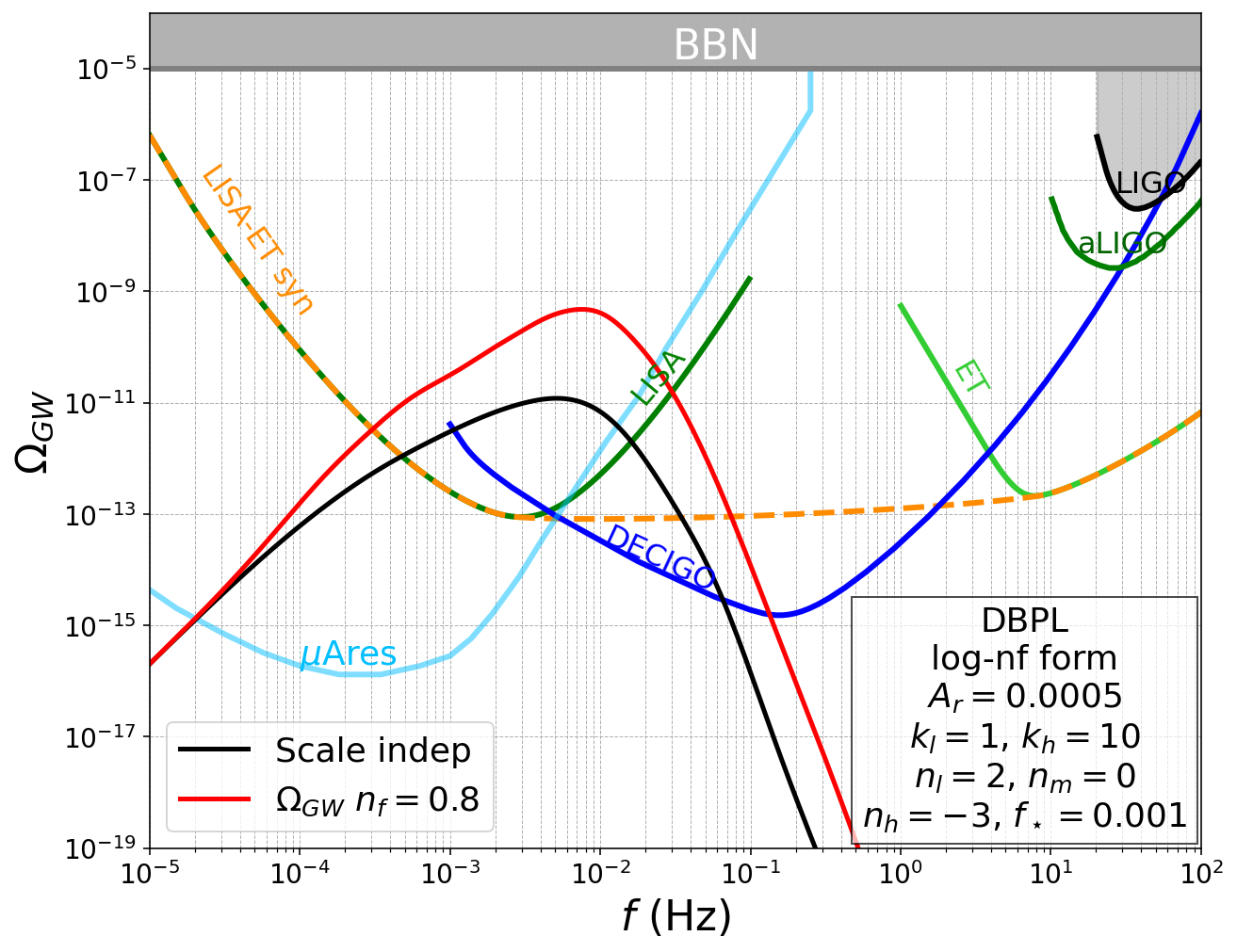}
    \end{subfigure}
    \hfill
    \begin{subfigure}{0.48\textwidth}
        \includegraphics[width=\textwidth]{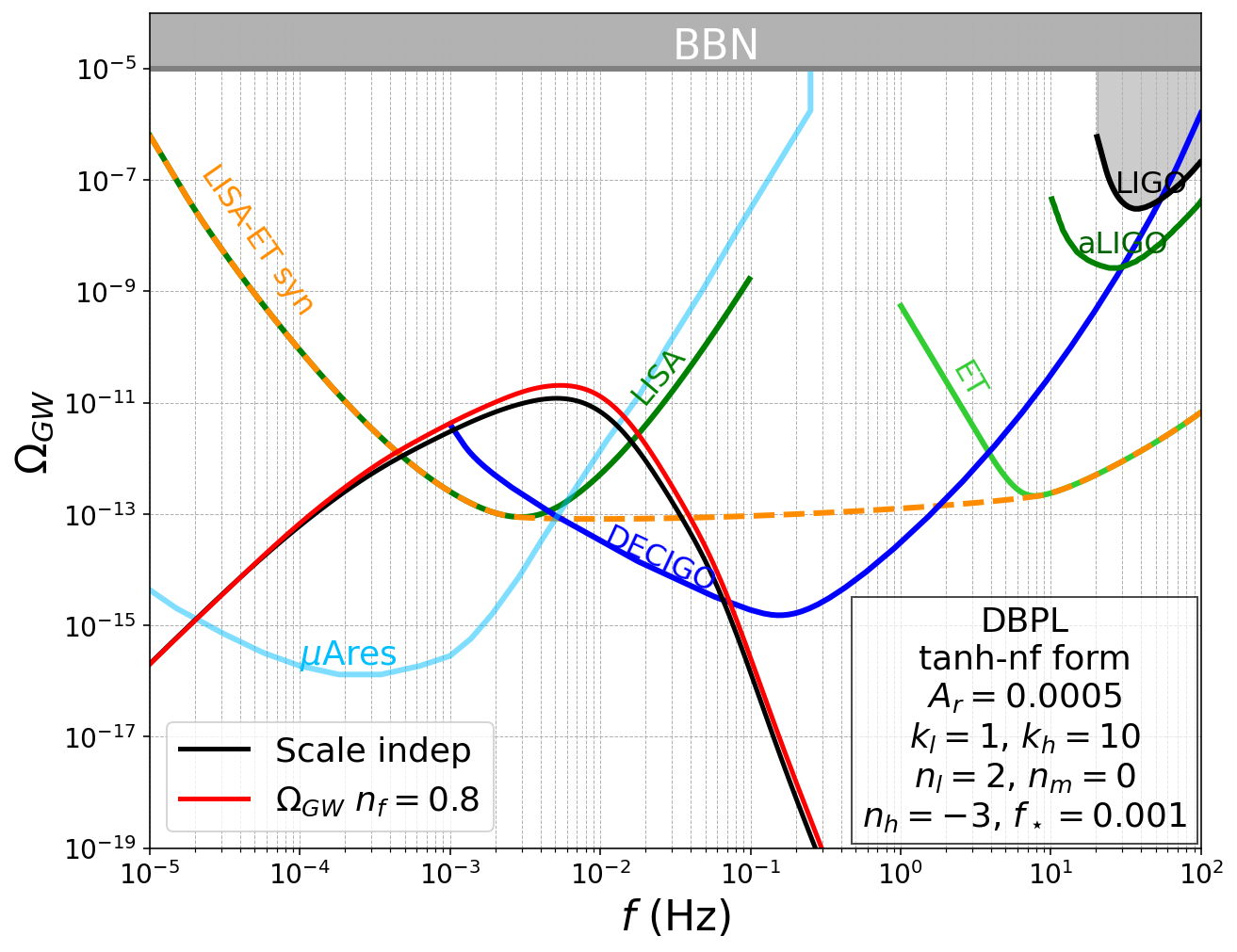}
    \end{subfigure}
    \caption{\it Total induced GW spectrum for the double-broken-power-law source with $(n_l,n_m,n_h)=(2,0,-3)$ and a peak placed in the LISA band.}
    \label{fig:app_dbpl_total_lisa_a}
\end{figure}

\begin{figure}[H]
    \centering
    \begin{subfigure}{0.48\textwidth}
        \includegraphics[width=\textwidth]{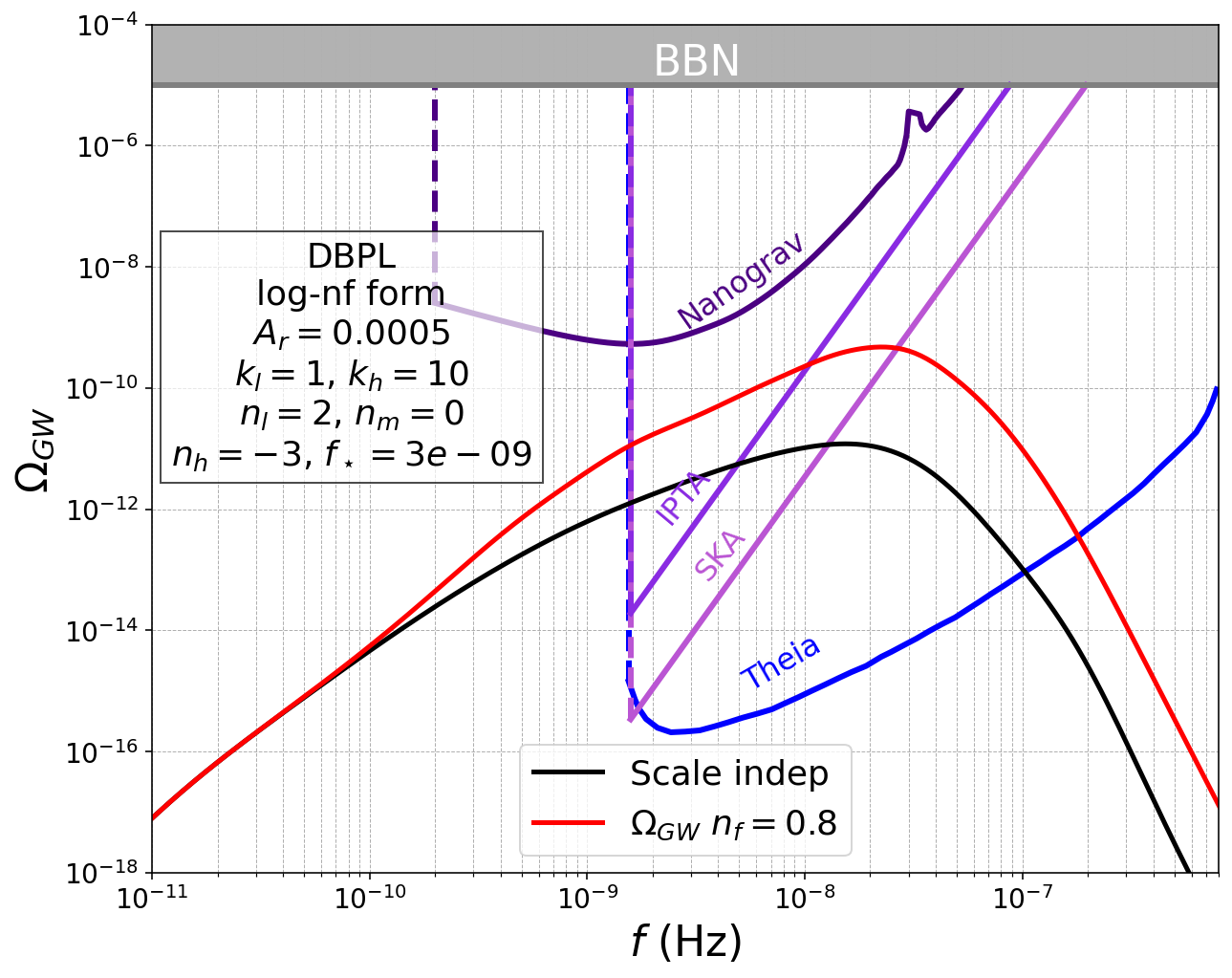}
    \end{subfigure}
    \hfill
    \begin{subfigure}{0.48\textwidth}
        \includegraphics[width=\textwidth]{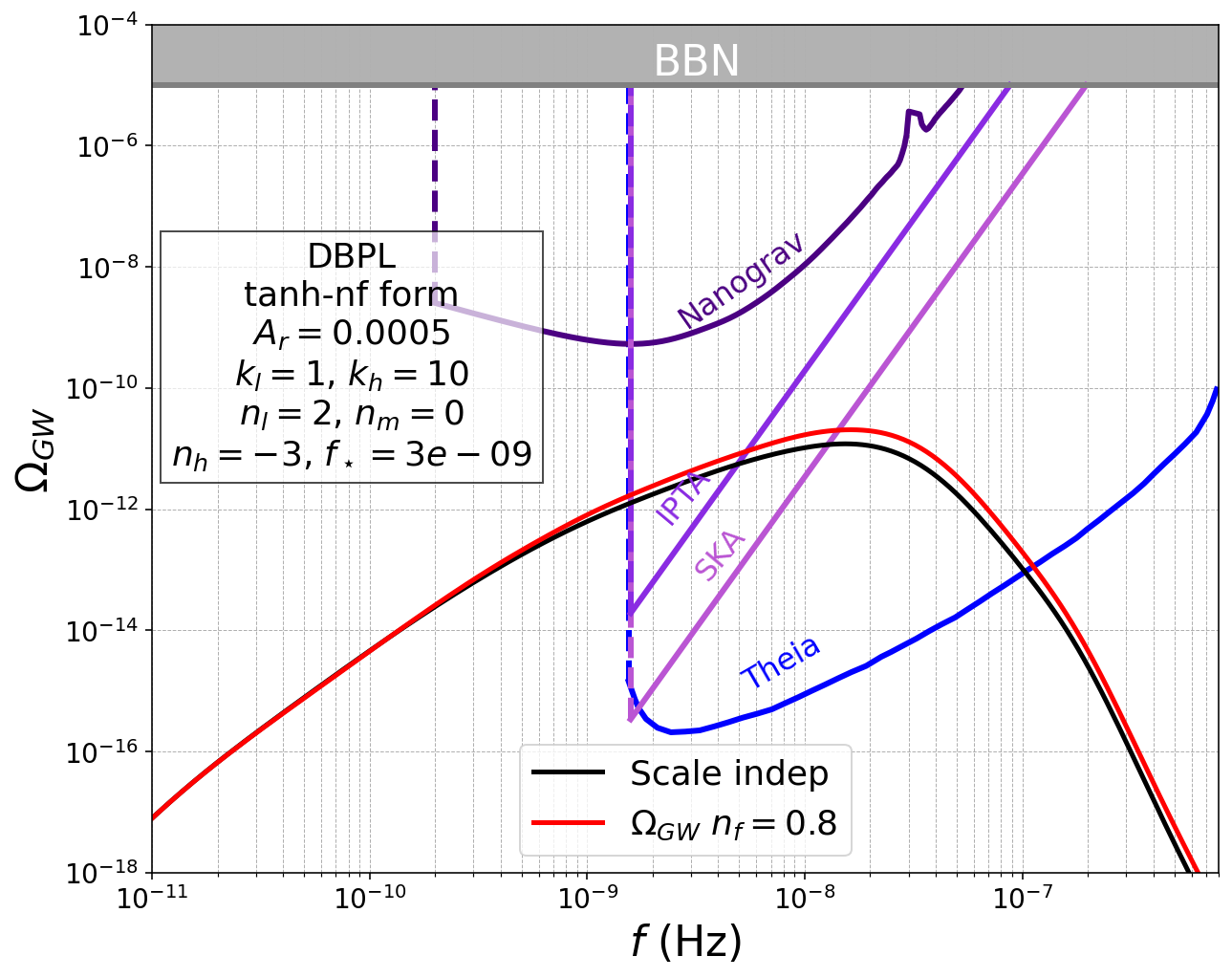}
    \end{subfigure}
    \caption{\it Same as Fig.~\ref{fig:app_dbpl_total_lisa_a}, but with the peak tuned to PTA frequencies.}
    \label{fig:app_dbpl_total_pta_a}
\end{figure}

\begin{figure}[H]
    \centering
    \begin{subfigure}{0.48\textwidth}
        \includegraphics[width=\textwidth]{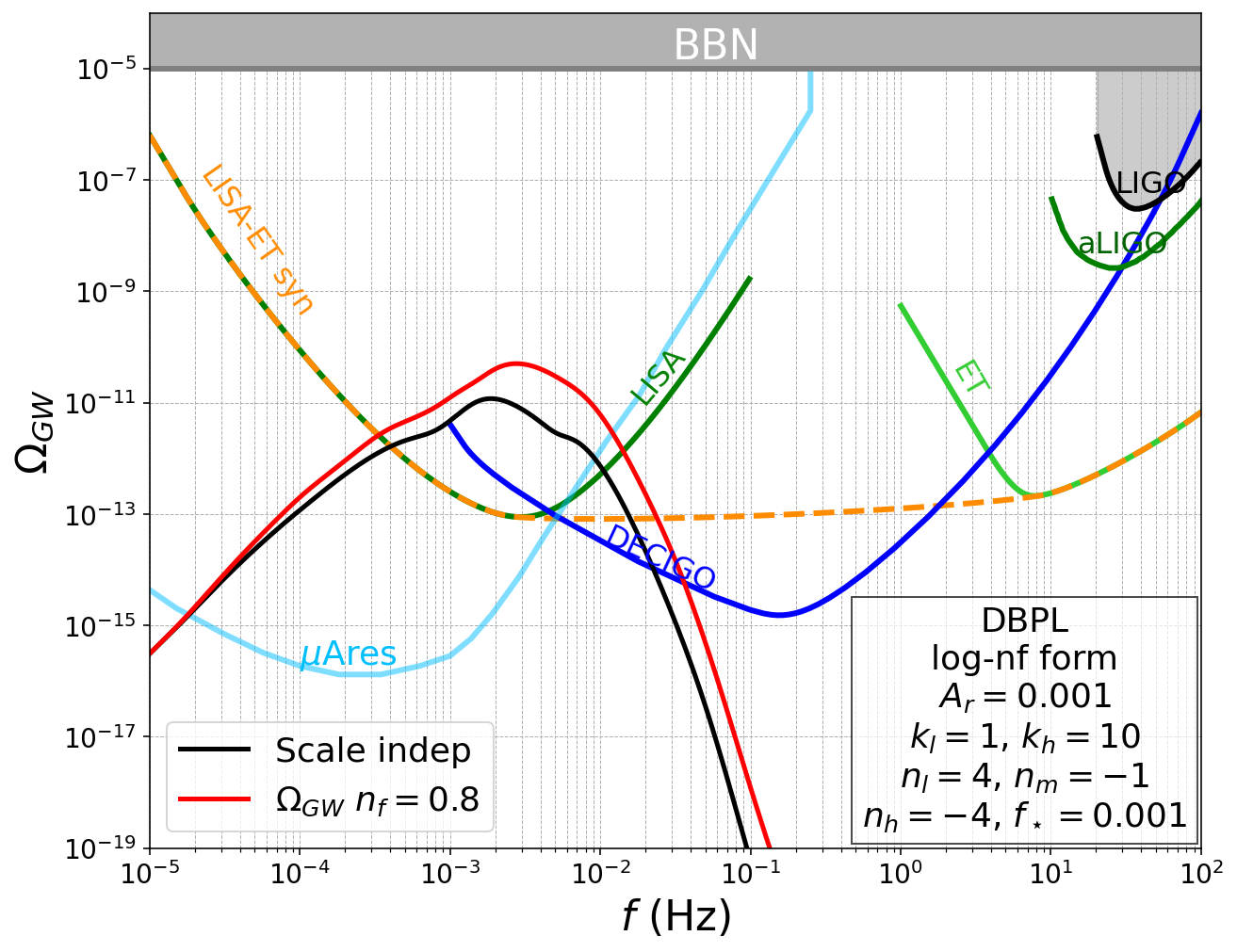}
    \end{subfigure}
    \hfill
    \begin{subfigure}{0.48\textwidth}
        \includegraphics[width=\textwidth]{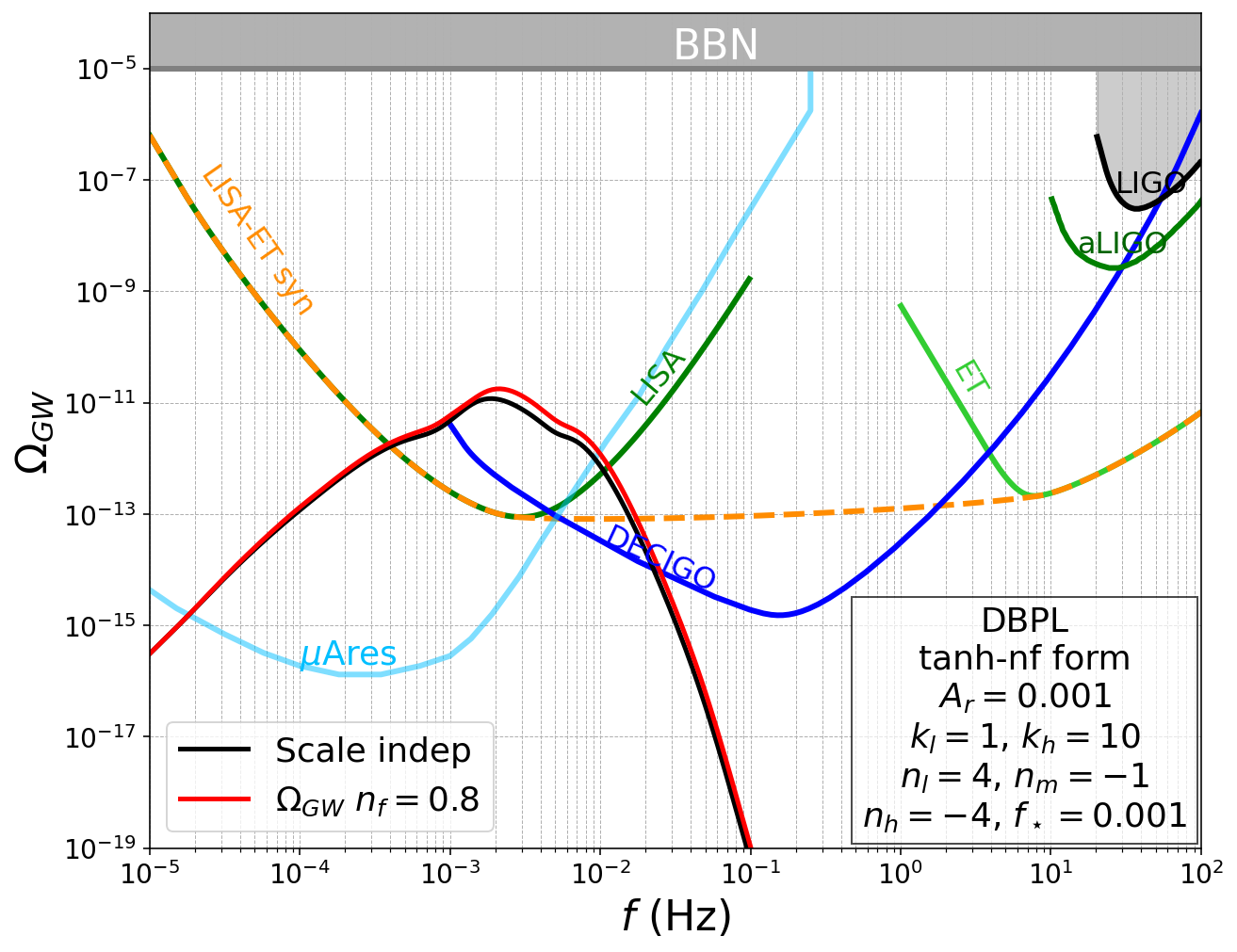}
    \end{subfigure}
    \caption{\it Total induced GW spectrum for the double-broken-power-law source with $(n_l,n_m,n_h)=(4,-1,-4)$ and a peak placed in the LISA band.}
    \label{fig:app_dbpl_total_lisa_b}
\end{figure}

\begin{figure}[H]
    \centering
    \begin{subfigure}{0.48\textwidth}
        \includegraphics[width=\textwidth]{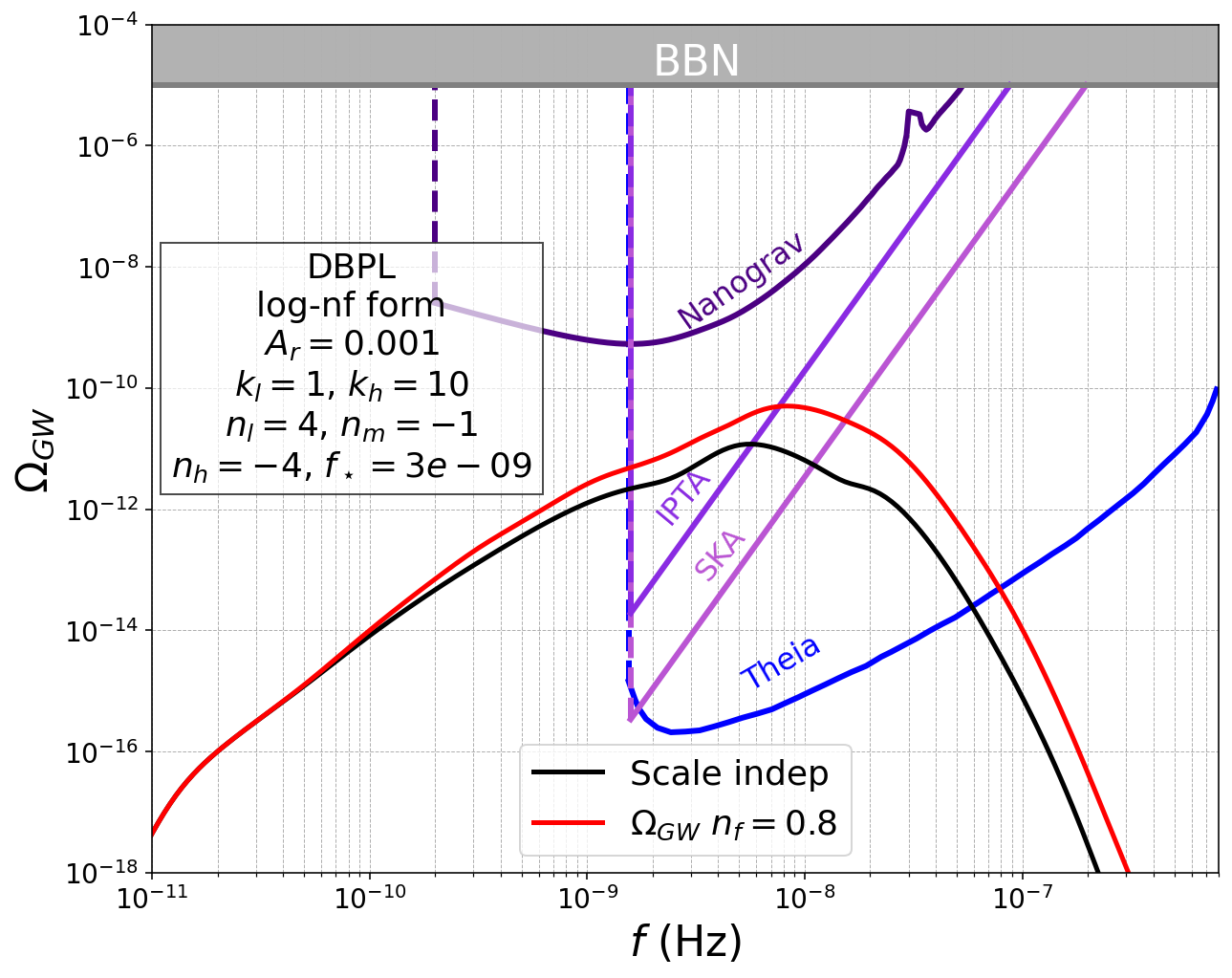}
    \end{subfigure}
    \hfill
    \begin{subfigure}{0.48\textwidth}
        \includegraphics[width=\textwidth]{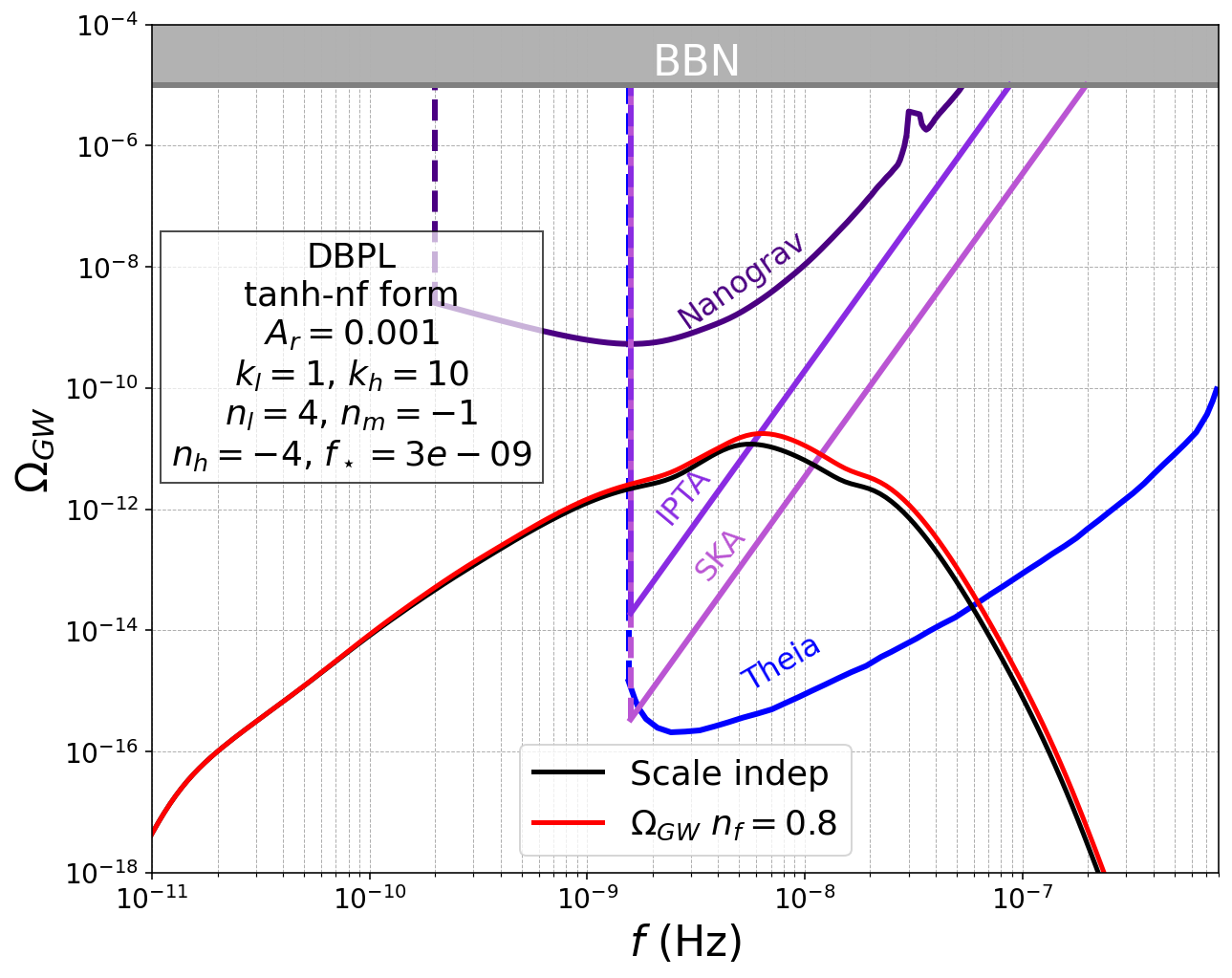}
    \end{subfigure}
    \caption{\it Same as Fig.~\ref{fig:app_dbpl_total_lisa_b}, but with the peak tuned to PTA frequencies.}
    \label{fig:app_dbpl_total_pta_b}
\end{figure}

\subsection*{Synthesis across all scalar-source benchmarks}
\label{sec:extra_summary}

{Section~\ref{sec:extra_summary} synthesizes the Gaussian bump (Sec.~\ref{sec:gaussian_bump_extra}), cutoff source (Sec.~\ref{sec:powerlaw_cutoff_extra}), broken power law (Sec.~\ref{sec:broken_power_law_extra}), and double-broken power law (Sec.~\ref{sec:dbpl_extra}), which reinforce the following conclusions:} First, the dominant template-level observable effect of the scale-dependent internal-leg separable kernel is generically a distortion of the \emph{shape} of the induced GW spectrum rather than a uniform rescaling. Second, power-law running and UV-matched tanh running lead to qualitatively different phenomenology: the former tends to enhance the ultraviolet side of the spectrum and can broaden the peak or shoulder structure, {whereas the latter is UV matched, suppresses the sub-pivot and pivot regions, and then saturates to the constant-weight normalization at high momentum.} Third, the detailed morphology of the scalar source (single-peak, asymmetric, or multi-slope) strongly controls where in frequency space these deformations are most visible.

For practical applications, the extended benchmark set shows how compact bumps, asymmetric inflationary peaks, smooth cutoffs, and multi-slope spectra imprint different shape deformations on $\Omega_{\rm GW}(f)$. The numerical size of the deformation is template-dependent, but the qualitative lesson is common: running non-Gaussianity carries information that is absent from an amplitude-only or constant-$f_{\rm NL}$ analysis.

\section{Phenomenology}
\label{sec:phenom}

The scale-dependent distortions in the SIGW spectrum \(\Omega_{\rm GW}(f)\) induced by running local non-Gaussianity have direct observational implications across multiple frequency bands. In this section we discuss detectability with current and future gravitational-wave observatories, the interplay with primordial black hole (PBH) constraints, and the infrared behavior of the spectrum.

\subsection{Gravitational-wave detector sensitivity and shape distortions}

The comparison between the estimates of Secs.~\ref{sec:analytic_estimates}-\ref{sec:analytic_numeric_implications} and the numerical spectra suggests that experimental searches should treat running PNG as a shape parameter rather than an amplitude nuisance parameter. A constant rescaling of \(A_r\) or \(\bar f_{\rm NL}\) changes the overall height of \(\Omega_{\rm GW}\), while running changes the relative power in the infrared tail, peak, shoulder, and UV side. This distinction is central for upcoming experiments because different frequency bands may sample different parts of the same induced spectrum.

{Current and planned GW experiments span many orders of magnitude in frequency.} {Pulsar timing arrays (PTAs)} probe the nHz band \cite{NANOGrav:2023gor,EPTA:2023fyk}, while space-based interferometers such as LISA \cite{LISACosmologyWorkingGroup:2022kbp,Baker:2019nia}, ET \cite{Punturo:2010zz,Hild:2010id}, BBO, and DECIGO target complementary bands from the mHz range up to sub-Hz and higher frequencies. Power-law integrated (PLI) sensitivity curves provide a convenient way to assess detectability for spectra with approximate power-law behavior \cite{Thrane:2013oya}.

For the representative scalar templates studied in Sec.~\ref{sec:results}, the Gaussian SIGW spectrum typically peaks near the present-day frequency \(f_\star\) associated with the comoving scale \(k_\star\) through Eq.~\eqref{eq:k_to_f_conversion}, with a high-frequency fall-off and an infrared tail \(\Omega_{\rm GW} \propto f^3\) (up to mild logarithmic corrections). The scale-dependent separable kernel modifies this shape in characteristic ways:

\begin{itemize}
\item \textbf{Power-law running} (\(n_f > 0\)): Enhances the high-frequency tail and broadens the peak by up to \(\mathcal O(10)\) near \(k \sim k_\star\), while producing a milder modification in the deep infrared. This can push parts of the spectrum into the LISA sensitivity window even for lower \(A_r\).
\item \textbf{Tanh transition}: Introduces a delayed onset of the non-Gaussian weighting around the pivot scale \(k_\star\). Relative to the scale-independent case, the UV-matched tanh template suppresses the sub-pivot and pivot regions, with $f_{
m tanh}(k_\star)=	anh(1)$, and approaches the constant-weight normalization as $k\gg k_\star$. This can create step-like or shoulder-like template features in \(\Omega_{\rm GW}(f)\).

\end{itemize}
These template-level shape modifications (peak shift, broadening, and asymmetric tails) are more informative than a pure amplitude rescaling and {can help reduce degeneracies with constant-$f_{\rm NL}$ templates}. The analytical peak-shift estimate in Eq.~\eqref{eq:peak_shift_estimate} shows why a positive power-law running tends to move the effective maximum or shoulder toward higher frequencies, while the UV-matched tanh template produces a localized transition feature. In practice, detector-facing frequency labels in the plots should always be read as the present-day frequencies obtained from the underlying comoving scales via Eq.~\eqref{eq:k_to_f_conversion}. For PTA bands (\(f \sim 10^{-9}\)-\(10^{-8}\) Hz), the infrared tail with power-law running may contribute to stochastic backgrounds consistent with recent NANOGrav/EPTA data \cite{NANOGrav:2023gor,Figueroa:2023zhu,Ellis:2023oxs}, provided \(k_\star\) is appropriately tuned. In the LISA band, high-frequency distortions provide a natural target for comparison with constant-kernel templates; quantitative measurability requires a dedicated likelihood or signal-to-noise analysis \cite{LISACosmologyWorkingGroup:2022jok}.

{Quantitative forecasts require specified values of \(A_r\), \(k_\star\), \(\bar f_{\rm NL}\), and \(n_f\).} The benchmarks shown here should be read primarily as perturbative shape studies: for the lower amplitudes displayed in several figures, $A_r\bar f_{\rm NL}^2$ is well below unity and the $\mathcal O(\bar f_{\rm NL}^4)$ sector is correspondingly a subleading diagnostic. Statements that the fourth-order sector must be retained for perturbative consistency apply to the larger-amplitude PBH-relevant regime $A_r\bar f_{\rm NL}^2\gtrsim1$.

\subsection{{Conditional PBH implications and consistency tests}}
\label{sec:pbh_running_estimates}
{
The same small-scale curvature perturbations that source scalar-induced gravitational waves can also produce primordial black holes (PBHs) at horizon re-entry. The observables probe complementary statistics: $\Omega_{\rm GW}$ depends polynomially on weighted scalar correlators, whereas the PBH abundance depends exponentially on the far tail of the smoothed density or compaction distribution~\cite{Byrnes:2012yxPBH,Young:2013oia,DeLuca:2019qsy,Pi:2024pbhreview}. We do not calculate a PBH abundance or mass function. Instead, we record the conditional scale mapping implied by the same internal-leg separable kernel in a narrow-support approximation.

During radiation domination, the characteristic PBH mass satisfies
\begin{equation}
 M(k)=\gamma M_H(k),
 \qquad
 M(k)\propto k^{-2},
 \qquad
 \frac{d\ln M}{d\ln k}=-2,
 \label{eq:pbh_mass_k_scaling}
\end{equation}
where $\gamma$ is the collapse efficiency. Critical collapse replaces the single horizon-mass assignment by
\begin{equation}
 M_{\rm PBH}=K_{\rm cc}M_H(\delta-\delta_c)^{\gamma_{\rm cc}},
 \label{eq:critical_collapse_scaling}
\end{equation}
with $\gamma_{\rm cc}\simeq0.36$ in radiation domination, broadening and shifting the mass spectrum~\cite{Niemeyer:1997mt,Green:1999yh,Kuhnel:2015vtw,Yoo:2022kcd}.

For a density contrast smoothed on $R\simeq k^{-1}$,
\begin{equation}
 \sigma_R^2(k)=\int d\ln q\,W^2(qR)\,\mathcal T_\delta^2(qR)\,\mathcal P_\zeta(q).
 \label{eq:pbh_density_variance}
\end{equation}
The Gaussian formation fraction is
\begin{equation}
 \beta_G(M)=\frac12\operatorname{erfc}\!\left(\frac{\nu_c}{\sqrt2}\right),
 \qquad
 \nu_c\equiv\frac{\delta_c}{\sigma_R},
 \label{eq:pbh_beta_gaussian_exact}
\end{equation}
and in the rare-peak limit
\begin{equation}
 \beta_G(M)\simeq\frac{1}{\sqrt{2\pi}\nu_c}\exp\!\left(-\frac{\nu_c^2}{2}\right).
 \label{eq:pbh_beta_gaussian_asymptotic}
\end{equation}
This exponential sensitivity distinguishes PBH formation from the polynomial response of SIGWs.

The tree-level bispectrum of the internal-leg separable kernel is
\begin{align}
 B_\zeta(k_1,k_2,k_3)=2\bar f_{\rm NL}\big[&f(k_1)f(k_2)\mathcal P_g(k_1)\mathcal P_g(k_2)
 +f(k_2)f(k_3)\mathcal P_g(k_2)\mathcal P_g(k_3)\nonumber\\
 &+f(k_3)f(k_1)\mathcal P_g(k_3)\mathcal P_g(k_1)\big],
 \label{eq:pbh_separable_bispectrum}
\end{align}
up to the momentum-conserving delta function and Fourier normalization. If the scalar spectrum and smoothing kernel are localized near $k_R\simeq R^{-1}$, the normalized skewness scales schematically as
\begin{equation}
 S_3(R)\equiv\frac{\langle\delta_R^3\rangle}{\sigma_R^4}
 \simeq\mathcal C_3(R)\bar f_{\rm NL}f^2(k_R),
 \label{eq:pbh_skewness_running}
\end{equation}
where $\mathcal C_3(R)$ contains the window, transfer, source-shape, cyclic, and convention-dependent factors. For power-law running,
\begin{equation}
 f^2(k_R)=\left(\frac{k_R}{k_\star}\right)^{2n_f}
 =\left(\frac{M}{M_\star}\right)^{-n_f}.
 \label{eq:pbh_running_weight_M}
\end{equation}
An Edgeworth estimate then gives
\begin{equation}
 \ln\!\left(\frac{\beta_{\rm NG}}{\beta_G}\right)
 \simeq\frac{\mathcal C_3(R)}{6}\bar f_{\rm NL}f^2(k_R)
 \frac{\delta_c^3}{\sigma_R^2}.
 \label{eq:pbh_running_beta_ratio_general}
\end{equation}
This expression is used only to display parametric dependence; an Edgeworth expansion is not a precision description of the far tail.

The present-day PBH fraction inherits the radiation-era redshift enhancement and scales schematically as
\begin{equation}
 f_{\rm PBH}(M)\propto M^{-1/2}\beta(M),
 \label{eq:pbh_fraction_scaling}
\end{equation}
up to the standard factors involving $\gamma$, the effective relativistic degrees of freedom, and the mass-function convention. Consequently, even a moderate scale dependence in the skewness can generate a much larger mass dependence in the abundance ratio $\beta_{\rm NG}/\beta_G$. This exponential response is the reason that a visually modest deformation of $\Omega_{\rm GW}$ need not correspond to a modest PBH effect.

A complementary single-scale interpretation follows by defining
\begin{equation}
 F_{\rm NL}^{\rm eff}(R)\equiv\bar f_{\rm NL}f^2(k_R).
 \label{eq:pbh_effective_fnl}
\end{equation}
At fixed $R$, this quantity may be matched heuristically to a local quadratic map in order to illustrate whether the effective Gaussian threshold is raised or lowered. The identification is not an exact real-space representation of the internal-leg separable kernel, because the smoothed skewness involves integrations over triangle configurations, window functions, and transfer factors. It is used only to summarize the leading scale dependence in a narrow-support limit.

For positive $\bar f_{\rm NL}$ and $n_f$, Eq.~\eqref{eq:pbh_running_weight_M} increases the effective positive-PNG weighting toward smaller masses. Negative $n_f$ reverses this trend, while negative $\bar f_{\rm NL}$ changes the direction in which the rare tail is skewed. The sign of a bispectrum parameter alone is not sufficient to determine the final abundance for broad spectra, because nonlinear density conversion, smoothing, higher cumulants, and the detailed collapse variable also enter. The useful comparison is therefore between mass-dependent abundance ratios calculated with the same primordial kernel and the same collapse prescription.

The UV-matched tanh weight gives a different qualitative scaling. For $k_R \ll k_{\star}$, the effective coupling inherits an approximate power law, while for $k_R \gg k_\star$ it approaches a constant proportional to $1$. In mass space, the modification is therefore concentrated around the range corresponding to the transition scale rather than extending as an indefinitely running low-mass trend. Critical collapse broadens this transition and prevents a feature in the primordial weighting from mapping to a single PBH mass.

SIGWs and PBHs consequently constrain different projections of the same statistics. The SIGW spectrum integrates weighted two-, three-, and four-point information over internal scalar momenta and is most directly sensitive to frequency-space morphology. PBH formation probes the extreme tail of a smoothed real-space variable and is sensitive to the full cumulant hierarchy. A joint analysis can therefore reject combinations of scalar-source shape and primordial kernel that fit one observable but predict an incompatible structure in the other. Such a comparison should use the full SIGW spectrum and the full PBH mass function rather than only a peak frequency and a characteristic mass.

Equations~\eqref{eq:pbh_mass_k_scaling} and \eqref{eq:pbh_running_weight_M} imply the conditional correspondence used in this work: positive power-law running preferentially weights larger scalar momenta in the SIGW convolution and, for the same scalar-source family, smoothing prescription, and collapse criterion, strengthens positive-PNG weighting toward lower horizon masses. The UV-matched tanh weight instead produces a transition-localized mass dependence and then approaches the constant-weight normalization at high momentum. Neither statement is a one-to-one prediction of the PBH mass function.

A quantitative joint SIGW-PBH analysis must derive the non-Gaussian smoothed density or compaction distribution from the same primordial kernel, include the nonlinear curvature-to-density relation, critical collapse, and the chosen window and threshold prescription~\cite{DeLuca:2019qsy,Yoo:2022kcd,Pi:2024pbhreview}. Nonperturbative probability distributions, peak theory, threshold statistics, or model-specific simulations are required in the far tail. The present PBH result is therefore a conditional scaling relation that motivates a separate mass-function calculation rather than a PBH constraint.
}

\subsection{Infrared behavior and logarithmic corrections}

\begingroup
An important feature is the infrared (IR) scaling of $\Omega_{\rm GW}(k)$. For compact or sufficiently localized scalar sources with finite weighted moments, the asymptotic deep-IR form is written safely as
\begin{equation}
\Omega_{\rm GW}^{(X)}(k)=k^3L_X(k),
\label{eq:ir_slow_log_function}
\end{equation}
where $X$ labels a diagrammatic sector and $L_X(k)$ is a slowly varying, dimensionless logarithmic function. Depending on the source and sector, $L_X$ may admit an expansion such as $C_X[1+\alpha_X\ln(k_c/k)+\cdots]$ over a restricted interval or may contain powers of logarithms. We do not use the notation $k^{3+a\ln(bk)}$, because it represents a different resummed functional form and can obscure the distinction between a leading power law and logarithmic modulation.

The numerical spectra are consistent with a leading $k^3$-type behavior for the localized benchmarks in the range where the deep-IR regime is reached, with diagram-dependent logarithmic modulation and potentially stronger pre-asymptotic variation in the $Z$ and nonplanar terms. A detector band or a finite numerical grid may instead probe an intermediate IR regime, especially for finite-width or extended scalar spectra. In that case the measured local slope need not yet equal the asymptotic value. For source templates with extended power-law support, the source slope and the momentum boundaries of the convolution can also modify the IR behavior, so Eq.~\eqref{eq:ir_slow_log_function} is not asserted without the localization and finite-moment assumptions stated in Sec.~\ref{sec:analytic_estimates}. Recent assessments further indicate that logarithmic terms require careful treatment of regularization and cutoff independence~\cite{Yuan:2019wwo,Yuan:2020iwf,Domenech:2025bvr}.
\endgroup

These IR features are potentially relevant for PTA data analyses and may help distinguish SIGW explanations from other stochastic backgrounds.

In summary, the scale-dependent internal-leg separable kernel introduces robust, template-level shape distortions in \(\Omega_{\rm GW}(f)\) across detector bands, with important implications for PBH phenomenology and IR modeling. These effects provide new discriminatory power beyond constant-\(f_{\rm NL}\) studies.

\section{Conclusions and discussion}
\label{sec:concl}
The small-scale Universe remains largely inaccessible to conventional cosmological probes. Scalar-induced gravitational waves offer a way to recover this information, but extracting it requires moving beyond amplitude-only templates. This work shows that, for a specified internal-leg separable quadratic kernel, the spectral morphology of the induced background retains direct information about the momentum dependence of primordial non-Gaussianity.

We have calculated the Gaussian, reducible, and connected primordial-statistics contributions through $\mathcal O(\bar f_{\rm NL}^4)$ within the second-order tensor source for
\[
\mathcal F_{\rm NL}(\mathbf{k};\mathbf q,\mathbf{k}-\mathbf q)=\bar f_{\rm NL}f(q)f(\lVert\mathbf{k}-\mathbf q\rVert).
\]
The explicit kernel assignment is central to the interpretation. The weighted spectrum $\mathcal P^S(p)=f^2(p)\mathcal P_g(p)$ is a bookkeeping object generated by contractions of weighted Gaussian legs, not a universal replacement rule and not an external-output prescription.

Three principal conclusions emerge. First, scale dependence produces genuine spectral morphology rather than a uniform normalization shift. Power-law running changes the effective slopes and momentum support sampled by the non-Gaussian sectors, generating asymmetric peak deformation and a persistent uplift or suppression of the ultraviolet flank. The UV-matched tanh weight produces its largest shape variation near the transition and approaches the constant-weight normalization at high scalar momentum. The contrast provides a physically transparent discriminator between growing and saturating primordial kernels.

Second, scalar-source morphology controls how strongly the running is expressed. Finite-width, asymmetric, and multi-slope spectra expose the redistribution of internal momentum support most clearly. In the exact monochromatic limit, all weighted Gaussian legs are fixed at $k_\star$: a pivot-normalized power law leaves both sector amplitudes and shapes unchanged, whereas the UV-matched tanh weight leaves each sector shape unchanged but suppresses the non-Gaussian sector amplitudes by powers of $	anh(1)$. The nontrivial narrow-source figures are therefore finite-width effects. This limit is both a conceptual result and a stringent consistency check on the numerical implementation.

Third, higher-order non-Gaussian sectors are especially informative shape probes. The analytical hierarchy estimate shows that additional weighted Gaussian factors amplify sensitivity to running, while the local-slope and log-normal saddle estimates explain the ultraviolet deformation and displacement of effective support. At the same time, for compact or sufficiently localized sources with finite weighted moments, smooth running preserves the leading $k^3$ deep-infrared behavior up to diagram-dependent logarithmic corrections. These results identify where in frequency space a measurement is most likely to distinguish primordial momentum dependence from an amplitude change.

The observational lesson is concrete: future stochastic-background analyses should fit peak location, curvature, turnover, shoulder structure, and ultraviolet slope jointly. A multi-band measurement can sample different regions of the same primordial spectrum and test whether an apparent excess is compatible with a growing power-law weight, a saturating transition, or a change in the underlying scalar source. The normalized shape statistic in Eq.~\eqref{eq:shape_diagnostic} provides a starting point, but quantitative discrimination requires a detector-level likelihood treatment including source-shape and thermal-history degeneracies.

The PBH connection strengthens the motivation while demanding care. In a narrow-support approximation, the same positive power-law weight that emphasizes high scalar momenta in the SIGW convolution enhances the effective positive-PNG weighting toward smaller horizon masses because $M\propto k^{-2}$. This is a conditional scale-mapping statement, not a prediction of the PBH abundance. A quantitative joint analysis must calculate the non-Gaussian smoothed density or compaction distribution from the same separable kernel, include critical collapse, and compare full mass-function shapes rather than a single characteristic mass.

Several extensions follow naturally. The separable kernel should be matched to concrete inflationary dynamics and compared with alternative momentum assignments and fully triangle-dependent shapes. Numerical convergence information and detector-level parameter inference will be needed to establish measurability. Generalizations including cubic primordial interactions, nonstandard expansion histories, and joint SIGW-PBH likelihoods would determine how robust the morphology diagnostics remain in a broader theory space.

The central result is that scale-dependent primordial non-Gaussianity leaves a structured, physically interpretable imprint on the induced gravitational-wave spectrum. Relative to constant-kernel diagrammatic calculations~\cite{Adshead:2021hnm,Li:2023qua,Li:2023tup}, the present analysis identifies how a specified internal-leg momentum dependence reorganizes every retained sector. Relative to general studies of running non-Gaussianity~\cite{Sefusatti:2009xu,Byrnes:2010ft,Becker:2010nm}, it provides the corresponding frequency-space observable and shows explicitly how source morphology controls it. Relative to recent all-order approaches~\cite{Zeng:2025allorder}, it supplies transparent finite-order decompositions and analytic shape diagnostics that can be used in parameter inference and as benchmarks for more complete calculations. Peak displacement, asymmetric shoulders, and ultraviolet-tail deformation are not secondary details; within the specified kernel they are the observables that carry information about primordial momentum dependence. This makes SIGW spectral shape a promising route to physics on scales beyond the reach of the CMB and large-scale structure.

\section*{{Data and code availability}}
{
The numerical spectra, scalar-template parameter files, frequency grids, integration maps, QMC sample counts, scramble seeds, and convergence-test outputs underlying the figures will be deposited in a public archival repository upon publication. The archived configurations will specify the software environment and will reproduce each displayed benchmark without manual parameter changes. During peer review, the corresponding data and code are available from the corresponding authors upon reasonable request.
}

\section*{Acknowledgements}
{The authors thank Caner \"Unal Arko Bhaumik for comments on the manuscript.} A.G. acknowledges the support from the Royal Society, UK, Funding Reference: NIF\ R1\ 253963.
GT is partially funded by the STFC grant ST/X000648/1, by the Royal Society grant IES  R3  43186 and by the Leverhulme Trust grant RF-2026-166 9.

\appendix

\section{Toy multifield realization of the internal-leg separable kernel}
\label{app:toy_internal_leg_realization}

{This appendix supplies the detailed field-theory and $\delta N$ construction underlying Sec.~\ref{sec:multifield_kernel_motivation}.} The purpose is not to claim a unique ultraviolet completion of Eqs.~\eqref{eq:internal_leg_kernel_main} and \eqref{eq:zeta_running_ansatz}, but to make explicit the dynamical assumptions behind the input-leg momentum routing. In particular, the construction realizes scale-dependent linear transfer before local quadratic conversion and shows how the compact main-text relations arise from a representative two-field action.

\subsection*{Linear multifield transfer}

Consider an adiabatic fluctuation $Q$ and a spectator or entropy fluctuation $s$ during inflation. A representative quadratic action is
\begin{equation}
S_2=\frac12\int dt\,d^3x\,a^3\left[
\dot Q^2-\frac{c_Q^2}{a^2}(\boldsymbol\nabla Q)^2
+\dot s^2-\frac{c_s^2}{a^2}(\boldsymbol\nabla s)^2-m_s^2(t)s^2
+2\rho(t)\dot Q\,s
\right],
\label{eq:toy_quadratic_action}
\end{equation}
where $\rho(t)$ describes linear adiabatic-entropy mixing. Actions of this form occur for a turning multifield background or an evolving spectator transfer, and scale-dependent transfer is generic when $m_s/H$, $\rho/H$, or the background trajectory varies during the range of horizon-exit times sampled by the source~\cite{Byrnes:2010ft,Chen:2010xka,Achucarro:2014msa}. We assume that, over the band of interest, the solution is dominated by one Gaussian seed $X_{\bf k}$ and write the spectator fluctuation on a conversion hypersurface $t_c$ as
\begin{equation}
s_{\bf k}(t_c)=A_s\,T_s(k)X_{\bf k},
\qquad
\zeta_{g,{\bf k}}=A_\zeta X_{\bf k}.
\label{eq:toy_linear_transfer}
\end{equation}
The assumption of one dominant seed is a rank-one, or fully correlated, limit of the linear transfer matrix. Defining
\begin{equation}
\mathcal T_s(k)\equiv\frac{T_s(k)}{T_s(k_\star)},
\qquad
\mathcal A_s\equiv\frac{A_sT_s(k_\star)}{A_\zeta},
\label{eq:toy_transfer_normalization}
\end{equation}
Eq.~\eqref{eq:toy_linear_transfer} becomes
\begin{equation}
s_{\bf k}(t_c)=\mathcal A_s\mathcal T_s(k)\zeta_{g,{\bf k}},
\label{eq:toy_filtered_spectator}
\end{equation}
Thus $s$ is local at the level of the underlying two-field dynamics, but after linear evolution it is a filtered version of the Gaussian curvature seed.

For illustration, let $N_k$ denote horizon exit, $k=a(N_k)H(N_k)$, and suppose the superhorizon transfer obeys
\begin{equation}
\frac{d\ln T_s}{dN_k}\simeq n_f
\label{eq:toy_transfer_running}
\end{equation}
over a finite interval. Since $N_k-N_\star\simeq\ln(k/k_\star)$ in quasi-de Sitter expansion, Eq.~\eqref{eq:toy_transfer_running} gives
\begin{equation}
\mathcal T_s(k)\simeq\left(\frac{k}{k_\star}\right)^{n_f}.
\label{eq:toy_powerlaw_transfer}
\end{equation}
A mixing rate localized in time instead gives a transfer that changes only across the corresponding range of $N_k$ and then approaches a constant. This provides a field-theory motivation for a bounded smooth-step envelope. The precise tanh profile used in the numerical study remains a phenomenological parametrization of this second possibility rather than a unique solution of Eq.~\eqref{eq:toy_quadratic_action}.

For the normalized transfer $\mathcal T_s(k_\star)=1$, a pivot-matched bounded profile would be obtained by dividing by its value at the pivot. The numerical analysis instead adopts the UV-matched convention $f_{\rm tanh}(k)=\tanh[(k/k_\star)^{n_f}]$, for which the ultraviolet plateau equals unity and $f_{\rm tanh}(k_\star)=\tanh(1)$. These conventions are related by a momentum-independent rescaling of the transferred field and therefore by a compensating redefinition of $\mathcal A_s$ and $\bar f_{\rm NL}$. The physical distinction studied in the paper is the scale dependence and ultraviolet saturation, not this constant normalization choice.

\subsection*{Local quadratic conversion}

On the conversion hypersurface, take the number of e-folds to depend locally and evenly on the spectator,
\begin{equation}
\zeta({\bf x})=\delta N({\bf x})
=N_Q Q({\bf x})
+\frac12N_{ss}\left[s^2({\bf x})-\langle s^2\rangle\right]
+\cdots,
\label{eq:toy_deltaN_expansion}
\end{equation}
where the background is chosen at a symmetry point with $N_s=0$. Such a dependence can arise, for example, from an even spectator dependence of a reheating rate, $\Gamma(s)=\Gamma_0[1+c_s s^2/\Lambda^2+\cdots]$, or of the end-of-inflation hypersurface. The conversion is local in real space. The momentum dependence enters only through the prior linear transfer in Eq.~\eqref{eq:toy_filtered_spectator}.

Identifying the linear term with $\zeta_g$ and Fourier transforming the quadratic term gives, apart from the unobservable zero mode,
\begin{align}
\zeta({\bf k})
&=\zeta_g({\bf k})
+\frac12N_{ss}\mathcal A_s^2
\int\frac{d^3q}{(2\pi)^{3/2}}
 \mathcal T_s(q)\mathcal T_s(\lVert{\bf k}-{\bf q}\rVert)
 \zeta_g({\bf q})\zeta_g({\bf k}-{\bf q})+\cdots
\nonumber\\
&=\zeta_g({\bf k})
+\bar f_{\rm NL}
\int\frac{d^3q}{(2\pi)^{3/2}}
 \mathcal T_s(q)\mathcal T_s(\lVert{\bf k}-{\bf q}\rVert)
 \zeta_g({\bf q})\zeta_g({\bf k}-{\bf q})+\cdots,
\label{eq:toy_kernel_derivation}
\end{align}
with
\begin{equation}
\bar f_{\rm NL}\equiv\frac12N_{ss}\mathcal A_s^2.
\label{eq:toy_fbar_identification}
\end{equation}
Equation~\eqref{eq:toy_kernel_derivation} is exactly the internal-leg separable ansatz used in the numerical calculation. At tree level it produces
\begin{align}
B_\zeta(k_1,k_2,k_3)=2\bar f_{\rm NL}\big[&
 \mathcal T_s(k_1)\mathcal T_s(k_2)\mathcal P_g(k_1)\mathcal P_g(k_2)
+\mathcal T_s(k_2)\mathcal T_s(k_3)\mathcal P_g(k_2)\mathcal P_g(k_3)
\nonumber\\
&+\mathcal T_s(k_3)\mathcal T_s(k_1)\mathcal P_g(k_3)\mathcal P_g(k_1)
\big],
\label{eq:toy_bispectrum}
\end{align}
up to the momentum-conserving delta function and the Fourier convention. Each factor of $\mathcal T_s(k_i)$ therefore has a direct origin as a linear transfer function attached to an incoming Gaussian mode. The general leg weight $f(k)$ used in the numerical calculation may differ from $\mathcal T_s(k)$ by a momentum-independent normalization; this freedom implements either pivot normalization or the UV-matched convention without changing the input-leg momentum routing.

\subsection*{Relation to standard $\delta N$ scale dependence and limitations}

In the usual local construction, momentum dependence is often summarized by evaluating slowly varying $N$ derivatives at the horizon-exit time of an external scale. That organization naturally resembles an output-scale-dependent coefficient~\cite{Byrnes:2010ft,Becker:2010nm}. The construction above realizes a different but consistent ordering:
\begin{equation}
\text{Gaussian seed}
\ \xrightarrow{\;T_s(k)\;}\
\text{linearly filtered spectator}
\ \xrightarrow{\;N_{ss}s^2/2\;}\
\text{local quadratic curvature perturbation}.
\label{eq:toy_ordering}
\end{equation}
Because the local quadratic map acts on two transferred fields, the two input momenta carry separate factors $T_s(q)$ and $T_s(\lVert{\bf k}-{\bf q}\rVert)$.

The realization also states clearly what has been truncated. An independent Gaussian component of $s$, imperfect correlation between $Q$ and $s$, intrinsic spectator self-interactions, or nonlinear transfer would generate additional power spectra and mixed kernels. A complete ultraviolet model would determine $m_s(t)$, $\rho(t)$, the conversion mechanism, the independent-mode content, and the resulting function $T_s(k)$ simultaneously. The present paper isolates the rank-one, weakly nonlinear limit in which those additional structures are subleading. Within that controlled limit, the internal-leg separable kernel is not merely an arbitrary momentum assignment: it is the direct image of scale-dependent linear multifield transfer followed by local quadratic conversion.

\section{Review of SIGWs and Gaussian Formalism}
\label{sec:A}


{We begin with a perturbed FLRW spacetime in conformal Newtonian gauge}
\begin{equation}
    \mathrm{d}s^2 = a(\tau)^2\left\{ -(1+2\Phi)\mathrm{d}\tau^2 + \left[ (1-2\Phi)\delta_{ij} + \frac{1}{2}h_{ij} \right]\mathrm{d}x^{i}\mathrm{d}x^{j} \right\},
\end{equation}
where we neglect first-order tensor perturbations, vector perturbations, and the anisotropic stress, and then $\Phi = \Psi$ follows from the Newtonian gauge. $\Phi$ is the first-order scalar perturbation and $h_{ij}$ is the second-order tensor perturbation. {We expand the tensor perturbation in Fourier space as}
\begin{equation}
    h_{ij}(\tau,\mathbf{x}) = \sum_{\lambda = + , \times} \int\frac{\mathrm{d}^3\mathbf{k}}{(2\pi)^{3/2}}e^{i\mathbf{k}\cdot\mathbf{x}}\epsilon_{ij}^{\lambda}(\mathbf{k})h_{\lambda}(\tau,\mathbf{k})
\end{equation}
with the polarization tensors defined by
\begin{subequations}
\begin{align}
    \epsilon_{ij}^{+}(\mathbf{k}) &= \frac{1}{\sqrt{2}}[\epsilon_{i}(\mathbf{k})\epsilon_{j}(\mathbf{k}) - \bar{\epsilon}_{i}(\mathbf{k})\bar{\epsilon}_{j}(\mathbf{k})], \\
    \epsilon_{ij}^{\times}(\mathbf{k}) &= \frac{1}{\sqrt{2}}[\epsilon_{i}(\mathbf{k})\bar{\epsilon}_{j}(\mathbf{k}) + \bar{\epsilon}_{i}(\mathbf{k})\epsilon_{j}(\mathbf{k})],
\end{align}
\end{subequations}
where $\epsilon_{i,j}(\mathbf{k})$ and $\bar{\epsilon}_{i,j}(\mathbf{k})$ form an orthogonal basis transverse to $\mathbf{k}$. We define the tensor-mode power spectrum as
\begin{equation}
    \langle h_{\lambda_{1}}(\tau, \mathbf{k})h_{\lambda_{2}}(\tau, \mathbf{q}) \rangle = \delta^3(\mathbf{k} + \mathbf{q})\delta^{\lambda_{1}\lambda_{2}} \mathcal P_{\lambda_{1}}(\tau, k),
\end{equation}
and the corresponding dimensionless tensor spectrum as
\begin{equation}
    \langle h_{\lambda_{1}}(\tau, \mathbf{k})h_{\lambda_{2}}(\tau, \mathbf{q}) \rangle = \delta^3(\mathbf{k} + \mathbf{q})\delta^{\lambda_{1}\lambda_{2}} \frac{2\pi^2}{k^3}\Delta_{\lambda_{1}}^2(\tau, k).
\end{equation}
The energy density spectrum is defined as
\begin{equation}\label{eq:GWspectrum}
    \Omega_{\mathrm{GW}}(\tau, k) = \frac{1}{48}\left( \frac{k}{\mathcal H}\right)^2 \sum_{\lambda = +, \times} \overline{\Delta_{\lambda}^2(\tau, k)},
\end{equation}
where the overline denotes a time, or oscillation, average. The energy-density spectrum is the fraction of the total energy density stored in gravitational waves per logarithmic wavenumber.

The equation of motion for the tensor perturbations can be derived from the second-order perturbed Einstein equations,
\begin{equation}
    h_{\lambda}^{\prime\prime}(\tau, \mathbf{k}) + 2\mathcal Hh_{\lambda}^{\prime}(\tau, \mathbf{k}) + k^2 h_{\lambda}(\tau, \mathbf{k}) = 4S_{\lambda}(\tau, \mathbf{k}),
\end{equation}
and in the absence of entropy perturbations, the equation of motion for $\Phi$ is
\begin{equation}\label{eq:EOMPhi}
    \Phi^{\prime\prime}_{\mathbf{k}} + 3\mathcal H(1+c_{s}^2)\Phi^{\prime}_{\mathbf{k}} + (2\mathcal H^{\prime}+(1 + 3c_{s}^2 )\mathcal H^2+c_{s}^2k^2)\Phi_{\mathbf{k}} = 0.
\end{equation}
To connect with primordial physics, we usually use a transfer function $\Phi(k\tau)$ to express the gravitational potential in terms of the primordial curvature perturbation $\zeta$ via the equation-of-state parameter $\omega$ as
\begin{equation}
    \Phi(\tau, \mathbf{k}) = \frac{3+3\omega}{5+3\omega}\Phi(k\tau)\zeta_{\mathbf{k}}.
\end{equation}
{Expressed in terms of the primordial curvature perturbation and transfer function, the source is}
\begin{equation}\label{eq:source}
    S_{\lambda}(\tau, \mathbf{k}) = \int \frac{\mathrm{d}^3\mathbf{q}}{(2\pi)^{3/2}}Q_{\lambda}(\mathbf{k}, \mathbf{q}){\mathcal K_{\rm src}(|\mathbf{k}-\mathbf{q}|,q,\tau)}\zeta_{\mathbf{q}}\zeta_{\mathbf{k}-\mathbf{q}},
\end{equation}
where {$\mathcal K_{\rm src}(|\mathbf{k}-\mathbf{q}|,q,\tau)$ is}
\begin{equation}
\begin{aligned}
    {\mathcal K_{\rm src}(p,q,\tau)} = \frac{3(1+\omega)}{(5+3\omega)^2}&[2(5+3\omega)\Phi(p\tau)\Phi(q\tau) + \tau^2(1+3\omega)^2\Phi^{\prime}(p\tau)\Phi^{\prime}(q\tau) \\
    & +2\tau(1+3\omega)(\Phi(p\tau)\Phi^{\prime}(q\tau)+\Phi^{\prime}(p\tau)\Phi(q\tau))],
\end{aligned}
\end{equation}
and the projection factor is
\begin{equation}
    Q_{\lambda}(\mathbf{k}, \mathbf{q}) \equiv \epsilon_{ij}^{\lambda}(\mathbf{k})q_{i}q_{j}.
\end{equation}

The equation of motion for $h_{ij}$ can be solved by the Green function method as
\begin{equation}\label{eq:hijsol}
    h_{\lambda}(\tau, \mathbf{k}) = \frac{4}{a(\tau)}\int_{\tau_{0}}^{\tau}d\tau^{\prime}G_{\mathbf{k}}(\tau, \tau^{\prime})a(\tau^{\prime})S_{\lambda}(\tau^{\prime}, \mathbf{k}),
\end{equation}
where the Green function satisfies
\begin{equation}\label{eq:EOMGreen}
    \partial_{\tau}^2G_{\mathbf{k}}(\tau, \tau^{\prime}) + \left( k^2 -\frac{a^{\prime\prime}(\tau)}{a(\tau)}\right)G_{\mathbf{k}}(\tau, \tau^{\prime}) = \delta(\tau - \tau^{\prime}).
\end{equation}
{Combining Eqs.~\eqref{eq:source} and \eqref{eq:hijsol} gives the tensor two-point function}
\begin{equation}\label{eq:GWspectrumGreen}
\begin{aligned}
    \langle h_{\mathbf{k_{1}}}^{\lambda_{1}}h_{\mathbf{k}_{2}}^{\lambda_{2}} \rangle = 16\int \frac{\mathrm{d}^3q_{1}}{(2\pi)^{3/2}}\frac{\mathrm{d}^3q_{2}}{(2\pi)^{3/2}}&\langle \zeta_{\mathbf{q_1}}\zeta_{\mathbf{k_1-q_1}}\zeta_{\mathbf{q_2}}\zeta_{\mathbf{k_2-q_2}} \rangle Q_{\lambda_{1}}(\mathbf{k}_1, \mathbf{q}_1)Q_{\lambda_2}(\mathbf{k}_2, \mathbf{q}_2)\\
    & \times I(|\mathbf{k}_1-\mathbf{q}_1|,q_1, \tau_1)I(|\mathbf{k}_2-\mathbf{q}_2|,q_2, \tau_2),
\end{aligned}
\end{equation}
{where}
\begin{equation}
    I(p,q,\tau) = \int_{\tau_{0}}^{\tau}\mathrm{d}\tau^{\prime}G_{\mathbf{k}}(\tau, \tau^{\prime})\frac{a(\tau^{\prime})}{a(\tau)}{\mathcal K_{\rm src}(p,q,\tau^{\prime})}.
\end{equation}
In the radiation-dominated era, we can take $c_{s}^2 = \omega$, and then the equation of motion~\eqref{eq:EOMPhi} becomes
\begin{equation}
    \Phi(k\tau)^{\prime\prime} + \frac{6(1+\omega)}{(1+3\omega)\tau}\Phi(k\tau)^{\prime} + \omega k^2\Phi(k\tau) = 0,
\end{equation}
{which, together with Eq.~\eqref{eq:EOMGreen}, yields in radiation domination}
\begin{equation}\label{eq:RDGreen}
    kG_{\mathbf{k}}(\tau, \tau^{\prime}) = \sin(x - x^{\prime}),\quad x = k\tau, \, x^{\prime} = k\tau^{\prime}
\end{equation}
and
\begin{equation}\label{eq:RDPhi}
    \Phi(k\tau) = \frac{9}{x^2}\left( \frac{\sin(x/\sqrt{3})}{x/\sqrt{3}} -\cos(x/\sqrt{3})\right).
\end{equation}

For Gaussian scalar perturbations, Wick contraction and the symmetries of the integral give
\begin{equation}\label{eq:Wick}
    \langle \zeta_{\mathbf{q_1}}\zeta_{\mathbf{k_1-q_1}}\zeta_{\mathbf{q_2}}\zeta_{\mathbf{k_2-q_2}} \rangle = 2\langle \zeta_{\mathbf{q}_1}\zeta_{\mathbf{q}_2} \rangle \langle \zeta_{\mathbf{k}_1-\mathbf{q}_1} \zeta_{\mathbf{k}_2-\mathbf{q}_2} \rangle.
\end{equation}
It is conventional to introduce two new variables $u = |k - q|/k$ and $v = q/k$ to simplify the above equations. Then, combining~\eqref{eq:GWspectrum}, \eqref{eq:GWspectrumGreen}, \eqref{eq:RDGreen}, and~\eqref{eq:Wick} and considering the limit $x\rightarrow\infty$ as we are interested in the GW spectrum observed today, {we obtain}~\cite{Kohri:2018awv,Adshead:2021hnm,Li:2023qua}
\begin{equation}
    \Omega^{\mathrm{G}}_{\mathrm{GW}}(k) = \frac{2}{3}\int_{0}^{\infty}dv\int_{|1-v|}^{1+v}du\overline{J^{2}(u,v,u,v,x\rightarrow \infty)}\frac{\Delta_{g}^2(vk)}{v^2}\frac{\Delta_{g}^2(uk)}{u^2},
\end{equation}
where $\Delta_{g}^2(k)\equiv \frac{k^3}{2\pi^2}\langle \zeta_{g}(k)\zeta_{g}(k) \rangle$ with $\zeta_{g}(k)$ the Gaussian curvature perturbation, and
\begin{equation}\label{eq:J2}
\begin{aligned}
\overline{J^2(u_1,v_1,u_2,v_2,x\rightarrow\infty)} &= \frac{x^2}{64}[(v_1+u_1)^2-1][1-(v_1-u_1)^2][(v_2+u_2)^2-1]\\
&\times [1-(v_2-u_2)^2]\overline{I_{\mathrm{RD}}(u_1,v_1,x\rightarrow\infty)I_{\mathrm{RD}}(u_2,v_2,x\rightarrow\infty)}
\end{aligned}
\end{equation}
with
\begin{equation}
\begin{aligned}
    \overline{I_{\mathrm{RD}}(u_1,v_1,x\rightarrow\infty)I_{\mathrm{RD}}(u_2,v_2,x\rightarrow\infty)} = &\frac{I_A(u_1,v_1)I_A(u_2,v_2)}{2x^2}[I_B(u_1,v_1)I_{B}(u_2,v_2)\\
    &+\pi^2I_{C}(u_1,v_1)I_C(u_2,v_2)],
\end{aligned}
\end{equation}
as well as
\begin{subequations}
\begin{align}
    I_A(u,v) &= \frac{3(u^2+v^2-3)}{4u^3v^3}\\
    I_{B}(u,v) &= -4uv +(u^2+v^2-3)\ln\left| \frac{3-(u+v)^2}{3-(u-v)^2} \right|\\
    I_C(u,v) &= (u^2+v^2-3)\Theta(u+v-\sqrt{3}),
\end{align}
\end{subequations}
where the overline denotes averaging over oscillations.

\section{Non-Gaussian Four-Point Function and Diagrammatic Contributions}\label{app:ng_4_pt_diagram}

In the non-Gaussian case, the simple Wick splitting in Eq.~\eqref{eq:Wick} no longer applies. We therefore consider local-type non-Gaussianity 
up to the second order,
\begin{equation}
\label{eq_znga}
    \zeta(\mathbf{x})= \zeta_{g}(\mathbf{x})+F_{\mathrm{NL}}\left( \zeta_{g}^2(\mathbf{x}) -\langle \zeta_{g}^2(\mathbf{x})\rangle  \right),
\end{equation}
where $F_{\mathrm{NL}}$ denotes the local-type non-Gaussian parameter. After Fourier transformation, we obtain
\begin{equation}\label{eq:localNG}
    \zeta(\mathbf{k}) = \zeta_{g}(\mathbf{k}) + F_{\mathrm{NL}}\int\frac{\mathrm{d}^3\mathbf{q}}{(2\pi)^{3/2}}\zeta_{g}(\mathbf{q})\zeta_{g}(\mathbf{k}-\mathbf{q}),
\end{equation}
where a delta-function term has been dropped. Substituting Eq.~\eqref{eq:localNG} into Eq.~\eqref{eq:GWspectrumGreen} and performing the Wick contractions gives seven diagrammatic contributions. Three are disconnected contributions, with $\langle\zeta\zeta\zeta\zeta\rangle_d=\sum\langle\zeta\zeta\rangle\langle\zeta\zeta\rangle$; the first is the Gaussian term,
\begin{equation}
    \Delta^{2,\mathrm{G}}_{h_\lambda}(k) = 2^5\int\frac{\mathrm{d}^3\mathbf{q}}{(2\pi)^{3}}I^2(|\mathbf{k}-\mathbf{q}|,q,\tau)Q_{\lambda}^2(\mathbf{k},\mathbf{q})\Delta_{g}^2(q)\Delta^2_{g}(|\mathbf{k}-\mathbf{q}|),
\end{equation}
while the other two are denoted as the hybrid and reducible terms,
\begin{equation}
\begin{aligned}
    \Delta^{2,\mathrm{H}}_{h_\lambda}(k) = 2^7F_{\mathrm{NL}}^2\int&\frac{\mathrm{d}^3\mathbf{q}_1}{(2\pi)^{3}}\frac{\mathrm{d}^3\mathbf{q}_2}{(2\pi)^{3}}I^2(|\mathbf{k}-\mathbf{q_1}|,q_1,\tau)Q_{\lambda}^2(\mathbf{k},\mathbf{q}_1)\\
    &\times\Delta_{g}^2(|\mathbf{k}-\mathbf{q}_1|)\Delta^2_{g}(q_2)\Delta^2_{g}(|\mathbf{q}_1-\mathbf{q}_2|)
\end{aligned}
\end{equation}
\begin{equation}
\begin{aligned}
     \Delta^{2,\mathrm{R}}_{h_\lambda}(k) = 2^7F_{\mathrm{NL}}^4\int&\frac{\mathrm{d}^3\mathbf{q}_1}{(2\pi)^{3}}\frac{\mathrm{d}^3\mathbf{q}_2}{(2\pi)^{3}}\frac{\mathrm{d}^3\mathbf{q}_3}{(2\pi)^{3}}I^2(|\mathbf{k}-\mathbf{q_1}|,q_1,\tau)Q_{\lambda}^2(\mathbf{k},\mathbf{q}_1)\\
     &\times\Delta_{g}^2(|\mathbf{k}-\mathbf{q}_1-\mathbf{q}_3|)\Delta^2_{g}(q_2)\Delta^2_{g}(q_3)\Delta^2_{g}(|\mathbf{q}_1-\mathbf{q}_2|),
\end{aligned}
\end{equation}
and the remaining four integrals are connected terms with the connected correlation function $\langle \zeta\zeta\zeta\zeta \rangle_{c}$ denoted as ``C'', ``Z'', ``planar'' and ``nonplanar'' terms,
\begin{equation}
\begin{aligned}
    \Delta^{2,\mathrm{C}}_{h_\lambda}(k) = 2^8F_{\mathrm{NL}}^2\int&\frac{\mathrm{d}^3\mathbf{q}_1}{(2\pi)^{3}}\frac{\mathrm{d}^3\mathbf{q}_2}{(2\pi)^{3}}I(|\mathbf{k}-\mathbf{q_1}|,q_1,\tau)Q_{\lambda}(\mathbf{k},\mathbf{q}_1)I(|\mathbf{k}-\mathbf{q_2}|,q_2,\tau)Q_{\lambda}(\mathbf{k},\mathbf{q}_2)\\
    &\times\Delta_{g}^2(|\mathbf{k}-\mathbf{q}_2|)\Delta^2_{g}(q_2)\Delta^2_{g}(|\mathbf{q}_1-\mathbf{q}_2|),
\end{aligned}
\end{equation}
\begin{equation}
\begin{aligned}
    \Delta^{2,\mathrm{Z}}_{h_\lambda}(k) = 2^8F_{\mathrm{NL}}^2\int&\frac{\mathrm{d}^3\mathbf{q}_1}{(2\pi)^{3}}\frac{\mathrm{d}^3\mathbf{q}_2}{(2\pi)^{3}}I(|\mathbf{k}-\mathbf{q_1}|,q_1,\tau)Q_{\lambda}(\mathbf{k},\mathbf{q}_1)I(|\mathbf{k}-\mathbf{q_2}|,q_2,\tau)Q_{\lambda}(\mathbf{k},\mathbf{q}_2)\\
    &\times\Delta_{g}^2(|\mathbf{k}-\mathbf{q}_1|)\Delta^2_{g}(q_2)\Delta^2_{g}(|\mathbf{q}_1-\mathbf{q}_2|),
\end{aligned}
\end{equation}
\begin{equation}
\begin{aligned}
    \Delta^{2,\mathrm{P}}_{h_\lambda}(k) = 2^9F_{\mathrm{NL}}^4\int&\frac{\mathrm{d}^3\mathbf{q}_1}{(2\pi)^{3}}\frac{\mathrm{d}^3\mathbf{q}_2}{(2\pi)^{3}}\frac{\mathrm{d}^3\mathbf{q}_3}{(2\pi)^{3}}I(|\mathbf{k}-\mathbf{q_1}|,q_1,\tau)Q_{\lambda}(\mathbf{k},\mathbf{q}_1)I(|\mathbf{k}-\mathbf{q_2}|,q_2,\tau)Q_{\lambda}(\mathbf{k},\mathbf{q}_2)\\
    &\times\Delta_{g}^2(|\mathbf{k}-\mathbf{q}_3|)\Delta^2_{g}(q_3)\Delta^2_{g}(|\mathbf{q}_1-\mathbf{q}_3|)\Delta^2_{g}(|\mathbf{q}_2-\mathbf{q}_3|),
\end{aligned}
\end{equation}
\begin{equation}
\begin{aligned}
    \Delta^{2,\mathrm{N}}_{h_\lambda}(k) = 2^8F_{\mathrm{NL}}^4\int&\frac{\mathrm{d}^3\mathbf{q}_1}{(2\pi)^{3}}\frac{\mathrm{d}^3\mathbf{q}_2}{(2\pi)^{3}}\frac{\mathrm{d}^3\mathbf{q}_3}{(2\pi)^{3}}I(|\mathbf{k}-\mathbf{q_1}|,q_1,\tau)Q_{\lambda}(\mathbf{k},\mathbf{q}_1)I(|\mathbf{k}-\mathbf{q_2}|,q_2,\tau)Q_{\lambda}(\mathbf{k},\mathbf{q}_2)\\
    &\times\Delta_{g}^2(|\mathbf{k}-\mathbf{q}_3|)\Delta^2_{g}(|\mathbf{q}_1-\mathbf{q}_3|)\Delta^2_{g}(|\mathbf{q}_2-\mathbf{q}_3|)\Delta_{g}^2(|\mathbf{q}_1+\mathbf{q}_2-\mathbf{q}_3|).
\end{aligned}
\end{equation}

The non-Gaussian contributions separate into two orders:

\begin{itemize}
    \item \textbf{$\mathcal O(F_{\mathrm{NL}}^2)$}: $\Delta^{2,\mathrm{H}}_{h_\lambda}$, $\Delta^{2,\mathrm{C}}_{h_\lambda}$, and $\Delta^{2,\mathrm{Z}}_{h_\lambda}$,
    \item \textbf{$\mathcal O(F_{\mathrm{NL}}^4)$}: $\Delta^{2,\mathrm{R}}_{h_\lambda}$, $\Delta^{2,\mathrm{P}}_{h_\lambda}$, and $\Delta^{2,\mathrm{N}}_{h_\lambda}$,
\end{itemize}

where the former set contributes to $\Omega_{\rm GW}^{(1)}$ and the latter contributes to $\Omega_{\rm GW}^{(2)}$.

\section{Change of variables and numerical form used in the code}
\label{app:variables}

{To prepare the integrals in Appendix~\ref{app:ng_4_pt_diagram} for numerical evaluation, we use the standard variable transformations} \cite{Adshead:2021hnm}. {These transformations map the diagrammatic integrals to domains suitable for quasi-Monte Carlo evaluation.}

For disconnected terms, we have

\begin{subequations}
    \begin{align}
        v_1 &= \frac{q_1}{k},& u_1 =& \frac{|\mathbf{k}-\mathbf{q}_1|}{k},\\
        v_2 &= \frac{q_2}{q_1},& u_2 =& \frac{|\mathbf{q}_1-\mathbf{q}_2|}{q_1},\\
        v_3 &= \frac{q_3}{|\mathbf{k}-\mathbf{q}_1|},& u_3 =& \frac{|\mathbf{k}-\mathbf{q}_1-\mathbf{q}_3|}{|\mathbf{k}-\mathbf{q}_1|},
    \end{align}
\end{subequations}
{and define}
\begin{subequations}
    \begin{align}
        s_{i} &= u_i - v_i, \\
        t_{i} &= u_i + v_i - 1.
    \end{align}
\end{subequations}
\begingroup
The compact expressions in Appendix~\ref{app:scale_dependent_extension} use
\begin{equation}
a_i\equiv t_i+1=u_i+v_i,
\qquad
b_i\equiv s_i=u_i-v_i,
\label{eq:ab_variable_definition}
\end{equation}
so that
\begin{equation}
u_i=\frac{a_i+b_i}{2},
\qquad
v_i=\frac{a_i-b_i}{2},
\qquad
a_i\in[1,\infty),
\qquad b_i\in[-1,1].
\label{eq:ab_inverse_definition}
\end{equation}
The Jacobian is $du_i\,dv_i=\tfrac12 da_i\,db_i$, and $u_iv_i=(a_i^2-b_i^2)/4$. These identities account for the measure factors $16/(a_i^2-b_i^2)^2$ appearing in Eqs.~\eqref{eq:omega_gw_fnl2} and \eqref{eq:omega_gw_fnl4}.
\endgroup
then using~\eqref{eq:GWspectrum}, {the disconnected terms become}
\begin{align}
    \Omega^{\mathrm{G}}_{\mathrm{GW}}(k) &= \frac{1}{3}\int_{0}^{\infty}\mathrm{d}t\int_{-1}^{1}\mathrm{d}s\overline{J^{2}(u,v,u,v,x\rightarrow \infty)}\frac{\Delta_{g}^2(vk)}{v^2}\frac{\Delta_{g}^2(uk)}{u^2},\label{eq:OmegaGWG}\\
    \Omega^{\mathrm{H}}_{\mathrm{GW}}(k) &= \frac{1}{3}F_{\mathrm{NL}}^2\int_{0}^{\infty}\mathrm{d}t_{1,2}\int_{-1}^{1}\mathrm{d}s_{1,2}\overline{J^{2}(u_1,v_1,u_1,v_1,x\rightarrow \infty)}\nonumber\\
    &\qquad\times v_1^2\frac{\Delta_{g}^2(v_1v_2k)}{(v_1v_2)^2}\frac{\Delta_{g}^2(u_1k)}{u_1^2}\frac{\Delta_{g}^2(v_1u_2k)}{(v_1u_2)^2},\label{eq:OmegaGWH}\\
    \Omega^{\mathrm{R}}_{\mathrm{GW}}(k) &= \frac{1}{12}F_{\mathrm{NL}}^4\int_{0}^{\infty}\mathrm{d}t_{1,2,3}\int_{-1}^{1}\mathrm{d}s_{1,2,3}\overline{J^{2}(u_1,v_1,u_1,v_1,x\rightarrow \infty)}\nonumber\\
    &\qquad\times v_1^2u_1^2\frac{\Delta_{g}^2(v_1v_2k)}{(v_1v_2)^2}\frac{\Delta_{g}^2(v_1u_2k)}{(v_1u_2)^2}\frac{\Delta_{g}^2(u_1v_3k)}{(u_1v_3)^2}\frac{\Delta_{g}^2(u_1u_3k)}{(u_1u_3)^2},\label{eq:OmegaGWR}
\end{align}
where $\overline{J^{2}(u_1,v_1,u_2,v_2,x\rightarrow \infty)}$ is defined in~\eqref{eq:J2}. For the connected terms, we first introduce another three sets of variables $(u_i,v_i)$,
\begin{subequations}
    \begin{align}
        v_i &= \frac{q_i}{k},\\
        u_i &= \frac{|\mathbf{k}-\mathbf{q}_i|}{k},
    \end{align}
\end{subequations}
and next define $\phi_{ij} \equiv \phi_i - \phi_j$.
\begingroup
For clarity, define the polar cosine of $\mathbf q_i$ relative to the external momentum $\mathbf k$ by
\begin{equation}
\mu_i\equiv\widehat{\mathbf k}\cdot\widehat{\mathbf q}_i
=\frac{1+v_i^2-u_i^2}{2v_i}.
\label{eq:mu_i_definition}
\end{equation}
The dimensionless scalar product is then
\begin{equation}
y_{ij}\equiv\frac{\mathbf q_i\cdot\mathbf q_j}{k^2}
=v_iv_j\left[
\mu_i\mu_j+
\sqrt{1-\mu_i^2}\sqrt{1-\mu_j^2}\cos\phi_{ij}
\right].
\label{eq:yij_correct_definition}
\end{equation}
This expression follows directly from the spherical decomposition of the two internal momenta and is symmetric under $i\leftrightarrow j$. It replaces the previous $a_i,b_i$ expression, whose mixed indices were typographical errors. The variables $a_i,b_i$ remain related to $u_i,v_i$ by Eqs.~\eqref{eq:ab_variable_definition} and \eqref{eq:ab_inverse_definition}.
\endgroup
After defining the following two quantities for later convenience,
\begin{align}
    \omega_{ij} &\equiv \frac{|\mathbf{q}_i-\mathbf{q}_j|}{k} =\sqrt{v_i^2+v_j^2-2y_{ij}},\\
    \omega_{123} &\equiv \frac{|\mathbf{q}_1+\mathbf{q}_2-\mathbf{q}_3|}{k} =\sqrt{v_1^2+v_2^2+v_3^2+2y_{12}-2y_{13}-2y_{23}},
\end{align}
{the connected terms are}
\begin{align}
    \Omega^{\mathrm{C}}_{\mathrm{GW}}(k) =& \frac{1}{3\pi}F_{\mathrm{NL}}^2\int_{0}^{\infty}\mathrm{d}t_{1,2}\int_{-1}^{1}\mathrm{d}s_{1,2}\int_{0}^{2\pi}\mathrm{d}\phi_{12}\,\cos2\phi_{12}\overline{J^{2}(u_1,v_1,u_2,v_2,x\rightarrow \infty)}\nonumber\\
    &\times u_1v_1u_2v_2\frac{\Delta_{g}^2(v_2k)}{v_2^3}\frac{\Delta_{g}^2(u_2k)}{u_2^3}\frac{\Delta_{g}^2(\omega_{12}k)}{\omega_{12}^3} \label{2.43} \\
    \Omega^{\mathrm{Z}}_{\mathrm{GW}}(k) =& \frac{1}{3\pi}F_{\mathrm{NL}}^2\int_{0}^{\infty}\mathrm{d}t_{1,2}\int_{-1}^{1}\mathrm{d}s_{1,2}\int_{0}^{2\pi}\mathrm{d}\phi_{12}\,\cos2\phi_{12}\overline{J^{2}(u_1,v_1,u_2,v_2,x\rightarrow \infty)}\nonumber\\
    &\times u_1v_1u_2v_2\frac{\Delta_{g}^2(v_2k)}{v_2^3}\frac{\Delta_{g}^2(u_1k)}{u_1^3}\frac{\Delta_{g}^2(\omega_{12}k)}{\omega_{12}^3}\label{2.44} \\
    \Omega^{\mathrm{P}}_{\mathrm{GW}}(k) =& \frac{1}{12\pi^2}F_{\mathrm{NL}}^4\int_{0}^{\infty}\mathrm{d}t_{1,2,3}\int_{-1}^{1}\mathrm{d}s_{1,2,3}\int_{0}^{2\pi}\mathrm{d}\phi_{12}\mathrm{d}\phi_{23}\,\cos2\phi_{12}\overline{J^{2}(u_1,v_1,u_2,v_2,x\rightarrow \infty)}\nonumber\\
    &\times u_1v_1u_2v_2u_3v_3\frac{\Delta_{g}^2(v_3k)}{v_3^3}\frac{\Delta_{g}^2(u_3k)}{u_3^3}\frac{\Delta_{g}^2(\omega_{13}k)}{\omega_{13}^3}\frac{\Delta_{g}^2(\omega_{23}k)}{\omega_{23}^3}\label{2.45} \\
    \Omega^{\mathrm{N}}_{\mathrm{GW}}(k) =& \frac{1}{24\pi^2}F_{\mathrm{NL}}^4\int_{0}^{\infty}\mathrm{d}t_{1,2,3}\int_{-1}^{1}\mathrm{d}s_{1,2,3}\int_{0}^{2\pi}\mathrm{d}\phi_{12}\mathrm{d}\phi_{23}\,\cos2\phi_{12}\overline{J^{2}(u_1,v_1,u_2,v_2,x\rightarrow \infty)}\nonumber\\
    &\times u_1v_1u_2v_2u_3v_3\frac{\Delta_{g}^2(v_3k)}{v_3^3}\frac{\Delta_{g}^2(\omega_{13}k)}{\omega_{13}^3}\frac{\Delta_{g}^2(\omega_{23}k)}{\omega_{23}^3}\frac{\Delta_{g}^2(\omega_{123}k)}{\omega_{123}^3}.\label{2.46}
\end{align}

These are the variables used in the numerical implementation of the diagrammatic contributions.

\section{Extension to scale-dependent non-Gaussianity}
\label{app:scale_dependent_extension}
This appendix defines the kernel used numerically:
\begin{equation}
\zeta(\mathbf{k})=\zeta_g(\mathbf{k})+\bar f_{\rm NL}\int\frac{d^3q}{(2\pi)^{3/2}}f(q)f(\lVert\mathbf{k}-\mathbf q\rVert)\zeta_g(\mathbf q)\zeta_g(\mathbf{k}-\mathbf q).
\label{eq:appD_internal_kernel}
\end{equation}
\begingroup
Its contractions generate the weighted dimensionful and dimensionless spectra
\begin{align}\label{eq:ps_redefined}
P_g^S(k)&=f^2(k)P_g(k),
&
\mathcal P^S(k)&=f^2(k)\mathcal P_g(k),
\end{align}
with $\mathcal P_g(k)=k^3P_g(k)/(2\pi^2)$. For the dimensionful physical spectrum, the one-loop result is
\begin{align}\label{eq:ps_scale_dependent_loop}
P_\zeta(k)&=P_g(k)+2\bar f_{\rm NL}^2
\int\frac{d^3\mathbf q}{(2\pi)^3}
P_g^S(q)P_g^S(\lVert\mathbf k-\mathbf q\rVert),\\
&=P_g(k)+P_{ng}(k).
\end{align}
The dimensionless spectrum used elsewhere is $\mathcal P_\zeta(k)=k^3P_\zeta(k)/(2\pi^2)$. Thus Eqs.~\eqref{eq:ps_redefined} and \eqref{eq:ps_scale_dependent_loop} do not mix dimensionful and dimensionless power spectra.
\endgroup
The weighted spectrum is a bookkeeping device. Every $\mathcal P^S$ factor corresponds to a Gaussian momentum weighted by Eq.~\eqref{eq:appD_internal_kernel}. These equations, not an external-output prescription, define all numerical figures.
Using the compact notation from \cite{Ellis:2023oxs}, \footnote{See the compact expressions for the merged non-Gaussian contributions in that reference.}
we can merge the connected and disconnected contributions to form individual equations for $\mathcal O(\bar f_{\rm NL}^2)$

\begin{align}\label{eq:omega_gw_fnl2}
 \Omega_{\mathrm{GW}}^{(1)}(k) & =\frac{2 \bar f_{\rm NL}^2}{3} \prod_{i=1}^2\left[\int_1^{\infty} \mathrm{d} a_i \int_{-1}^1 \mathrm{~d} b_i \frac{16}{\left(a_i^2-b_i^2\right)^2}\right]\left\{\frac{1}{2} \overline{J^2\left(a_1, b_1\right)} \mathcal P^S\left(v_1 v_2 k\right) \mathcal P_g\left(u_1 k\right) \mathcal P^S\left(v_1 u_2 k\right)\right. \nonumber \\ & \left.+\int_0^{2 \pi} \frac{\mathrm{~d} \varphi_{12}}{2 \pi} \cos 2 \varphi_{12} \overline{J\left(a_1, b_1\right) J\left(a_2, b_2\right)} \frac{\mathcal P^S\left(v_2 k\right)}{v_2^3} \frac{\mathcal P^S\left(d_{12} k\right)}{d_{12}^3}\left[\frac{\mathcal P_g\left(u_2 k\right)}{u_2^3}+\frac{\mathcal P_g\left(u_1 k\right)}{u_1^3}\right]\right\}
\end{align}
and the $\mathcal O(\bar f_{\rm NL}^4)$ contribution is

\begin{align} \label{eq:omega_gw_fnl4}
\Omega_{\mathrm{GW}}^{(2)}(k) & =\frac{\bar f_{\rm NL}^4}{6} \prod_{i=1}^3\left[\int_1^{\infty} \mathrm{d} a_i \int_{-1}^1 \mathrm{~d} b_i \frac{16}{\left(a_i^2-b_i^2\right)^2}\right]\left\{\frac{1}{2} \overline{J^2\left(a_1, b_1\right)} \mathcal P^S\left(v_1 v_2 k\right) \mathcal P^S\left(v_1 u_2 k\right) \mathcal P^S\left(u_1 v_3 k\right) \mathcal P^S\left(u_1 u_3 k\right)\right. \nonumber \\ & \left.+\int_0^{2 \pi} \frac{\mathrm{~d} \varphi_{12}}{2 \pi} \frac{\mathrm{~d} \varphi_{23}}{2 \pi} \cos \left(2 \varphi_{12}\right) \overline{J\left(a_1, b_1\right) J\left(a_2, b_2\right)} \frac{\mathcal P^S\left(u_3 k\right)}{u_3^3} \frac{\mathcal P^S\left(d_{13} k\right)}{d_{13}^3} \frac{\mathcal P^S\left(d_{23} k\right)}{d_{23}^3}\left[\frac{\mathcal P^S\left(v_3 k\right)}{v_3^3}+\frac{\mathcal P^S\left(w_{123} k\right)}{w_{123}^3}\right]\right\}
\end{align}

where the integrated, time-averaged transfer functions encoding SIGW propagation are

\begin{align}\label{eq:int_transfer_fn}
& \overline{J\left(b_i, a_i\right) J\left(a_j, b_j\right)}=\frac{9}{8} \frac{\left(a_i^2-1\right)\left(b_j^2-1\right)\left(1-b_i^2\right)\left(1-a_j^2\right)\left(a_i^2+b_i^2-6\right)\left(b_j^2+a_j^2-6\right)}{\left(a_i^2-b_i^2\right)^3\left(b_j^2-a_j^2\right)^3} \nonumber \\ & {\left[\left(\left(a_i^2+b_i^2-6\right) \ln \left|\frac{a_i^2-3}{3-b_i^2}\right|-2\left(a_i^2-b_i^2\right)\right)\left(\left(b_j^2+a_j^2-6\right) \ln \left|\frac{b_j^2-3}{3-a_j^2}\right|-2\left(b_j^2-a_j^2\right)\right)\right.} \nonumber \\ & \left.+\pi^2 \Theta\left(a_i-\sqrt{3}\right) \Theta\left(b_j-\sqrt{3}\right)\left(a_i^2+b_i^2-6\right)\left(b_j^2+a_j^2-6\right)\right]~.
\end{align}


\section{Exact monochromatic limit of the internal-leg separable kernel}\label{app:exact_monochromatic_limit}

This appendix checks Appendix~\ref{app:scale_dependent_extension} and distinguishes the exact delta source from the finite-width numerical regularization. For the exact monochromatic source,
\begin{equation}
  \Pg(k)=A_{\cal R}\,\delta\!\left[\ln\left(\frac{k}{k_\star}\right)\right] .
  \label{eq:mono_source_snippet}
\end{equation}
The derivation assumes the internal-leg separable kernel,
\begin{equation}
  \zeta({\bf k})=\zeta_g({\bf k})+\fbar
  \int \frac{d^3q}{(2\pi)^{3/2}}
  f(q)f(|{\bf k}-\mathbf{q}|)
  \zeta_g({\bf q})\zeta_g({\bf k}-\mathbf{q}) ,
  \label{eq:internal_kernel_snippet}
\end{equation}
and therefore introduces the weighted Gaussian spectrum
\begin{equation}
  \PS(k)\equiv f^2(k)\Pg(k).
  \label{eq:weighted_spectrum_snippet}
\end{equation}
This is the crucial simplification of the exact monochromatic limit: because every weighted Gaussian leg is forced onto the support of the delta function, every occurrence of \(f(k)\) is evaluated at \(k=k_\star\). Hence the running does not generate a new normalized spectral shape in the exact delta-source limit; it only multiplies individual non-Gaussian sectors by constant factors. Any residual shape dependence in a numerical ``monochromatic'' plot must therefore come from finite-width regularization, a non-pivot-normalized template, an output-momentum kernel, or numerical implementation details rather than from the exact internal-leg delta limit itself.

\subsection*{Collapse of the weighted spectrum}

Using Eq.~\eqref{eq:mono_source_snippet}, the weighted spectrum becomes
\begin{align}
  \PS(k)&=f^2(k)A_{\cal R}\delta\!\left[\ln\left(\frac{k}{k_\star}\right)\right] \\
  &=w_\star^2 A_{\cal R}\delta\!\left[\ln\left(\frac{k}{k_\star}\right)\right]
  =w_\star^2\Pg(k),
  \label{eq:weighted_delta_collapse}
\end{align}
where
\begin{equation}
  w_\star\equiv f(k_\star).
\end{equation}
The equality follows from the distributional identity \(g(k)\delta[\ln(k/k_\star)]=g(k_\star)\delta[\ln(k/k_\star)]\). Therefore each weighted scalar leg contributes one constant factor \(w_\star^2\).

\subsection*{Sector-by-sector scaling of the induced GW spectrum}

Write the induced GW spectrum as
\begin{equation}
  \Om(k)=\Omega_{\rm GW}^{(0)}(k)+\Omega_{\rm GW}^{(1)}(k)+\Omega_{\rm GW}^{(2)}(k),
\end{equation}
where \(\Omega_{\rm GW}^{(0)}\) is the Gaussian contribution, \(\Omega_{\rm GW}^{(1)}\propto \fbar^2\) denotes the reducible and connected second-order-in-\(\fbar\) sector, and \(\Omega_{\rm GW}^{(2)}\propto \fbar^4\) denotes the fourth-order-in-\(\fbar\) sector within the second-order tensor source. For an exact monochromatic source one may factor out the amplitude dependence as
\begin{align}
  \Omega_{{\rm GW},\delta}^{(0)}(k)&=A_{\cal R}^2\,{
  \cal K}_0(x),
  \label{eq:mono_gaussian_shape}\\
  \Omega_{{\rm GW},\delta}^{(1)}(k)&=\fbar^2 A_{\cal R}^3 w_\star^4\,{
  \cal K}_1(x),
  \label{eq:mono_second_shape}\\
  \Omega_{{\rm GW},\delta}^{(2)}(k)&=\fbar^4 A_{\cal R}^4 w_\star^8\,{
  \cal K}_2(x),
  \label{eq:mono_fourth_shape}
\end{align}
with \(x\equiv k/k_\star\). The functions \({\cal K}_0,{\cal K}_1,{\cal K}_2\) are the dimensionless monochromatic transfer-shape functions associated with the Gaussian, \(\fbar^2\), and \(\fbar^4\) sectors. They contain the radiation-era transfer kernels, angular factors, triangle constraints, and the support restriction inherited from the delta source. Importantly, \({\cal K}_i(x)\) are independent of the running template in the exact internal-leg monochromatic limit. The running enters only through \(w_\star^4\) and \(w_\star^8\).

It is useful to define the hierarchy parameters
\begin{align}
  R_1(x)&\equiv \frac{\Omega_{{\rm GW},\delta}^{(1)}(k)}{\Omega_{{\rm GW},\delta}^{(0)}(k)}
  = A_{\cal R}\fbar^2 w_\star^4\,\frac{{\cal K}_1(x)}{{\cal K}_0(x)},
  \label{eq:R1_mono}\\
  R_2(x)&\equiv \frac{\Omega_{{\rm GW},\delta}^{(2)}(k)}{\Omega_{{\rm GW},\delta}^{(1)}(k)}
  = A_{\cal R}\fbar^2 w_\star^4\,\frac{{\cal K}_2(x)}{{\cal K}_1(x)}.
  \label{eq:R2_mono}
\end{align}
Thus, up to order-one shape-kernel ratios, the effective expansion parameter is
\begin{equation}
  \epsilon_{\rm NG}^{\delta}\simeq A_{\cal R}\fbar^2 w_\star^4 .
  \label{eq:epsilon_delta}
\end{equation}
This is the exact monochromatic specialization of the more general weighted-moment estimate.

\subsection*{Power-law running}

For the power-law template
\begin{equation}
  f_{\rm PL}(k)=\left(\frac{k}{k_\star}\right)^{n_f},
\end{equation}
one has
\begin{equation}
  w_\star=f_{\rm PL}(k_\star)=1
  \quad \hbox{for all } n_f .
\end{equation}
Therefore
\begin{align}
  \Omega_{{\rm GW},\delta}^{(0)}(k)&=A_{\cal R}^2{\cal K}_0(x),\\
  \Omega_{{\rm GW},\delta}^{(1)}(k)&=\fbar^2A_{\cal R}^3{\cal K}_1(x),\\
  \Omega_{{\rm GW},\delta}^{(2)}(k)&=\fbar^4A_{\cal R}^4{\cal K}_2(x),
\end{align}
which are exactly the constant-kernel results sector by sector. Hence a pivot-normalized power-law running template cannot change the normalized SIGW shape for an exact monochromatic source in the internal-leg separable model. In particular, for the normalized shape ratio
\begin{equation}
  {\cal S}(k)
  \equiv
  \frac{\Omega_{{\rm GW},{\rm run}}(k)/\Omega_{{\rm GW},{\rm const}}(k)}
       {\Omega_{{\rm GW},{\rm run}}(k_{\rm pk})/\Omega_{{\rm GW},{\rm const}}(k_{\rm pk})},
\end{equation}
one obtains
\begin{equation}
  {\cal S}_{\rm PL}^{\delta}(k)=1
  \quad \hbox{for the exact delta source.}
  \label{eq:S_powerlaw_delta}
\end{equation}
This result is independent of \(n_f\). The running exponent becomes visible only once the scalar source has a finite width or when a different kernel assignment, such as output-momentum running, is used.
\subsection*{Tanh running}
For the UV-matched tanh template used in the numerical analysis,
\begin{equation}
f_{\rm tanh}(k)=\tanh\!\left[(k/k_\star)^{n_f}\right],
\end{equation}
one has $w_\star=f_{\rm tanh}(k_\star)=\tanh(1)$ for every $n_f$. The exact delta source fixes every weighted Gaussian leg at the pivot, and hence
\begin{align}
\Omega_{{\rm GW},\delta}^{(0)}(k)&=A_{\cal R}^2{\cal K}_0(x),\\
\Omega_{{\rm GW},\delta}^{(1)}(k)&=\fbar^2A_{\cal R}^3\tanh^4(1){\cal K}_1(x),\\
\Omega_{{\rm GW},\delta}^{(2)}(k)&=\fbar^4A_{\cal R}^4\tanh^8(1){\cal K}_2(x).
\end{align}
Thus the UV-matched tanh weight leaves each individual sector shape unchanged but suppresses the two non-Gaussian sectors by different constants. The normalized total shape need not equal the constant-kernel total when more than one sector contributes appreciably, because the relative sector weights have changed. At high momentum $f_{\rm tanh}\to1$, so both the logarithmic running and the absolute leg-weight mismatch vanish.
\subsection*{Benchmark hierarchy estimates}
The following values illustrate the exact-delta hierarchy parameter in Eq.~\eqref{eq:epsilon_delta}. For the power-law template, $w_\star^4=1$, whereas for the UV-matched tanh template $w_\star^4=\tanh^4(1)\simeq0.33643$.
\begin{table}[H]
\centering
\begin{tabular}{ccccc}
\toprule
$A_{\cal R}$ & $\fbar$ & template & $w_\star^4$ & $\epsilon_{\rm NG}^{\delta}\simeq A_{\cal R}\fbar^2w_\star^4$ \\
\midrule
$10^{-5}$ & 5 & power law & 1 & $2.5\times10^{-4}$ \\
$10^{-5}$ & 5 & UV-matched tanh & $0.33643$ & $8.41\times10^{-5}$ \\
$10^{-4}$ & 5 & power law & 1 & $2.5\times10^{-3}$ \\
$10^{-4}$ & 5 & UV-matched tanh & $0.33643$ & $8.41\times10^{-4}$ \\
$10^{-3}$ & 15 & power law & 1 & $2.25\times10^{-1}$ \\
$10^{-3}$ & 15 & UV-matched tanh & $0.33643$ & $7.57\times10^{-2}$ \\
$10^{-2}$ & 5 & power law & 1 & $2.5\times10^{-1}$ \\
$10^{-2}$ & 5 & UV-matched tanh & $0.33643$ & $8.41\times10^{-2}$ \\
$10^{-2}$ & 10 & power law & 1 & $1$ \\
$10^{-2}$ & 10 & UV-matched tanh & $0.33643$ & $0.33643$ \\
\bottomrule
\end{tabular}
\caption{Exact-delta hierarchy estimates for the running templates. For the finite-width regulator, the relevant diagnostic is $\epsilon_{\rm NG}(k)=A_{\cal R}\fbar^2f_{\rm eff}^4(k)$ evaluated from the weighted convolution moments.}
\label{tab:exact_delta_hierarchy_benchmarks}
\end{table}
The plotted narrow-source benchmark is not necessarily in the extremely small-hierarchy regime. When $\epsilon_{\rm NG}(k)$ approaches unity, the retained fourth-order sector becomes comparable to the second-order sector, while confidence in truncating the primordial expansion decreases. Retaining the $\mathcal O(\fbar^4)$ terms does not by itself establish perturbative control.
\subsection*{Finite-width correction and interpretation}

The results above are exact only for the distributional source in Eq.~\eqref{eq:mono_source_snippet}. If the delta function is implemented numerically by a finite-width regularization, the running exponent can re-enter through moments of the regulated distribution. For example, replacing the delta by a log-normal of width \(\sigma\), with \(x=\ln(k/k_\star)\), gives for one weighted scalar leg under power-law running
\begin{equation}
  f_{\rm PL}^2(k)\Pg(k)
  \propto
  \exp\!\left(2n_fx-\frac{x^2}{2\sigma^2}\right)
  =
  \exp\!\left[-\frac{(x-2n_f\sigma^2)^2}{2\sigma^2}\right]
  \exp(2n_f^2\sigma^2).
  \label{eq:finite_width_shift}
\end{equation}
Thus a finite-width source has its weighted support shifted by \(\Delta x=2n_f\sigma^2\) and its weighted integral enhanced by \(\exp(2n_f^2\sigma^2)\) per weighted leg. This correction vanishes as \(\sigma\to0\), recovering Eq.~\eqref{eq:S_powerlaw_delta}. For the UV-matched tanh template, an analogous expansion around $k=k_\star$, with $x=\ln(k/k_\star)$, gives
\begin{equation}
\ln f_{\rm tanh}(k)=\ln[\tanh(1)]+n_f\frac{\operatorname{sech}^2(1)}{\tanh(1)}x+\mathcal O(x^2).
\end{equation}
The constant term records the suppression already present at the pivot, while the linear term controls the finite-width redistribution across the source support. The redistribution vanishes in the exact zero-width limit, although the constant sector suppressions $\tanh^4(1)$ and $\tanh^8(1)$ remain.

\section{Numerical implementation and stability checks}
\label{app:numerical_stability}
{The integrals in Appendix~\ref{app:scale_dependent_extension} were evaluated by quasi-Monte Carlo (QMC) integration using scrambled Sobol sequences.} {The production sample counts were} $N=2^{18}$ for $\Omega^{(0)}$, and $N=2^{20}$ for both $\Omega^{(1,2)}$. {To assess convergence, the narrow-source and double-broken-power-law benchmarks are evaluated} at $N=2^{15},\; 2^{17},\;2^{18},\;2^{19},\;2^{20},\; 2^{21}$ with six independent Sobol {realizations} (seeds $1000$-$1005$, distinct
from the single seed $42$ used throughout the production integration) for each $N$. Both spectra are analyzed with both the log and tanh templates. {The scatter across realizations at fixed $N$ provides the uncertainty estimate.} The results are summarized in Fig. \ref{fig:convergence}{ and Tables~}\ref{tab:conv-log}, \ref{tab:conv-tanh}.\newline
\begin{figure}[H]
    \centering
    \includegraphics[width=1\linewidth]{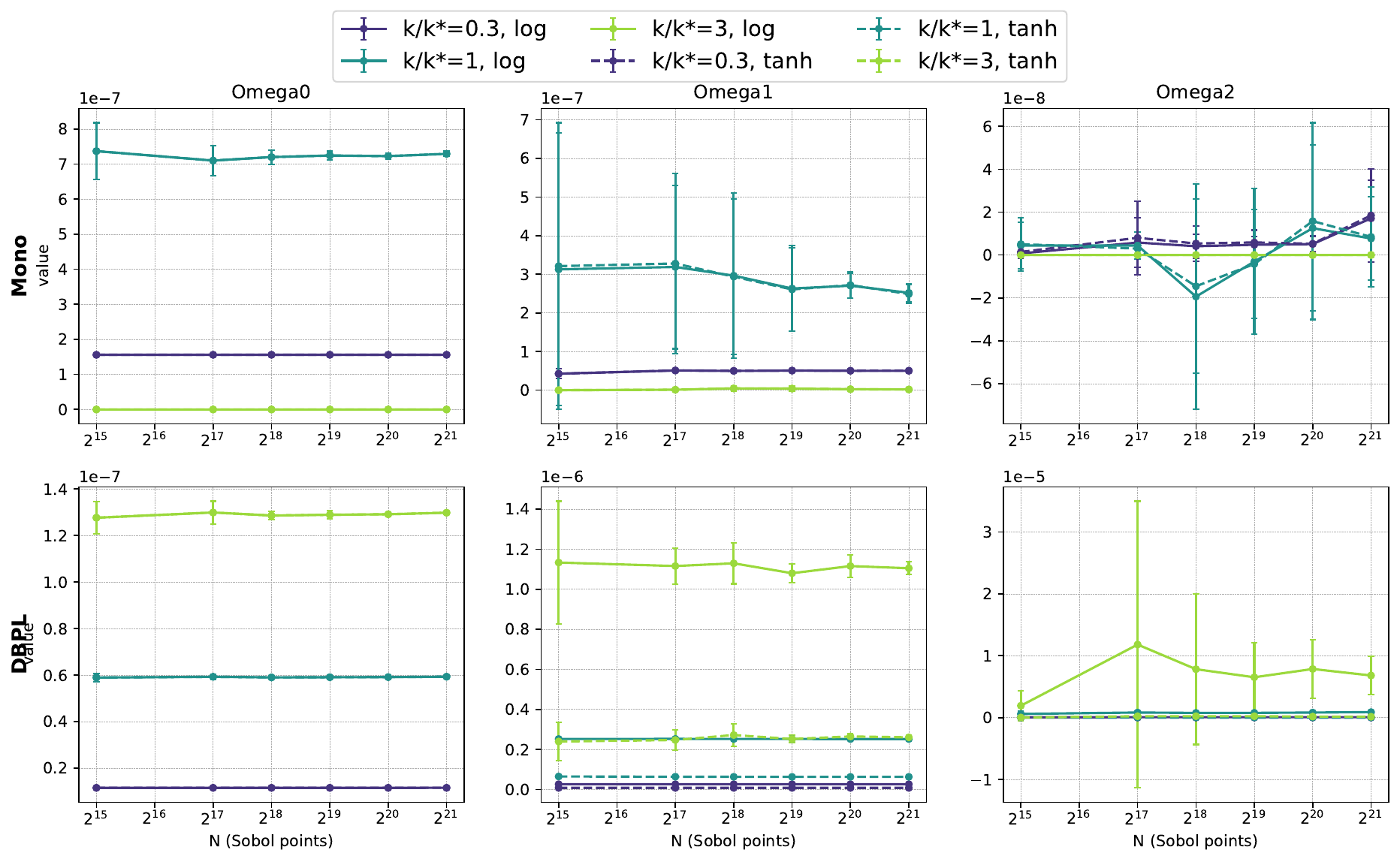}
    \caption{{\it Convergence of the Gaussian, $\mathcal O(\bar f_{\rm NL}^{2})$, and $\mathcal O(\bar f_{\rm NL}^{4})$ SIGW sectors under increasing randomized Sobol sample count. Columns show the narrow regularized source and the double-broken-power-law benchmark; rows show $\Omega_{\rm GW}^{(0)}$, $\Omega_{\rm GW}^{(1)}$, and $\Omega_{\rm GW}^{(2)}$. Colors identify the sampled tensor wavenumbers $k/k_\star=0.3$, $1$, and $3$. Solid curves denote power-law running and dashed curves denote the UV-matched tanh weight. The horizontal axis gives $N_{\rm QMC}=2^{15},2^{17},2^{18},2^{19},2^{20},2^{21}$. At each sample count, the central marker is the mean across six independently scrambled Sobol realizations and the error bar shows the realization-to-realization standard deviation. The overlapping zeroth-order curves provide a template-independent Gaussian implementation check. The broad DBPL source converges rapidly, whereas the narrow source shows slower convergence at $\mathcal O(\bar f_{\rm NL}^{2})$ and an unresolved $\mathcal O(\bar f_{\rm NL}^{4})$ noise floor over the tested range. Quantitative statements are restricted to intervals for which both the scramble scatter and the sample-doubling drift are small compared with the plotted signal.}}
    \label{fig:convergence}
\end{figure}
{
The zeroth-order power-law and {UV-matched tanh} curves overlap, as expected because $\Omega_{\rm GW}^{(0)}$ contains only Gaussian statistics. For the narrow-source benchmark, $\Omega_{\rm GW}^{(0)}$ converges across the tested range. The scatter in $\Omega_{\rm GW}^{(1)}$ decreases with increasing $N$, although the central estimate retains visible sample-count dependence at selected wavenumbers. The narrow-source $\Omega_{\rm GW}^{(2)}$ sector remains unresolved within the tested range: at $k=k_\star$ the central estimate changes sign between (positive at $N=2^{15},2^{17}$; negative at $N=2^{19}$; positive again at $N=2^{21}$), and the scatter remains large for both running templates.
}
For the DBPL benchmark, $\Omega^{(0)}_{\rm GW}$ and $\Omega^{(1)}_{\rm GW}$ converge to sub-percent scatter already at the smallest tested $N$, for both running templates, and show no further drift with increasing $N$. $\Omega^{(2)}_{\rm GW}$ converges comparably well at $k=0.3\,k_\star$ and $k=k_\star$; however at $k=3\,k_\star$, scatter is substantial at intermediate $N$ for the log running template before decreasing
markedly by $N=2^{21}$. For the UV-matched tanh running template at this same point, the central estimate continues to drift by a comparable amount
even at $N=2^{21}$ and should not be treated as converged.
\newline
Overall $\Omega^{(2)}_{\rm GW}$ does not converge for narrow (near-monochromatic) source spectra within our tested range of sample sizes: relative scatter exceeds $100\%$ at every tested $N$, and the central value changes sign rather than stabilising. For broad spectra, $\Omega^{(2)}_{\rm GW}$ converges cleanly.

\begin{table}[H]
\centering
\resizebox{\linewidth}{!}{%
\begin{tabular}{lcc}
\hline\hline
{\textbf{Diagnostic (power-law running)}} & \textbf{Narrow (Mono)} & \textbf{Multi-slope (DBPL)} \\
\hline
Median rel.\ scatter, $\Omega^{(0)}$ & $0.01\%$ & $0.25\%$ \\
Maximum rel.\ scatter, $\Omega^{(0)}$ & $1.03\%$ & $0.45\%$ \\
Rel.\ scatter at production $N=2^{18}$, $\Omega^{(0)}$ & $1.47\%$ / $2.81\%$ ($x=1$) & $0.53\%$ / $1.30\%$ ($x=3$) \\
Median rel.\ scatter, $\Omega^{(1)}$ & $9.12\%$ & $0.29\%$ \\
Maximum rel.\ scatter, $\Omega^{(1)}$ & $76.59\%$ ($x=3$) & $2.83\%$ ($x=3$) \\
Rel.\ scatter at production $N=2^{20}$, $\Omega^{(1)}$ & $11.85\%$ / $130.48\%$\textsuperscript{b} ($x=3$) & $0.51\%$ / $5.03\%$ ($x=3$) \\
Median rel.\ scatter, $\Omega^{(2)}$ & $167.53\%$\textsuperscript{a} & $12.51\%$ \\
Maximum rel.\ scatter, $\Omega^{(2)}$ & $253.84\%$ ($x=1$) & $45.11\%$ ($x=3$) \\
Rel.\ scatter at production $N=2^{20}$, $\Omega^{(2)}$ & $224.48\%$\textsuperscript{a} / $308.68\%$\textsuperscript{a,b} ($x=1$) & $9.93\%$ / $60.63\%$ ($x=3$) \\
Stable interval ($x=k/k_\star$) & $x\in[0.3,3]$ $\Omega^{(0)}$; $\Omega^{(1)}$ stable only at $N\gtrsim2^{21}$\textsuperscript{b} & $x\in[0.3,1]$ $\Omega^{(2)}$ \\
\hline\hline
\end{tabular}%
}
\caption{{Numerical stability results for the power-law leg weight.} Values are relative scatter (median / maximum across $x=k/k_\star=0.3,1,3$) over $N_{\rm scr}=6$ independently-scrambled Sobol realizations, reported both at the largest tested $N=2^{21}$ and at the actual production sample count used for each sector ($N=2^{18}$ for $\Omega^{(0)}$, $N=2^{20}$ for $\Omega^{(1)},\Omega^{(2)}$). $^a$Large relative scatter reflects a numerical noise floor rather than a physical instability. $^b$Flagged sign-unstable: $|{\rm mean}|<{\rm std}$, i.e.\ the estimate is consistent with zero and the quoted percentage should not be read as a meaningful relative error.}
\label{tab:conv-log}
\end{table}

\begin{table}[H]
\centering
\resizebox{\linewidth}{!}{%
\begin{tabular}{lcc}
\hline\hline
\textbf{Diagnostic (UV-matched tanh running)} & \textbf{Narrow (Mono)} & \textbf{Multi-slope (DBPL)} \\
\hline
Median rel.\ scatter, $\Omega^{(0)}$ & $0.01\%$ & $0.25\%$ \\
Maximum rel.\ scatter, $\Omega^{(0)}$ & $1.03\%$ & $0.45\%$ \\
Rel.\ scatter at production $N=2^{18}$, $\Omega^{(0)}$ & $1.47\%$ / $2.81\%$ ($x=1$) & $0.53\%$ / $1.30\%$ ($x=3$) \\
Median rel.\ scatter, $\Omega^{(1)}$ & $9.57\%$ & $0.28\%$ \\
Maximum rel.\ scatter, $\Omega^{(1)}$ & $87.01\%$ ($x=3$) & $4.05\%$ ($x=3$) \\
Rel.\ scatter at production $N=2^{20}$, $\Omega^{(1)}$ & $12.41\%$ / $140.89\%$\textsuperscript{b} ($x=3$) & $0.57\%$ / $4.72\%$ ($x=3$) \\
Median rel.\ scatter, $\Omega^{(2)}$ & $191.87\%$\textsuperscript{a} & $10.66\%$ \\
Maximum rel.\ scatter, $\Omega^{(2)}$ & $276.21\%$ ($x=1$) & $38.20\%$ ($x=3$) \\
Rel.\ scatter at production $N=2^{20}$, $\Omega^{(2)}$ & $236.83\%$\textsuperscript{a} / $290.25\%$\textsuperscript{a,b} ($x=1$) & $14.29\%$ / $50.38\%$ ($x=3$) \\
Stable interval ($x=k/k_\star$) & $x\in[0.3,3]$ $\Omega^{(0)}$; $\Omega^{(1)}$ stable only at $N\gtrsim2^{21}$\textsuperscript{b} & $x\in[0.3,1]$ $\Omega^{(2)}$ \\
\hline\hline
\end{tabular}%
}
\caption{{Numerical stability results for the UV-matched tanh leg weight $f_{\rm tanh}(k)=\tanh[(k/k_\star)^{n_f}]$.} Values are relative scatter (median / maximum across $x=k/k_\star=0.3,1,3$) over $N_{\rm scr}=6$ independently-scrambled Sobol realizations, reported both at the largest tested $N=2^{21}$ and at the actual production sample count used for each sector ($N=2^{18}$ for $\Omega^{(0)}$, $N=2^{20}$ for $\Omega^{(1)},\Omega^{(2)}$). The protocol is otherwise the same as in Table~\ref{tab:conv-log}. $^a$Large relative scatter reflects a numerical noise floor rather than a physical instability; the affected sector is not used for quantitative interpretation. $^b$Flagged sign-unstable: $|{\rm mean}|<{\rm std}$.}
\label{tab:conv-tanh}
\end{table}

\subsection{Reproducibility information}
\label{app:reproducibility}

{The calculations used \texttt{JAX 0.5.3} and \texttt{scipy.stats.qmc.Sobol} from \texttt{SciPy 1.15.2} on an eight-core Apple M3 CPU. The production integrations used \texttt{seed=42}; the convergence study used the independent scramble seeds \texttt{1000-1005}.}

\medskip

\newpage

\bibliographystyle{JHEP}

\bibliography{ref_SIGW}

\end{document}